\documentclass[aps,prc,preprint, amsmath,showpacs,superscriptaddress,nofootinbib]{revtex4-1}
\usepackage{graphicx,epsfig,psfrag}
\newcommand{\beqy}{\begin{eqnarray}}
\newcommand{\eeqy}{\end{eqnarray}}
\newcommand{\bmlet}{\begin{subequations}}
\newcommand{\emlet}{\end{subequations}}
\usepackage{graphicx}
\usepackage{dcolumn}
\usepackage{hyperref}
\usepackage[colorinlistoftodos]{todonotes}

\def\apj{ApJ}  
\def\apjl{ApJ} 
\def\prl{Phys.~Rev.~Lett.} 
\def\pr{Phys.~Rev.} 
\def\prc{Phys.~Rev.~C} 
\def\prx{Phys.~Rev.~X} 
\def\aap{A\&A}   
\def\nphysa{Nucl.~Phys.~A}   
\def\mnras{MNRAS}             
\def\cpc{Chin.~Phys.~C}
\def\jcap{J. Cosmol. Astropart. P.} 

\newcommand{\ym}{\widetilde{y}}

\begin{document}

\title{Role of dense matter in tidal deformations of inspiralling neutron stars and in gravitational waveform with unified equations of state}

\author{L. Perot}
\affiliation{Institute of Astronomy and Astrophysics, Universit\'e Libre de Bruxelles, CP 226, Boulevard du Triomphe, B-1050 Brussels, Belgium}
\author{N. Chamel}
\affiliation{Institute of Astronomy and Astrophysics, Universit\'e Libre de Bruxelles, CP 226, Boulevard du Triomphe, B-1050 Brussels, Belgium}

\begin{abstract}
The role of dense-matter properties in the tidal deformability of a cold nonaccreted neutron star is further investigated. Using the set of Brussels-Montreal unified equations of state, 
we have computed the gravitoelectric Love numbers $k_\ell$ and the gravitomagnetic Love numbers $j_\ell$ up to $\ell=5$. Their relative importance and their sensitivity to the symmetry energy and the neutron-matter stiffness are numerically assessed. 
Their impact on the phase of the gravitational-wave signal emitted by binary neutron star inspirals is also discussed. 
\end{abstract}

\maketitle

\section{Introduction}
\label{sec:intro}

On-going and future observational gravitational-wave campaigns open new prospects for exploring the properties of dense matter through the coalescence of two neutron stars (NSs) or of a NS and a black hole~\cite{chaves2019}, as might have been recently observed~\cite{ligo2020b}. With the increasing sensitivity of the gravitational-wave interferometers, hundreds of such events are expected to be detected in the next few years~\cite{abbott2018}. 

The deformations of inspiralling NSs induced  by their mutual gravitational attractions leave a characteristic observable imprint on the gravitational waveform as compared to binary  black holes. The detection of the  gravitational-wave signal GW170817~\cite{ligo2017inspiral} has allowed to measure these tidal effects for the first time, thus shedding some light on the dense  matter constituting the interior of NSs. The initial analyses~\cite{ligo2017inspiral,de2018} made use of the analytic  gravitational-wave template `TaylorF2'~\cite{sathyaprakash1991,mikoczi2005,buonanno2009,arun2009,vines2011,bohe2013}, which is entirely based on the post-Newtonian (PN) theory~\cite{blanchet2014,blanchet2019}. This waveform model consists of a binary black-hole (BBH) baseline at 3.5 PN order to which are added finite-size corrections up to 6 PN order. These corrections depend on the dimensionless tidal deformability parameter of each star defined by 
\begin{equation}
    \label{eq:Lambda}
    \Lambda_{2}=\dfrac{2}{3} \,   k_2 \, \left(\dfrac{c^2 R}{GM}\right)^5\, ,
\end{equation} 
with $R$ the circumferential radius of the star, $M$ its mass, $k_2$ the second gravitoelectric Love number (characterizing the importance of the tidally induced mass quadrupole), $c$ the speed of light, and $G$ the gravitational constant. Subsequent analyses~\cite{ligo2018,radice2019} considered the phenomenological gravitational waveform models `PhenomPNRT'~\cite{schmidt2012,hannam2014,schmidt2015,husa2016,khan2016} and `PhenomDNRT'~\cite{dietrich2017,dietrich2019}. 
The BBH baseline was obtained by interpolating between PN  and numerical-relativity (NR) predictions over the entire evolution (inspiral-merger-ringdown), making also use of the solution of an effective one-body (EOB) problem to the real two-body dynamics~\cite{buonanno1999,buonanno2000} (see Ref.~\cite{damour2014} for a review). In both cases, tidal corrections were added phenomenologically using the `NRTidal' approximant~\cite{dietrich2017,dietrich2019}. Based on PN and EOB predictions, this approximant was calibrated to NR simulations of binary NS mergers for a few selected EoSs. Higher-order tidal effects were thus effectively taken into  account, but were still parametrized solely in terms of the gravitoelectric Love numbers $k_2$ of the two stars.

More recently, the gravitational-wave data have been reanalysed~\cite{ligo2019prx,ligo2020} using the `PhenomPNRT' waveform approximant and the following models based on the EOB approach: `SEOBNRT' (combining the BBH baseline referred to as  `SEOBNRv4$\_$ROM'~\cite{bohe2017,purrer2014} with the `NRTidal' approximant), as well as the time-domain waveform models  `TEOBResumS'~\cite{TEOBRESUMS} and `SEOBNRv4T'~\cite{SEOBNRV4}. Several variants of the `TaylorF2' template including higher-order tidal corrections up to 7.5PN have been also considered~\cite{damour2012,bini2012}. 
The `TEOBResumS' and `SEOBNRv4T' waveform approximants are parametrized not only in terms of $k_2$ but also of  $k_3$ and $k_4$. In the PN theory, these gravitoelectric Love numbers enter at 7 PN and 9 PN respectively. Incidentally, the PN expansion  at such high orders involves also contributions from gravitomagnetic Love numbers characterizing tidally induced current multipoles. 
In particular, the  second and third gravitomagnetic Love numbers $j_2$ and $j_3$ appear at 6 PN and 8 PN respectively. 

The continuous technological improvements of gravitational-wave detectors call for more realistic gravitational-waveform approximants from binary NS mergers. Indeed, as shown in  Ref.~\cite{samajdar2018}, the tidal deformability parameters~\eqref{eq:Lambda} inferred from gravitational-wave signals with strength similar to GW170817 but observed by the LIGO and Virgo interferometers at their design sensitivity could differ by about a factor of  two depending  on the choice of the  template (see also Refs.~\cite{narikawa2020,gamba2020}). 
The impact of the gravitomagnetic Love number $j_2$ was investigated in Ref.~\cite{jimenez2018}. Although the errors incurred by the neglect of $j_2$ amount to a few percent, these deviations turn out to be much larger than those due to the crust~\cite{piekarewicz2019,perot2020,gittins2020}. More importantly, they will be potentially observable with third-generation gravitational-wave detectors thus offering new possibilities to probe the interior of NS~\cite{maggiore2020}. 

Despite the introduction of higher-order tidal Love numbers in gravitational-waveform models and their potential importance for the analyses of upcoming gravitational-wave detections  of  binary NS mergers, little attention has been paid so far to their nuclear-physics aspects~\cite{Kumar2017}. In our previous work~\cite{perot2019}, we studied the role of the neutron-matter stiffness and of the symmetry energy in the gravitoelectric Love number $k_2$ and in the tidal deformability~\eqref{eq:Lambda} using the recent series of Brussels-Montreal unified EoSs~\cite{potekhin2013,pearson2018,pearson2019}, which provide a thermodynamically consistent description of all regions of a cold nonaccreted NS. 

In this paper, we pursue our investigation to assess the relative importance of higher-order gravitoelectric and gravitomagnetic tidal Love numbers and their sensitivity to nuclear-matter properties. Their impact on the phase of the gravitational-wave signal from binary NS inspirals is also discussed. After briefly reviewing the theory of tidal deformations in Sec.~\ref{sec:tidal}, numerical results are presented and discussed in Sec.~\ref{sec:results}.

\section{Tidal effects in binary neutron-star systems}
\label{sec:tidal}

We shall briefly review here the main results from the relativistic theory of tidal effects in compact binary systems, as developed in Refs.~\cite{flanagan08,hinderer08,damour2009,poisson2009,poisson2015,pani2018}. We shall also provide explicit expressions for the Love numbers. 

\subsection{Tidal deformabilities and Love numbers}
\label{sec:tidal-def}

Let us consider a star that is both static and spherically symmetric. In a close orbit with another compact stellar companion, the star will be tidally deformed by the mutual gravitational interactions. 
The tidal field can be decomposed into an ``electric'' (even parity or polar) component $\mathcal{E}_L$, where $L$ denotes a set of space indices $i_1 i_2\dotsi i_\ell$, and a ``magnetic'' (odd parity or axial) component $\mathcal{M}_L$ (which is absent in Newtonian theory), inducing inside the star a mass multipole moment $\mathcal{Q}_L$ and a current multipole moment $\mathcal{S}_L$ respectively (assuming internal motions are much faster than orbital motions $-$ so called adiabatic approximation). 
To leading order, these induced moments are given by 
\begin{align}
\mathcal{Q}_L &= \lambda_\ell\mathcal{E}_L\, ,  \\
\mathcal{S}_L &= \sigma_\ell\mathcal{M}_L \, ,
\end{align}
where the coefficients $\lambda_\ell$ and $\sigma_\ell$ are referred to as the gravitoelectric and gravitomagnetic tidal deformabilities of order $\ell$, respectively. These parameters are related to the dimensionless gravitoelectric and gravitomagnetic tidal Love numbers $k_\ell$ and $j_\ell$ through the following definitions:
\begin{align}
k_\ell &= \frac{1}{2}(2\ell-1)!!\frac{G\lambda_\ell}{R^{2\ell+1}}\, , \\
j_\ell &= 4(2\ell-1)!!\frac{G\sigma_\ell}{R^{2\ell+1}} \, .
\end{align}
Note that $\lambda_\ell$ was denoted by $\mu_\ell$ in Ref.~\cite{damour2009}, and our normalization of gravitomagnetic Love numbers differs from that of Ref.~\cite{damour2009}.  Our coefficient $j_\ell$ corresponds to $(\ell-1)/(\ell+2)j_\ell$ in the notations of Ref.~\cite{damour2009} (see also Ref.~\cite{pani2018} for further discussion about normalizations). Dimensionless tidal deformability coefficients, which can be potentially extracted from gravitational-wave signals as we shall discuss in Sec.~\ref{sec:tidal-gw-waveform}, are given by
\begin{align}
\Lambda_\ell &= \frac{2}{(2\ell-1)!!} k_\ell \Big(\frac{c^2 R}{G M}\Big)^{2\ell+1}\, , \\
\Sigma_\ell &= \frac{1}{4(2\ell-1)!!} j_\ell \Big(\frac{c^2 R}{G M}\Big)^{2\ell+1}\, .
\end{align}
Note  that these observable coefficients do not depend on the adopted normalization for the Love numbers. 

Given an EoS relating the pressure $P$ to the mass-energy density $\rho$, 
these tidal deformability coefficients can be computed by integrating the  Tolman-Oppenheimer-Volkoff (TOV) equations~\cite{tolman1939,oppenheimer1939} ($r$ is the radial coordinate and radial derivatives are denoted by a prime) 
\begin{equation}
P^\prime(r)= -\frac{G\, \rho(r)m(r)}{ r^2}
\biggl[1+\frac{P(r)}{\rho(r) c^2}\biggr] \biggl[1+\frac{4\pi P(r)r^3}{c^2m(r)}\biggr]\biggl[1-\frac{2Gm(r)}{c^2 r}\biggr]^{-1}\, ,
\label{tolP}
\end{equation}
\begin{equation}
m^\prime(r)= 4\pi \rho (r)r^2 \, ,
\label{tolM}
\end{equation}
from the stellar center at $r=0$ to the stellar surface at $r=R$ (where the pressure vanishes), 
simultaneously with the following differential equation for the functions $H_\ell(r)$ characterizing the small perturbations of the static metric~\cite{damour2009,poisson2009,poisson2015}:  
\begin{itemize}
    \item gravitoelectric perturbations
\begin{align}
&H_\ell''(r)+ H_\ell'(r) \biggl[1-\frac{2 G m(r)}{c^2r}\biggr]^{-1} \Bigg\{\frac{2}{r} - \frac{2G m(r)}{c^2r^2} - \frac{4 \pi G}{c^4}\, r \left[\rho(r)c^2 -  P(r)\right] \Bigg\} \nonumber\\& + H_\ell(r) \biggl[1-\frac{2G m(r)}{c^2r}\biggr]^{-1} \Bigg\{\frac{4\pi G}{c^4}\bigg[5\rho(r)c^2 + 9P(r) +c^2\dfrac{d\rho}{dP} \left[\rho(r)c^2  + P(r)\right] \bigg] \nonumber  \\& - \frac{\ell(\ell+1)}{r^2} - 4\biggl[1-\frac{2Gm(r)}{c^2r}\biggr]^{-1} \biggl[\frac{Gm(r)}{c^2r^2} + \frac{4\pi G}{c^4}\, r\,  P(r)\biggr]^2 \Bigg\} = 0 \,,
\label{H_eq}
\end{align}
    \item gravitomagnetic perturbations
\begin{align}
&\widetilde{H}_{\ell}''(r)- \widetilde{H}_{\ell}'(r) \biggl[1-\frac{2 G m(r)}{c^2r}\biggr]^{-1} \frac{4\pi G}{c^4} r \left[ P(r) + \rho(r)c^2 \right] \nonumber \\ &- \widetilde{H}_{\ell}(r) \biggl[1-\frac{2 G m(r)}{c^2r}\biggr]^{-1} \left\{ \frac{\ell(\ell+1)}{r^2} - \frac{4Gm(r)}{c^2r^3} + \theta\frac{8 \pi G}{c^4} \left[ P(r) + \rho(r)c^2 \right] \right\} = 0\, ,
\label{h_eq}
\end{align}
where $\theta=+1$ for a strictly static fluid and $\theta=-1$ for an irrotational fluid (note that gravitoelectric perturbations are not affected by fluid motions). As discussed in Ref.~\cite{pani2018}, $\theta=-1$ is expected to be more realistic, the irrotational fluid motion being driven by the tides. 
\end{itemize}

The explicit expressions of the first gravitoelectric and gravitomagnetic Love numbers up to $\ell=5$ are as follows:   
\begin{align}
k_2 =& \, \frac{8}{5} C^5 (1-2C)^2 \big[ 2(y_2-1)C - y_2 + 2 \big] \nonumber \\ &\times \Big\{ 2C \big[ 4(y_2+1)C^4 + 2(3y_2-2)C^3 - 2(11y_2-13)C^2 + 3(5y_2-8)C - 3(y_2-2) \big] \nonumber \\ &+ 3(1-2C)^2 \big[ 2(y_2-1)C-y_2+2 \big] \log(1-2C) \Big\}^{-1} \, , 
\label{eq:k2}
\end{align}
\begin{align} 
k_3 =& \, \frac{8}{7} C^7 (1-2C)^2 \big[ 2(y_3-1)C^2 - 3(y_3-2)C + y_3 - 3 \big] \nonumber \\ &\times \Big\{ 2C \big[ 4(y_3+1)C^5 + 2(9y_3-2)C^4 - 20(7y_3-9)C^3 + 5(37y_3-72)C^2 - 45(2y_3-5)C \nonumber \\ &+ 15(y_3-3) \big] + 15(1-2C)^2 \big[ 2(y_3-1)C^2 - 3(y_3-2)C + y_3 - 3 \big] \log(1-2C) \Big\}^{-1}  \, , 
\label{eq:k3}
\end{align}
\begin{align}
k_4 =& \, \frac{32}{147} C^9 (1-2C)^2 \big[ 12(y_4-1)C^3 - 34(y_4-2)C^2 + 28(y_4-3)C - 7(y_4-4) \big] \nonumber \\ &\times \Big\{ 2C \big[ 8(y_4+1)C^6 + 4(17y_4-2)C^5 - 12(83y_4-107)C^4 + 40(55y_4-116)C^3 \nonumber \\ &- 10(191y_4-536)C^2 + 105(7y_4-24)C - 105(y_4-4) \big] + 15(1-2C)^2 \big[ 12(y_4-1)C^3 \nonumber \\ &- 34(y_4-2)C^2 + 28(y_4-3)C - 7(y_4-4) \big] \log(1-2C) \Big\}^{-1} \, ,
\label{eq:k4}
\end{align}
\begin{align}
k_5 =& \, \frac{32}{99} C^{11} (1-2C)^2 \big[ 4(y_5-1)C^4 - 18(y_5-2)C^3 + 26(y_5-3)C^2 - 15(y_5-4)C + 3(y_5-5) \big] \nonumber \\ &\times \Big\{ 2C \big[ 8(y_5+1)C^7 + 4(27y_5-2)C^6 - 56(47y_5-60)C^5 + 56(158y_5-345)C^4 \nonumber \\ &- 210(57y_5-170)C^3 + 105(75y_5-278)C^2 - 315(8y_5-35)C + 315(y_5-5) \big] \nonumber \\ &+105(1-2C)^2 \big[ 4(y_5-1)C^4 - 18(y_5-2)C^3 + 26(y_5-3)C^2 \nonumber \\ &- 15(y_5-4)C + 3(y_5-5) \big] \log(1-2C) \Big\}^{-1}\, ,
\label{eq:k5}
\end{align}
\begin{align}
j_2 =& \, \frac{24}{5} C^5 \big[ 2(\ym_2-2)C - \ym_2 + 3 \big] \nonumber \\ &\times \Big\{ 2C \big[ 2(\ym_2+1)C^3 + 2\ym_2 C^2 + 3(\ym_2-1)C - 3(\ym_2-3) \big] \nonumber \\ &+ 3 \big[ 2(\ym_2-2)C - \ym_2 + 3 \big] \log(1-2C) \Big\}^{-1} \, , 
\label{eq:j2} 
\end{align}
\begin{align}
j_3 =& \, \frac{64}{21} C^7 \big[ 8(\ym_3-2)C^2 - 10(\ym_3-3)C + 3(\ym_3-4) \big] \nonumber \\ &\times \Big\{ 2C \big[ 4(\ym_3+1)C^4 + 10\ym_3C^3 + 30(\ym_3-1)C^2 - 15(7\ym_3-18)C + 45(\ym_3-4) \big] \nonumber \\ &+ 15 \big[ 8(\ym_3-2)C^2 - 10(\ym_3-3)C + 3(\ym_3-4) \big] \log(1-2C) \Big\}^{-1} \, , 
\label{eq:j3} 
\end{align}
\begin{align}
j_4 =& \, \frac{80}{147} C^9 \big[ 40(\ym_4-2)C^3 - 90(\ym_4-3)C^2 + 63(\ym_4-4)C - 14(\ym_4-5) \big] \nonumber \\ &\times \Big\{ 2C \big[ 4(\ym_4+1)C^5 + 18\ym_4C^4 + 90(\ym_4-1)C^3 - 5(137\ym_4-334)C^2 \nonumber \\ &+ 105(7\ym_4-26)C - 210(\ym_4-5) \big] + 15 \big[ 40(\ym_4-2)C^3 - 90(\ym_4-3)C^2 \nonumber \\ &+ 63(\ym_4-4)C - 14(\ym_4-5) \big] \log(1-2C) \Big\}^{-1}\, , 
\label{eq:j4}
\end{align}
\begin{align}
j_5 =& \, \frac{128}{165} C^{11} \big[ 40(\ym_5-2)C^4 - 140(\ym_5-3)C^3 + 168(\ym_5-4)C^2 - 84(\ym_5-5)C + 15(\ym_5-6) \big] \nonumber \\ &\times \Big\{ 2C \big[ 8(\ym_5+1)C^6 + 56\ym_5C^5 + 420(\ym_5-1)C^4 - 210(27\ym_5-64)C^3 + 420(26\ym_5-93)C^2 \nonumber \\ &- 315(23\ym_5-110)C + 1575(\ym_5-6) \big] + 105 \big[ 40(\ym_5-2)C^4 - 140(\ym_5-3)C^3 \nonumber \\ &+ 168(\ym_5-4)C^2 - 84(\ym_5-5)C + 15(\ym_5-6) \big] \log(1-2C) \Big\}^{-1}  \, , 
\label{eq:j5}
\end{align}
where $C = GM/(Rc^2)$ is the compactness parameter (the gravitational mass being given by $M=m(R)$), $y_\ell = R H'_\ell(R)/H_\ell(R)$, and $\ym_\ell = R \widetilde{H}'_\ell(R)/\widetilde{H}_\ell(R)$. 
Note that the numerical values of the gravitomagnetic Love number $j_2$ calculated from Eq.~(\ref{eq:j2}) are identical to those obtained from Eq.~(73) of Ref.~\cite{damour2009} modulo a normalization factor $(\ell-1)/(\ell+2)=1/4$ for $\ell=2$. The apparent discrepancy in the  formal expressions stems from the fact that the authors of Ref.~\cite{damour2009} adopted a different definition of $\ym_\ell=R \psi'_\ell(R)/\psi_\ell(R)$, in which $\psi_\ell(r)$ is a combination of $\widetilde{H}(r)$ and $\widetilde{H}'(r)$. 
The explicit expressions for $k_5$, $j_3$, $j_4$, and $j_5$, which we could not find in the literature, were derived here following the formalism of Refs.~\cite{damour2009,poisson2009,poisson2015}.

\subsection{Tidal corrections to the gravitational-wave signal}
\label{sec:tidal-gw-waveform}

Tidal effects on the gravitational waveform from  inspiralling NSs  have been calculated within the post-Newtonian (PN) theory. The gravitoelectric coefficient $k_\ell$ leads to a correction of order $(2\ell+1)$PN to the phase of the gravitational-wave signal given by ($\delta_{\ell\ell'}$ is the Kronecker symbol and the index $i=1,2$ is used to distinguish the two stars of the binary system)~\cite{yagi2014}
\begin{align}\label{eq:Psil}
\Psi_{\ell} = - \sum_{i=1}^{2} \Biggl[\frac{5}{16}\frac{(2\ell-1)!!(4\ell+3)(\ell+1)}{(4\ell-3)(2\ell-3)} \Lambda_{\ell,i} X_i^{2\ell-1} x^{2\ell-3/2}+ \frac{9}{16}\delta_{\ell 2} \Lambda_{2,i} \frac{X_i^4}{\eta} x^{5/2} \Biggr] + \mathcal{O}(x^{2\ell-1/2})\, ,
\end{align}
where $x=(G \pi M f/c^3)^{2/3}$, $f$ is the frequency of the gravitational-wave signal, $M=M_1+M_2$, $X_i = M_i/M$ and $\eta = M_1 M_2/M^2$. The gravitomagnetic tidal coefficient $j_\ell$ contributes at higher order, namely  $(2\ell+2)$PN. The first correction at 6PN is given by~\cite{yagi2017}
\begin{align}\label{eq:Psi2tilde}
\widetilde{\Psi}_2 =& \sum_{i=1}^{2} \frac{5}{224} \Sigma_{2,i} \frac{X_i^4}{\eta} (X_i-1037 X_j) x^{7/2} + \mathcal{O}(x^{9/2}) \, .
\end{align}

\section{Numerical results}
\label{sec:results}

The Love numbers $k_2$, $k_3$, $k_4$, and $j_2$ were previously evaluated in Ref.~\cite{Kumar2017} using a few EoSs based on relativistic mean-field models. However, the adopted EoSs in the outer and inner regions of the crust were not specified although the crust has a strong impact on the values of the Love numbers~\cite{perot2020}.  The ad hoc matching of different EoSs may also lead to uncertainties as large as 20\% for one-solar mass  NSs~\cite{ferreira2020}. 
Moreover, one of the EoSs adopted in Ref.~\cite{Kumar2017}, namely NL3~\cite{lalazissis1997}, is found to be incompatible with experimental nuclear data~\cite{danielewicz2002}, see also Ref.~\cite{dutra2014}). More importantly, the calculations of Ref.~\cite{Kumar2017} for the gravitomagnetic Love number $j_2$ appear to be incorrect because the authors applied Eq.~(73) of Ref.~\cite{damour2009} using the numerical solution from the same differential equation as for the gravitoelectric Love number $k_2$. This may explain the large positive values they found for $j_2$ at variance with the comparatively small \emph{negative} values obtained in Ref.~\cite{damour2009}. In this section, we shall present results obtained using the set of Brussels-Montreal unified EoSs~\cite{potekhin2013,pearson2018,pearson2019}.

\subsection{Unified equations of state of dense matter in neutron stars}
\label{sec:EoS}

The Brussels-Montreal EoSs~\cite{pearson2018,pearson2019} we adopt here are all based on generalized Skyrme functionals, whose parameters were precision-fitted to the 2353 measured masses of atomic nuclei having proton number $Z\geq8$ and neutron number $N\geq8$ from the 2012 Atomic Mass Evaluation~\cite{AME2012} with a root mean square deviation of order 0.5-0.6 MeV~\cite{gcp2013}. A number of constraints were simultaneously imposed. The incompressibility coefficient $K_v$ of symmetric nuclear matter was restricted to lie in the experimental range $K_v=240\pm10$~MeV~\cite{colo2004}, while the symmetry-energy coefficient $J$ at normal density $n_0\approx 0.16$~fm$^{-3}$ was allowed to take one of these four different possible values: 29, 30, 31, and 32 MeV. 
To ensure reliable extrapolations to the highly neutron-rich matter of NSs,  these functionals were further constrained to yield realistic neutron-matter EoSs. 
The functionals BSk22, BSk23, BSk24 and BSk25 were all fitted to the rather stiff EoS labeled as `V18' by Li and Schulze~\cite{ls2008}, but differ in their value for the symmetry-energy coefficient $J$ = 32, 31, 30 and 29 MeV, respectively. The intermediate functional BSk23 will not be further considered in this  work. The functional BSk26 was fitted to the softer EoS labeled as `A18 + $\delta\,v$ + UIX$^*$' by Akmal, Pandharipande and Ravenhall~\cite{apr1998} with $J=30$~MeV. For the sake of comparison, we shall also consider the older unified EoS BSk19~\cite{pearson2011,pcgd2012}, whose eponymous functional~\cite{gcp2010} was fitted to the neutron-matter EoS of Ref.~\cite{Friedman1981}. Although this EoS is too soft at high densities to explain the existence of the most massive observed NSs~\cite{chamel2011}, it has found some support in the analyses of kaon and pion productions in heavy-ion collisions~\cite{fuchs2001,sturm2001,hartnack2006,xiao2009}. 
As shown in Fig.~1 of Ref.~\cite{perot2019}, these functionals remain consistent with more recent ab initio calculations. All functionals are also consistent with EoS constraints inferred from heavy-ion collisions~\cite{danielewicz2002,lynch2009} and with various empirical determinations of the symmetry energy at densities $n\leq n_0$ (see Fig.~2 of Ref.~\cite{perot2019}). The functionals mainly differ in their predictions for the density dependence of the symmetry energy at higher densities. As discussed in Refs.~\cite{chamel2011,pearson2018}, all functionals but BSk19 and BSk26 satisfy the causality requirement up to the highest densities that can be possibly reached in NSs. The violation of causality found for BSk26 in the central core of the most massive NSs can be traced back to that of the underlying microscopic EoS of Ref.\cite{apr1998} to which BSk26 was fitted.  However, as shown in Ref.~\cite{apr1998}, the 
reduction in the NS mass resulting from a restoration of causality 
is quite small. Moreover, the dimensionless tidal deformability coefficients are 
the smallest in magnitude (therefore the most challenging to measure) for the most massive NSs, as we shall show. Therefore, the calculation of these coefficients in the most compact NSs appears 
to be of less astrophysical relevance than for medium- and low- mass NSs.

All EoSs considered here are consistent with the tidal deformability constraints inferred from analyses of the gravitational-wave signal GW170817 (BSk22 being only marginally compatible)~\cite{perot2019}.

\subsection{Comparison between different EoSs}

To check our code, we have calculated the various Love numbers for the SLy EoS~\cite{douchin2001} and our results are found to be in excellent agreement with those obtained earlier in Ref.~\cite{damour2009} with the same EoS and for $k_2$, $k_3$, $k_4$ and $j_2$\footnote{We multiplied our coefficient $j_2$ by 4 so as to use the same normalization as in Ref.~\cite{damour2009}, as discussed in Sec.~\ref{sec:tidal-def}.}. 

\begin{figure}[h!]
\begin{center}
\includegraphics[width=\textwidth]{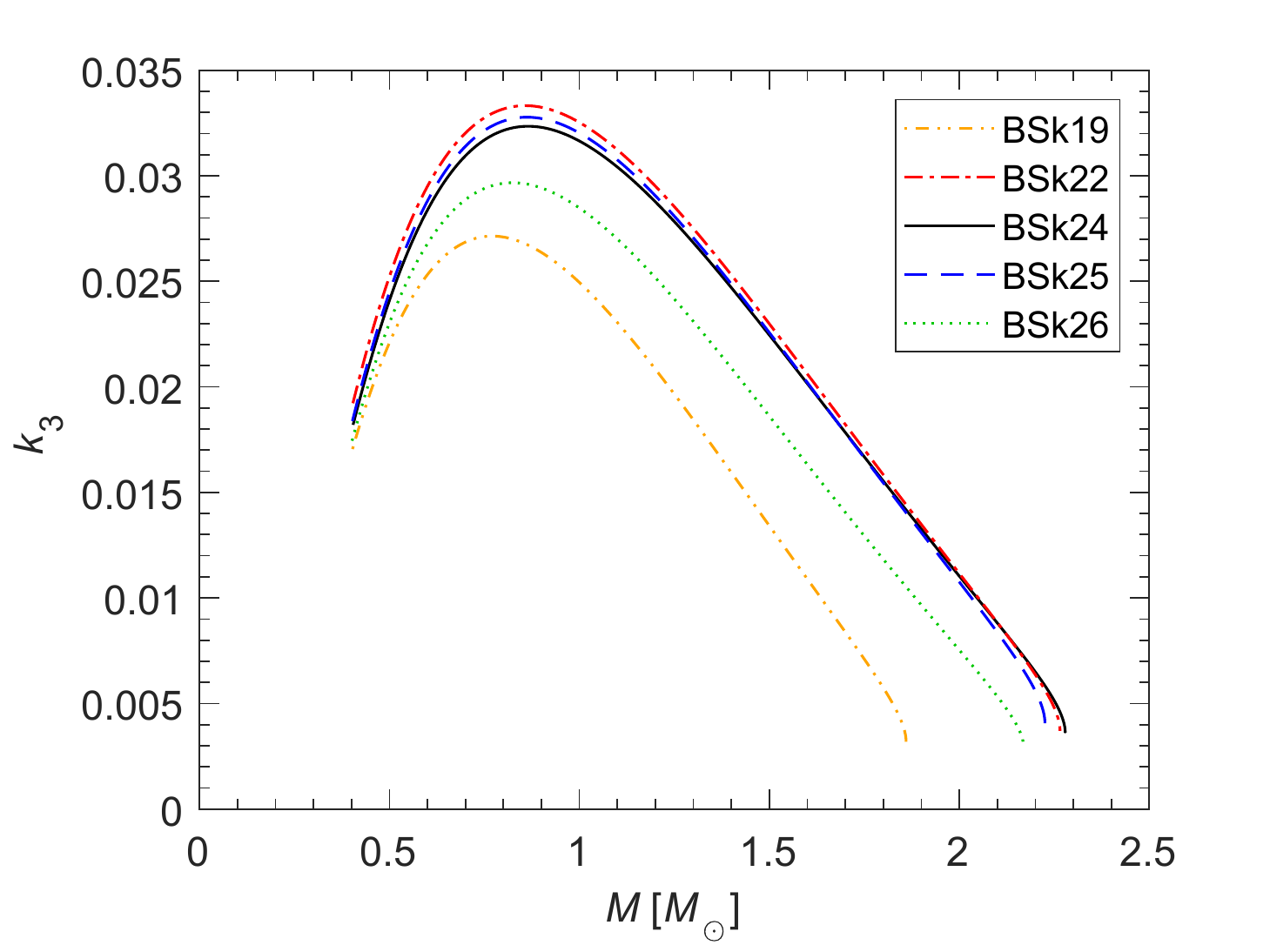}
\caption{(Color online) Gravitoelectric Love number $k_3$ as a function of NS mass computed with  Brussels-Montreal unified EoSs.}
\label{fig:k3-M}
\end{center}
\end{figure}

\begin{figure}[h!]
\begin{center}
\includegraphics[width=\textwidth]{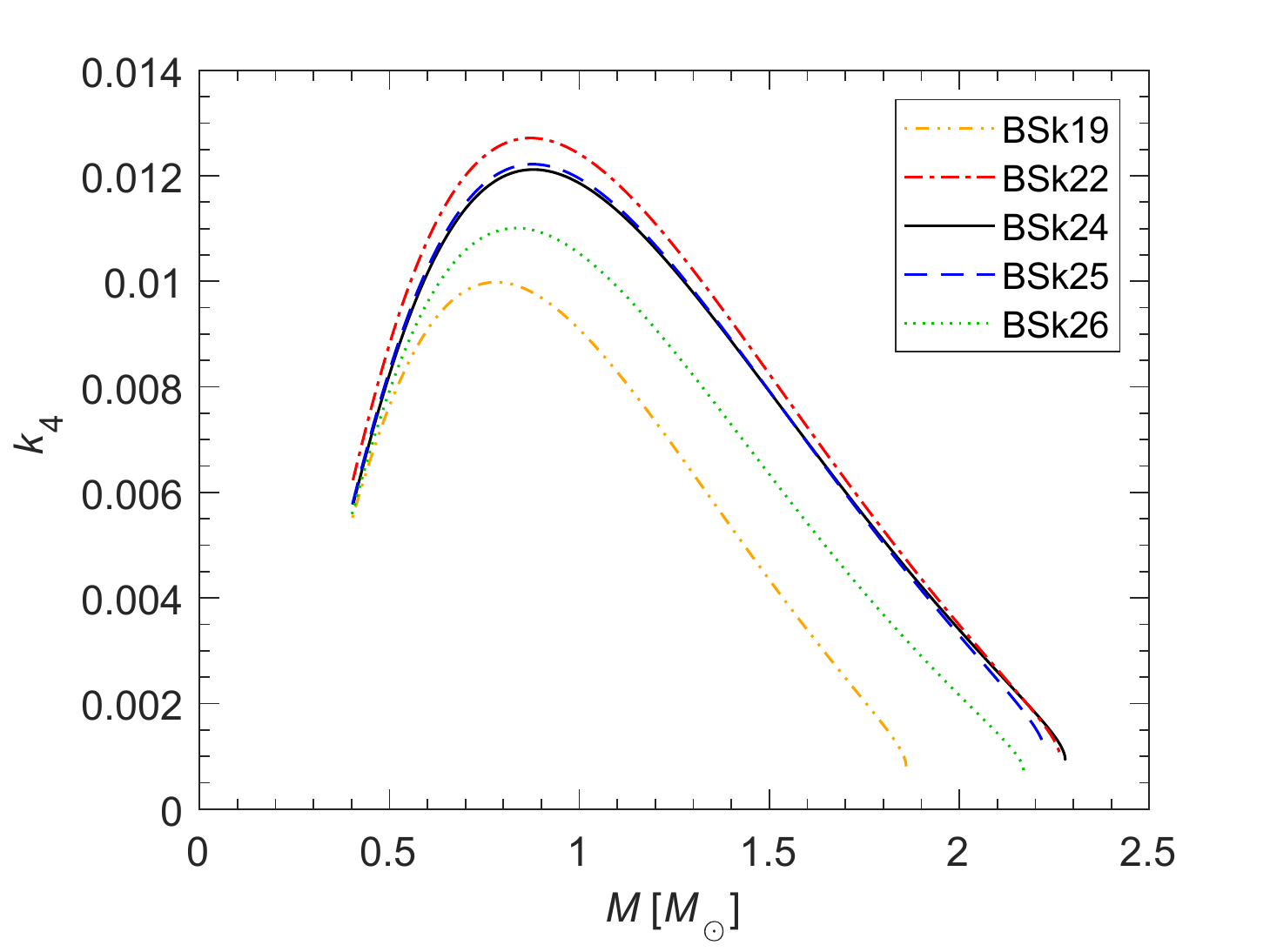}
\caption{(Color online) Gravitoelectric Love number $k_4$ as a function of NS mass computed with  Brussels-Montreal unified EoSs.}
\label{fig:k4-M}
\end{center}
\end{figure}

\begin{figure}[h!]
\begin{center}
\includegraphics[width=\textwidth]{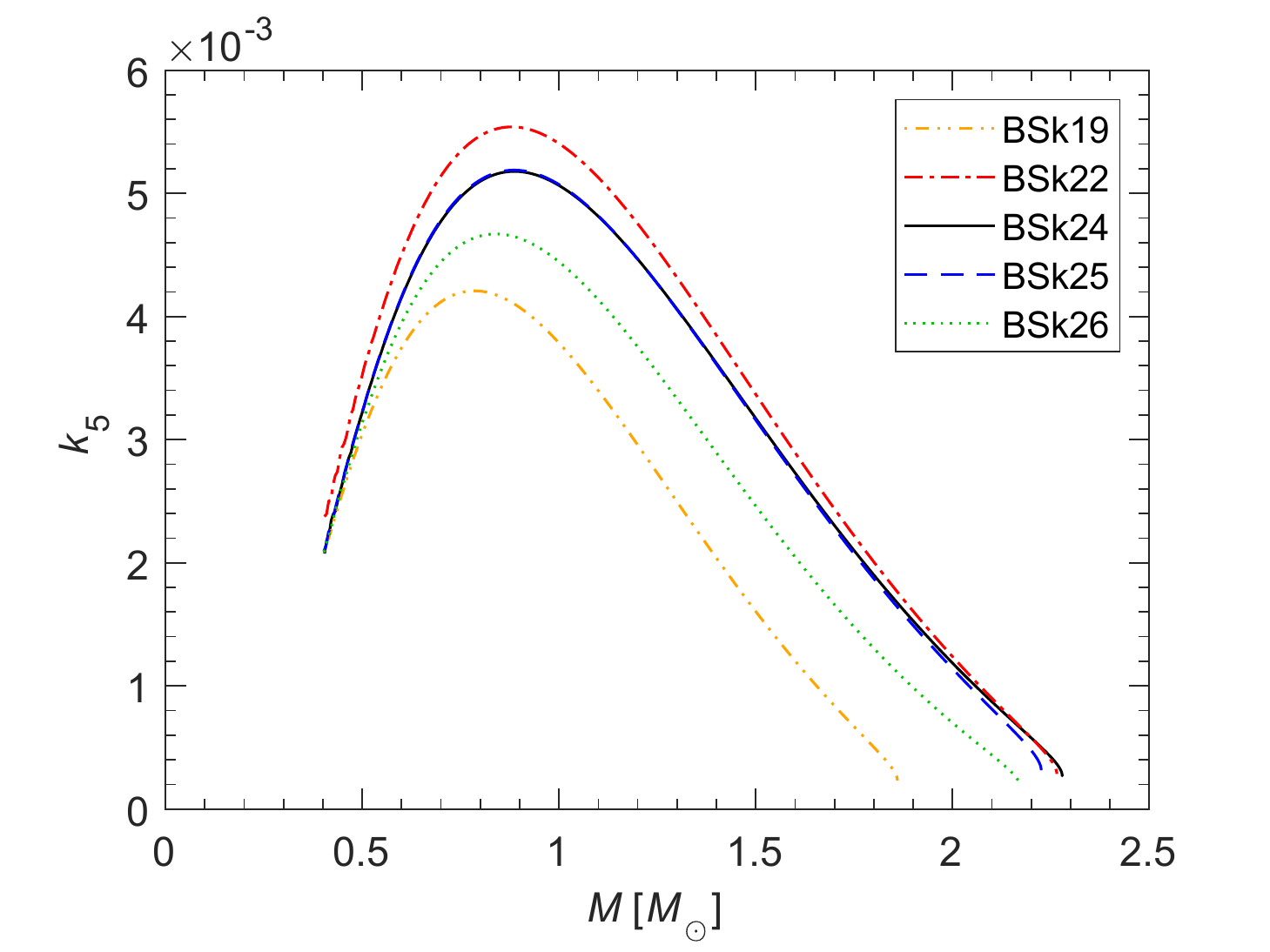}
\caption{(Color online) Gravitoelectric Love number $k_5$ as a function of NS mass computed with  Brussels-Montreal unified EoSs.}
\label{fig:k5-M}
\end{center}
\end{figure}

The gravitoelectric Love numbers calculated with the Brussels-Montreal EoSs are plotted in Figs.~\ref{fig:k3-M}, \ref{fig:k4-M}, and \ref{fig:k5-M}. Comparing results for BSk22, BSk24, and BSk25 shows that the symmetry energy plays a minor role although its impact on $k_\ell$ becomes more visible with increasing $\ell$. The key factor appears to be the stiffness of the neutron-matter EoS, as can be clearly seen by comparing results for BSk19, BSk26, and BSk24 (by increasing order of stiffness): the softer the EoS, the lower is the value for $k_\ell$ for a given NS mass. 

\begin{figure}[h!]
\begin{center}
\includegraphics[width=\textwidth]{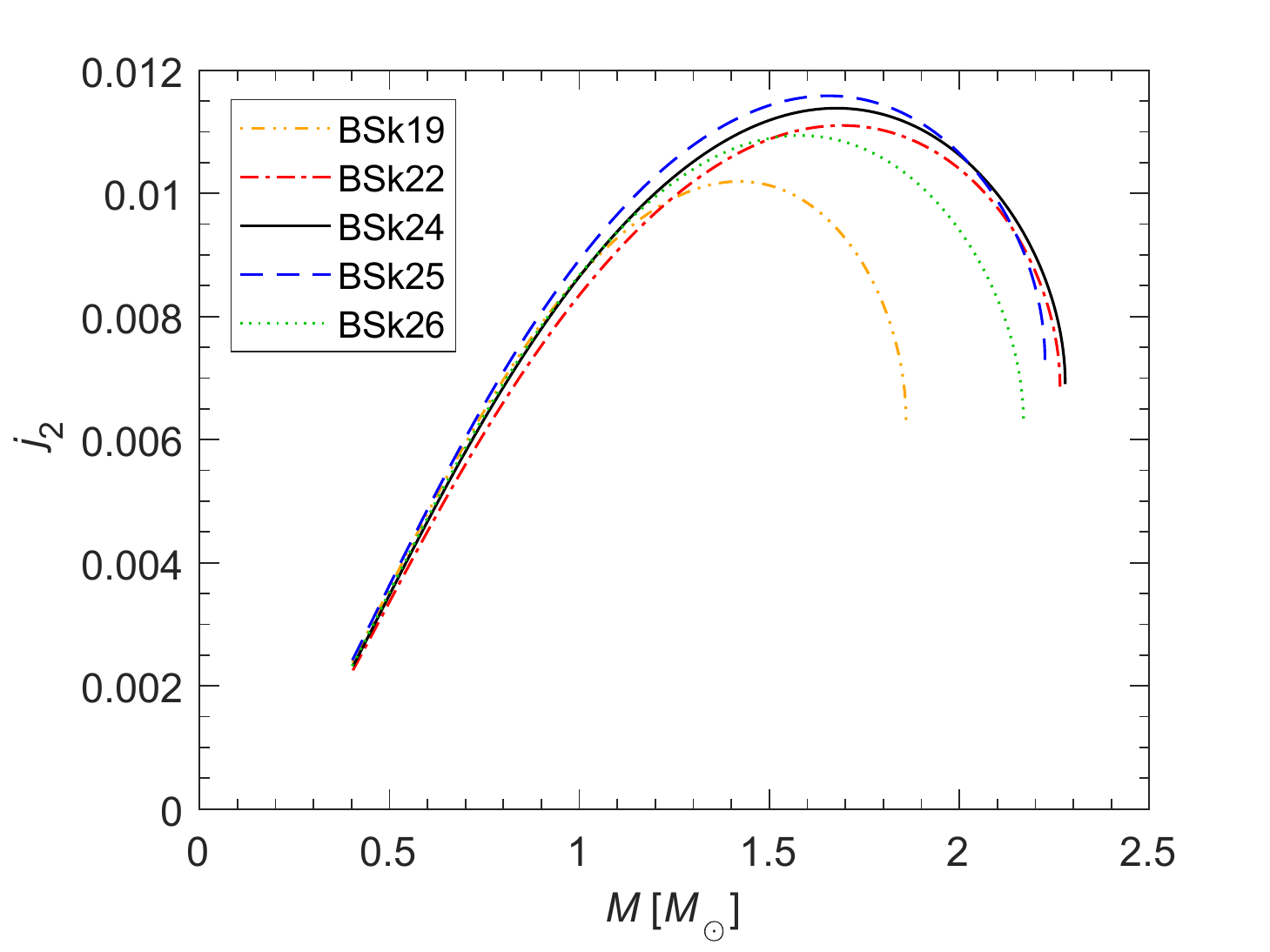}
\caption{(Color online) Gravitomagnetic Love number $j_2$ for a static fluid as a function of NS mass computed with  Brussels-Montreal unified EoSs.}
\label{fig:j2-M-stat}
\end{center}
\end{figure}

\begin{figure}[h!]
\begin{center}
\includegraphics[width=\textwidth]{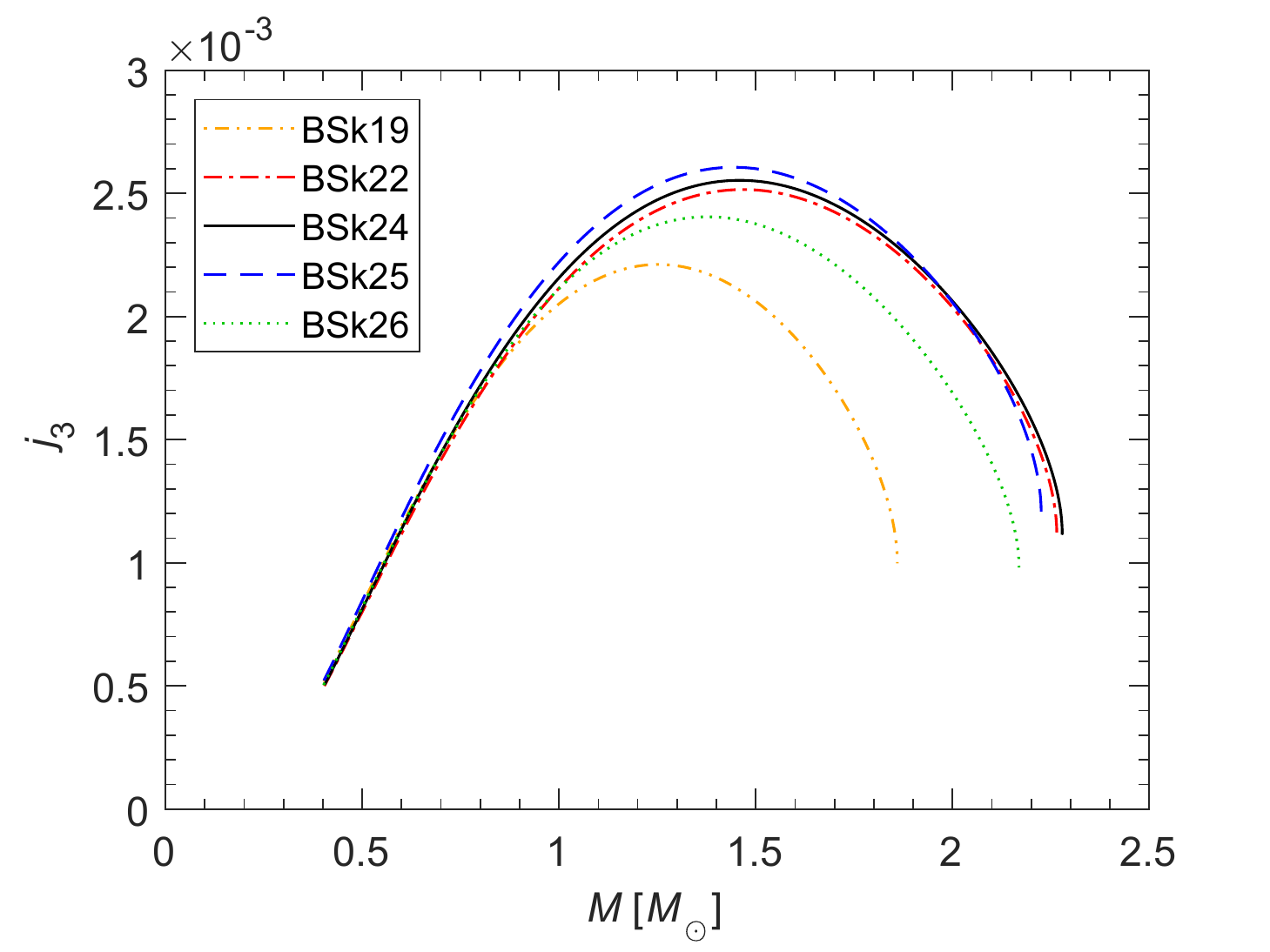}
\caption{(Color online) Gravitomagnetic Love number $j_3$ for a static fluid as a function of NS mass computed with the Brussels-Montreal unified EoSs.}
\label{fig:j3-M-stat}
\end{center}
\end{figure}

\begin{figure}[h!]
\begin{center}
\includegraphics[width=\textwidth]{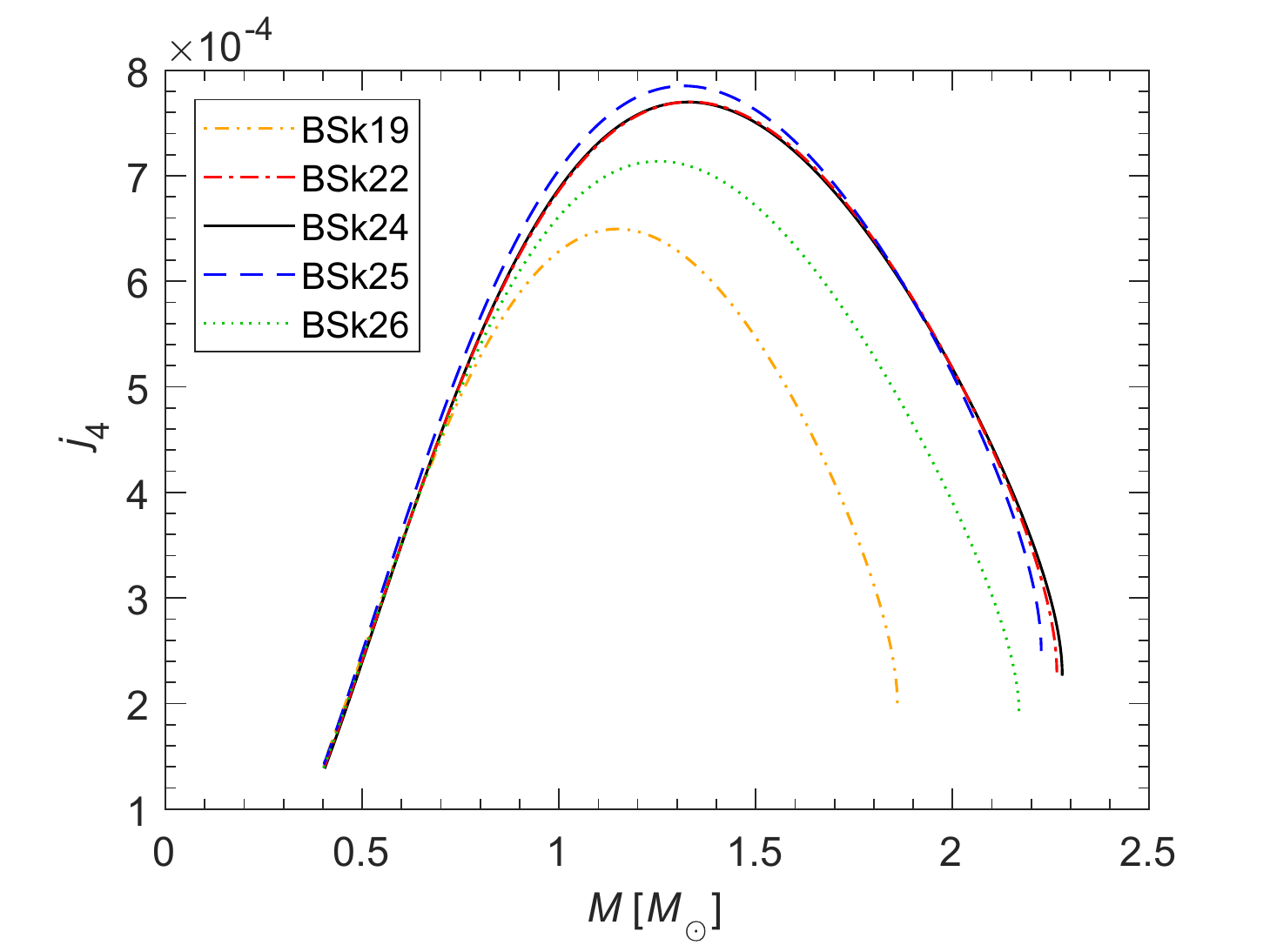}
\caption{(Color online) Gravitomagnetic Love number $j_4$ for a static fluid as a function of NS mass computed with Brussels-Montreal unified EoSs.}
\label{fig:j4-M-stat}
\end{center}
\end{figure}

\begin{figure}[h!]
\begin{center}
\includegraphics[width=\textwidth]{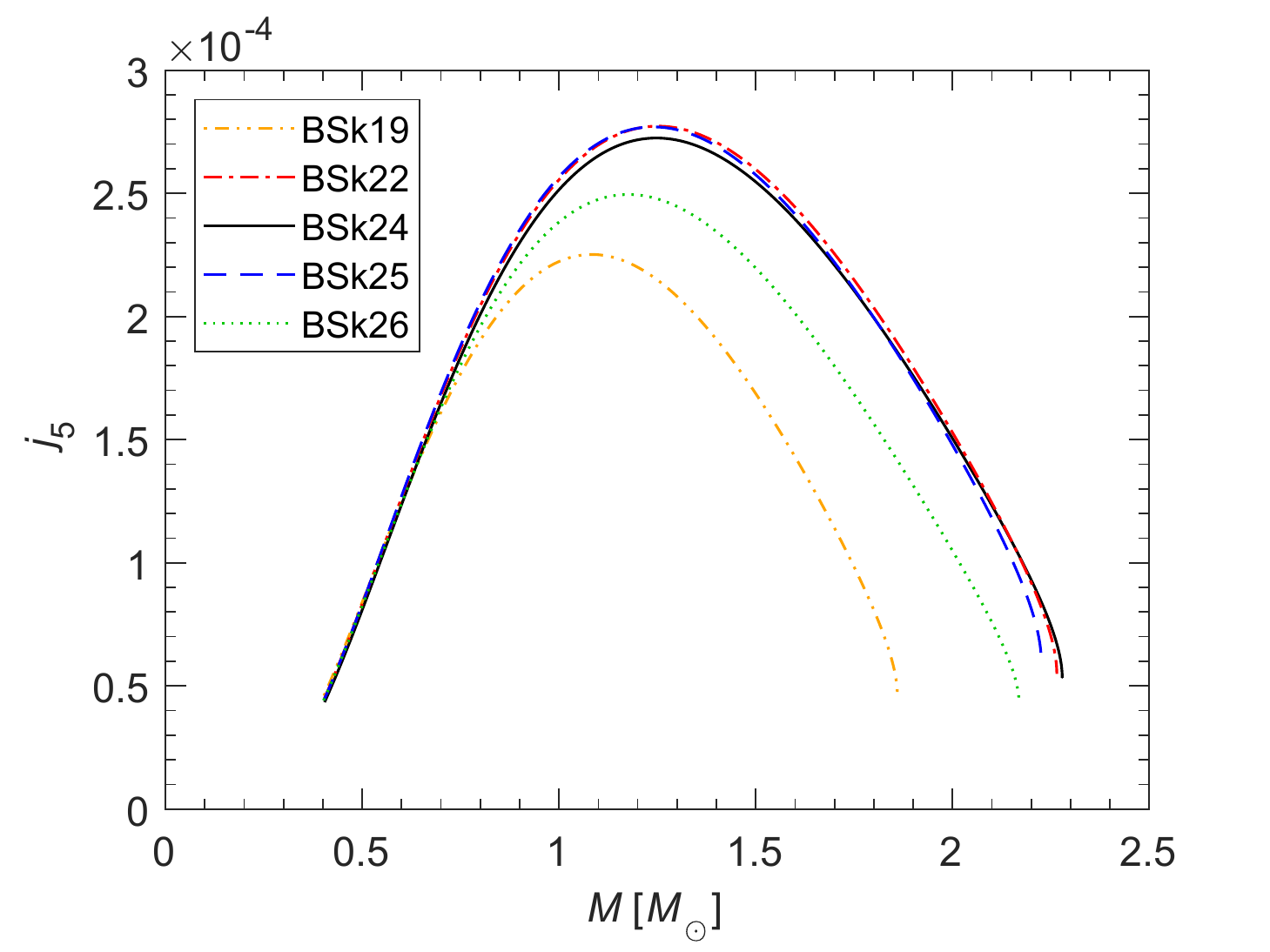}
\caption{(Color online) Gravitomagnetic Love number $j_5$ for a static fluid as a function of NS mass computed with the Brussels-Montreal unified EoSs.}
\label{fig:j5-M-stat}
\end{center}
\end{figure}

\begin{figure}[h!]
\begin{center}
\includegraphics[width=\textwidth]{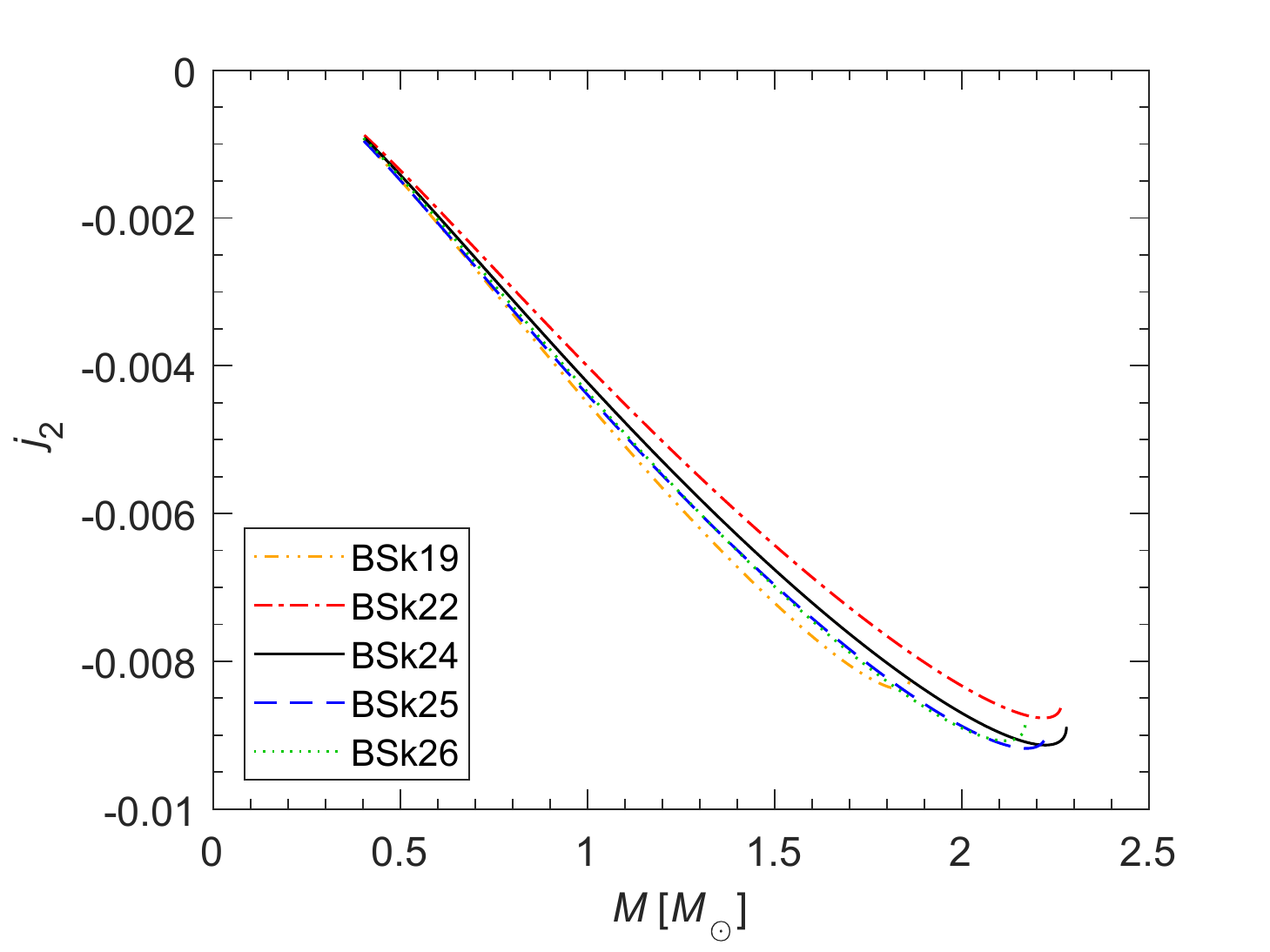}
\caption{(Color online) Gravitomagnetic Love number $j_2$ for an irrotational fluid as a function of NS mass computed with Brussels-Montreal unified EoSs.}
\label{fig:j2-M-irrot}
\end{center}
\end{figure}

\begin{figure}[h!]
\begin{center}
\includegraphics[width=\textwidth]{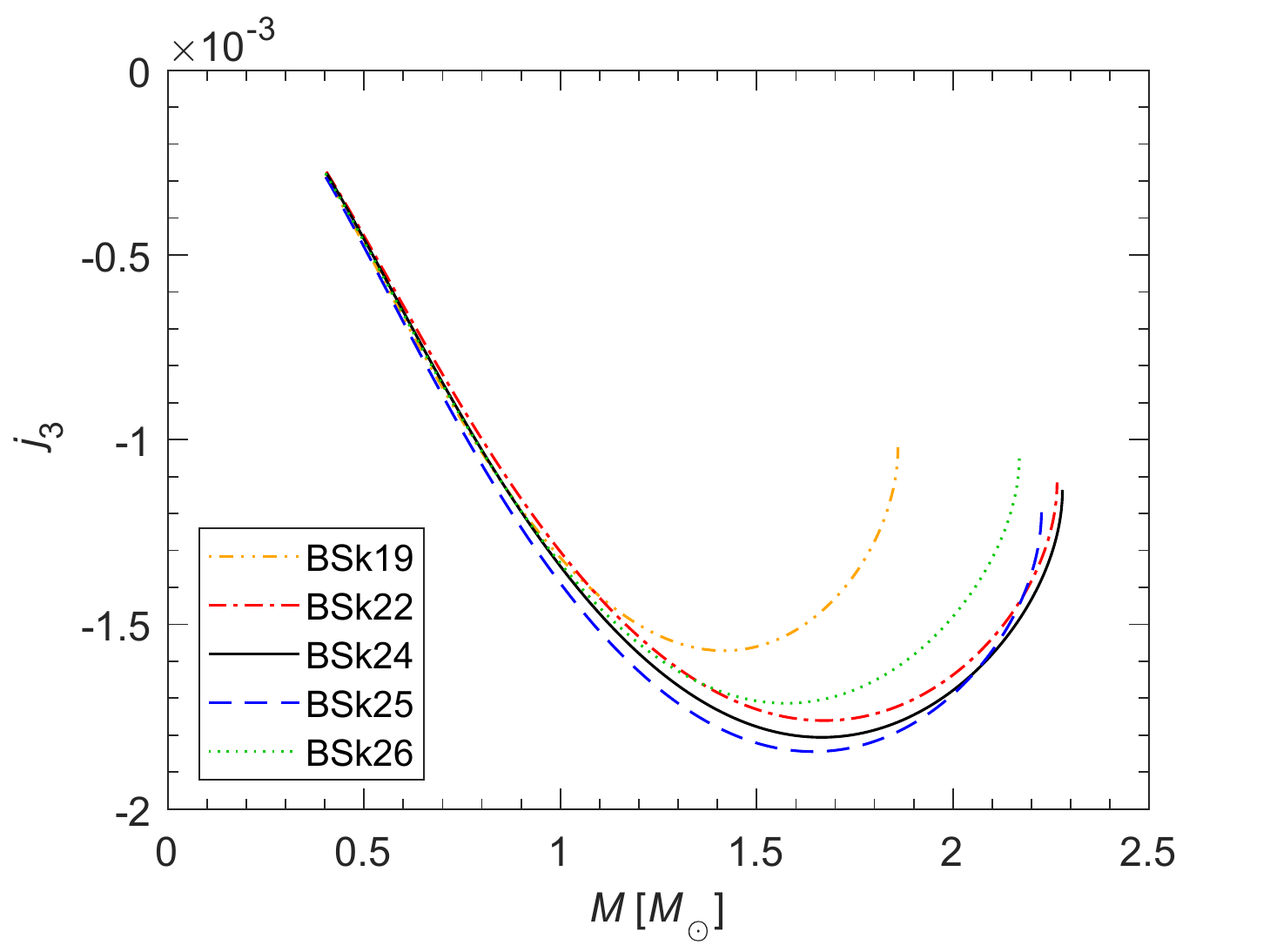}
\caption{(Color online) Gravitomagnetic Love number $j_3$ for an irrotational fluid as a function of NS mass computed with Brussels-Montreal unified EoSs.}
\label{fig:j3-M-irrot}
\end{center}
\end{figure}

\begin{figure}[h!]
\begin{center}
\includegraphics[width=\textwidth]{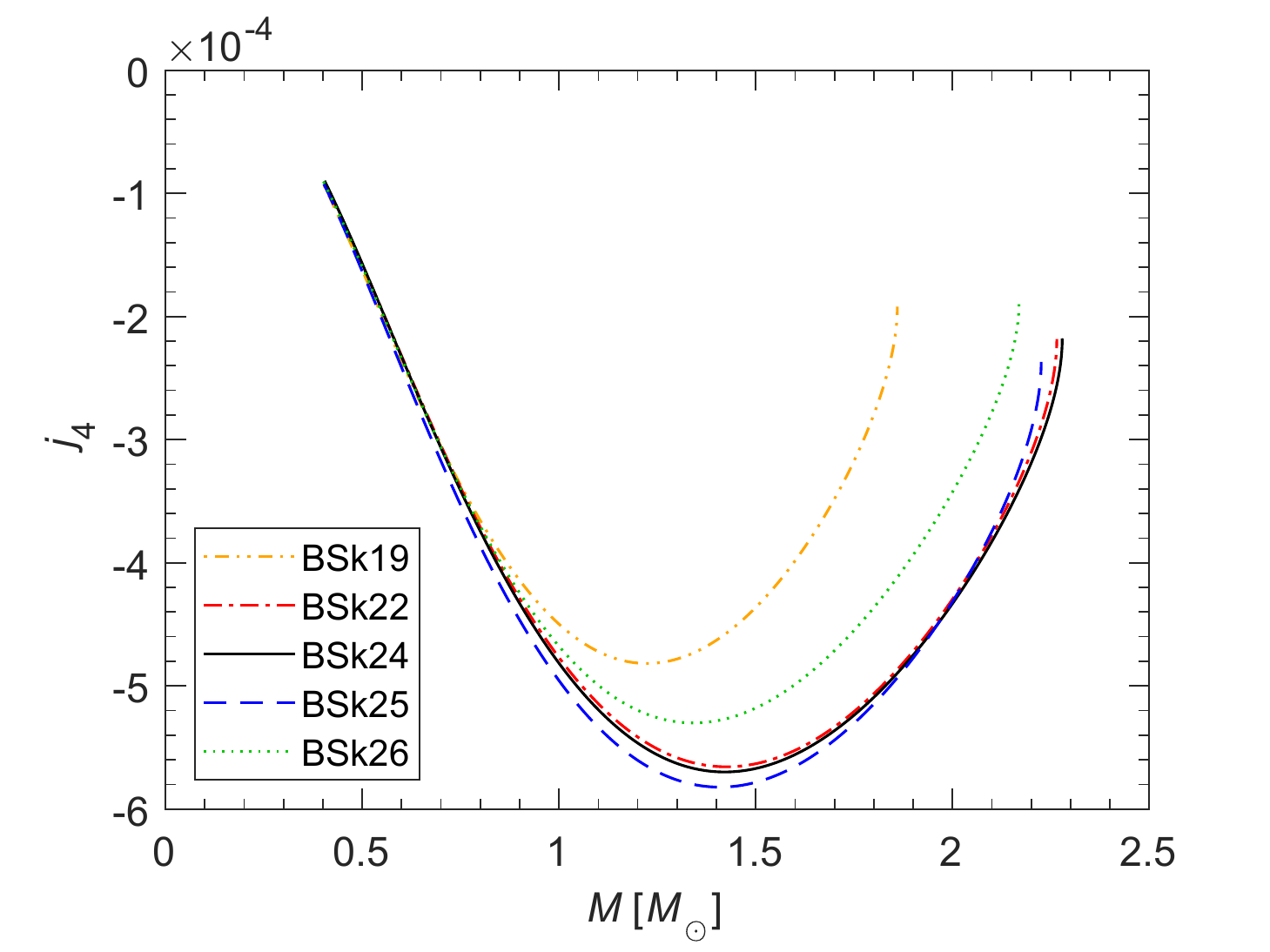}
\caption{(Color online) Gravitomagnetic Love number $j_4$ for an irrotational fluid as a function of NS mass computed with Brussels-Montreal unified EoSs.}
\label{fig:j4-M-irrot}
\end{center}
\end{figure}

\begin{figure}[h!]
\begin{center}
\includegraphics[width=\textwidth]{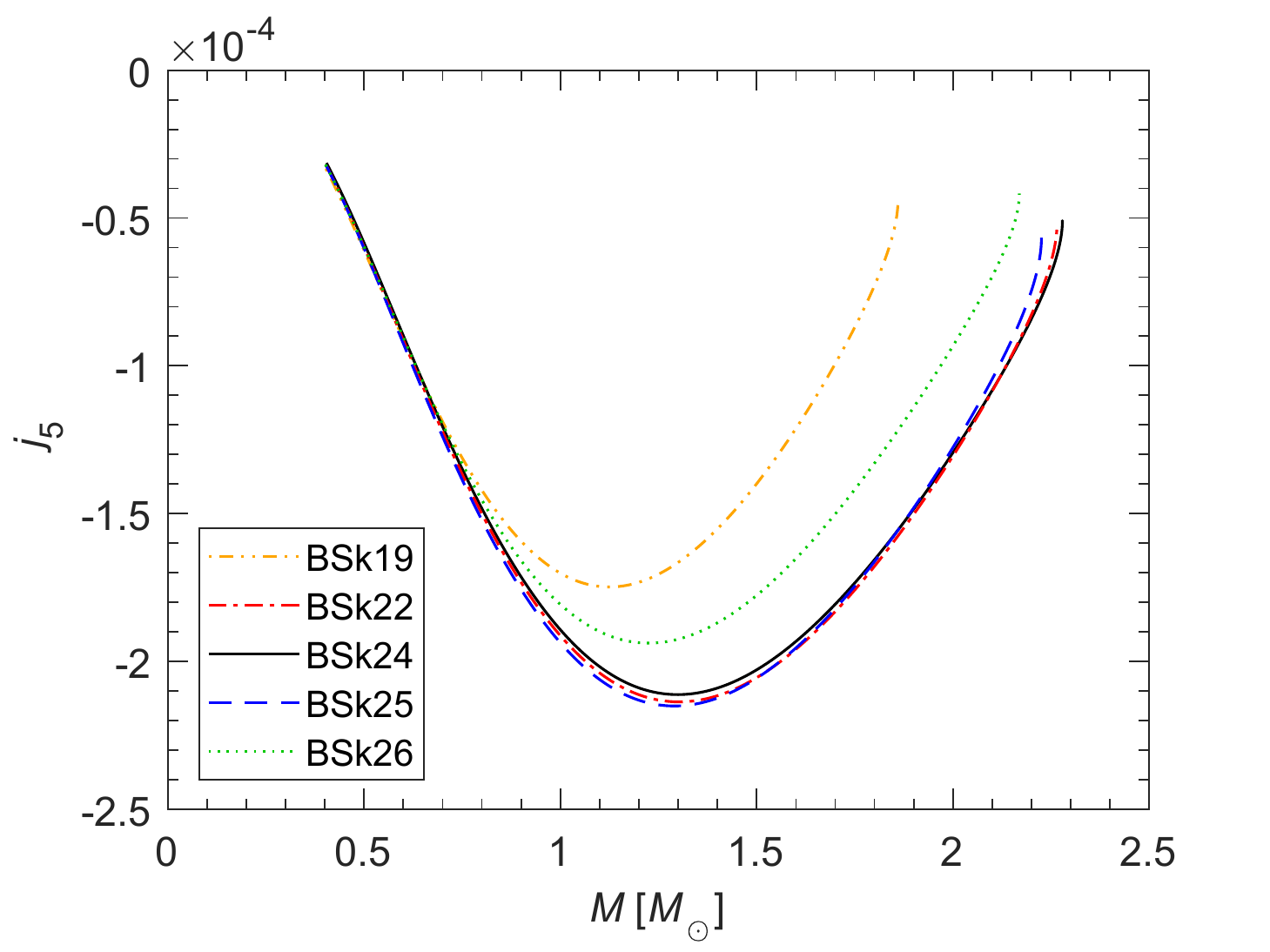}
\caption{(Color online) Gravitomagnetic Love number $j_5$ for an irrotational fluid as a function of NS mass computed with Brussels-Montreal unified EoSs.}
\label{fig:j5-M-irrot}
\end{center}
\end{figure}

The gravitomagnetic Love numbers are plotted in Figs.~\ref{fig:j2-M-stat}, \ref{fig:j3-M-stat}, \ref{fig:j4-M-stat}, \ref{fig:j5-M-stat} for static fluids and in Figs.~\ref{fig:j2-M-irrot}, \ref{fig:j3-M-irrot}, \ref{fig:j4-M-irrot}, \ref{fig:j5-M-irrot} for irrotational fluids. These two different assumptions lead to opposite tidal effects: $j_\ell$ are positive for the former but negative for the latter, as previously discussed in Ref.~\cite{poisson2015}. In either case, $j_\ell$ have similar absolute magnitudes and are found to be more sensitive to the neutron-matter EoS than to the symmetry energy. The influence of the symmetry energy is less and less pronounced as $\ell$ is increased, at variance with the behavior observed for $k_\ell$.

\subsection{Comparison between different orders}

\begin{figure}[ht!]
\begin{center}
\includegraphics[width=\textwidth]{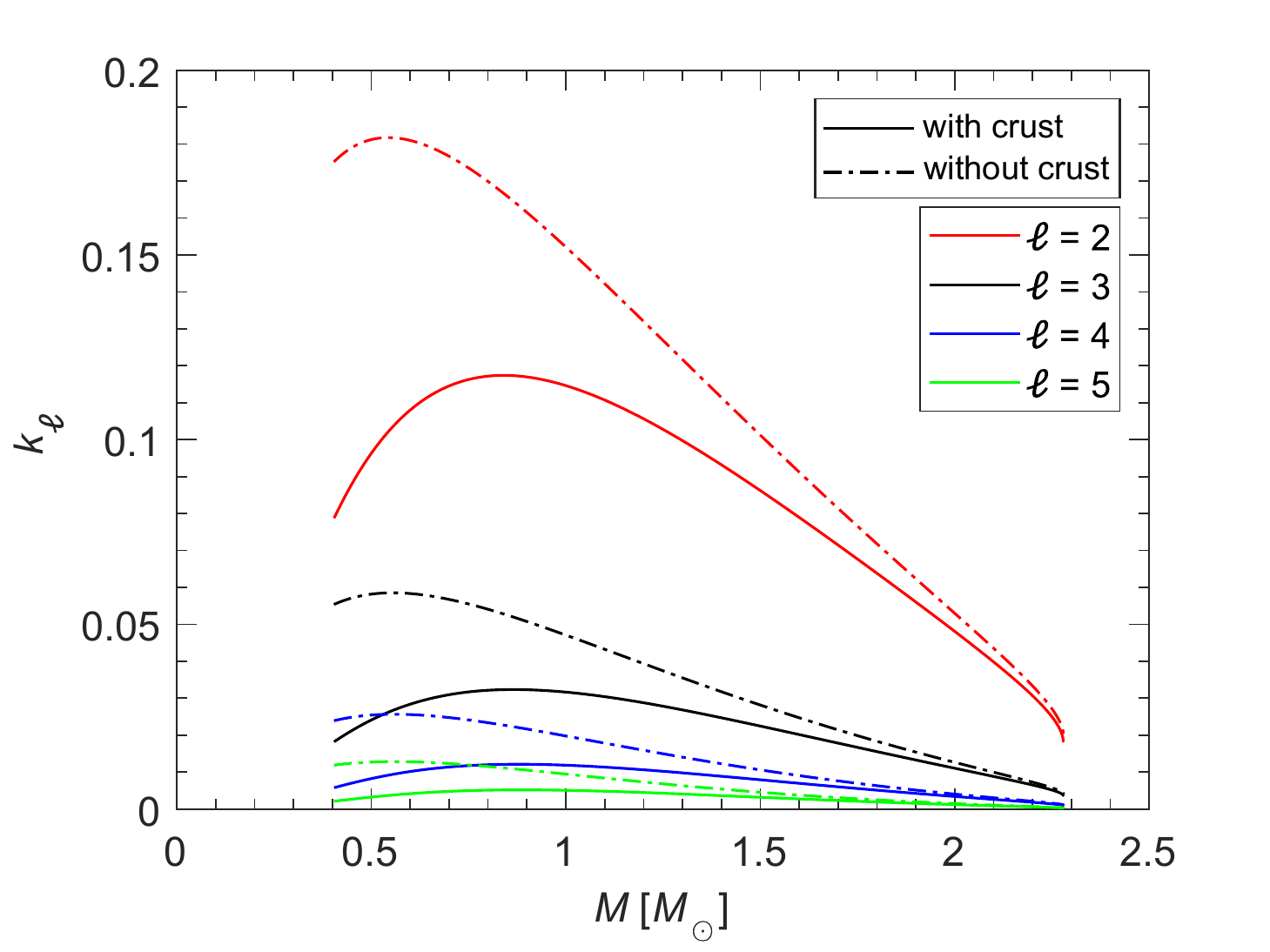}
\caption{(Color online) Gravitoelectric Love numbers $k_\ell$ as a function  of the gravitational mass $M$ for a NS with and without crust. Calculations were performed using the Brussels-Montreal nuclear energy-density functional BSk24.}
\label{fig:k2345}
\end{center}
\end{figure}

\begin{figure}[ht!]
\begin{center}
\includegraphics[width=\textwidth]{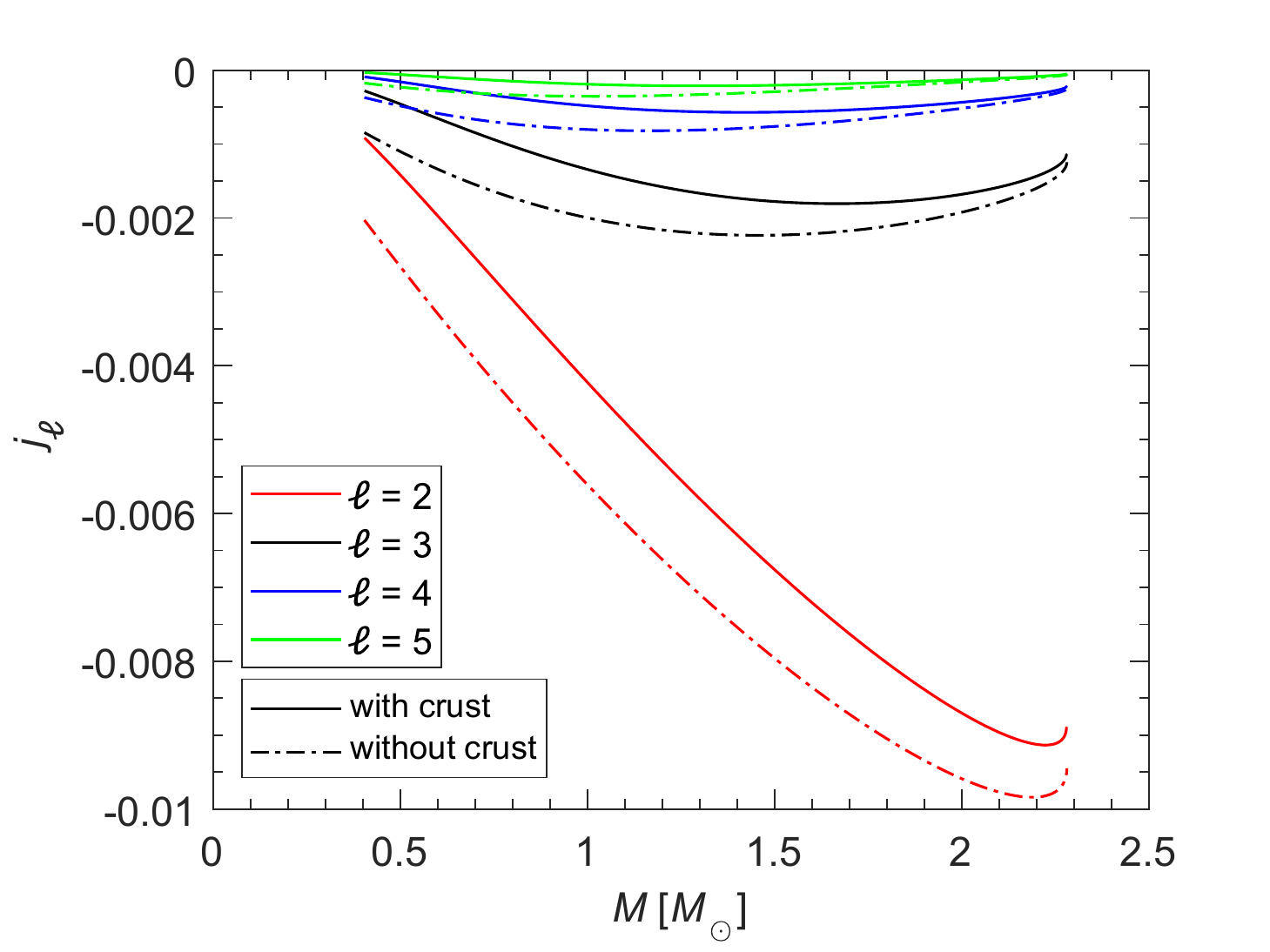}
\caption{(Color online) Gravitomagnetic Love numbers $j_\ell$ as a function  of the gravitational mass $M$ for a NS (irrotational fluid) with and without crust. Calculations were performed using the Brussels-Montreal nuclear energy-density functional BSk24.}
\label{fig:j2345}
\end{center}
\end{figure}

To assess the relative importance of the higher-order tidal effects and the convergence of the multipole expansion, we have calculated the Love numbers for the unified EoS BSk24, as this EoS appears to be the most favored by both nuclear and astrophysical data~\cite{gcp2013,pearson2018,perot2019}. As discussed in Ref.~\cite{poisson2015}, the assumption of static or irrotational fluid leads to qualitatively different gravitomagnetic Love numbers, however  their magnitude turns out to be comparable. The gravitomagnetic tidal coefficients are found to be about an order of magnitude smaller than their gravitoelectric counterpart. As shown in Figs.~\ref{fig:k2345} and \ref{fig:j2345}, the magnitude of both the gravitoelectric and gravitomagnetic coefficients decreases with increasing $\ell$. For a $1.4 M_\odot$ NS, the values of $k_5$ and $j_5$ represent only 4\% and 3\% of those of $k_2$ and $j_2$ respectively  (gravitomagnetic Love numbers were calculated for an irrotational fluid). 

\begin{figure}[ht!]
\begin{center}
\includegraphics[width=\textwidth]{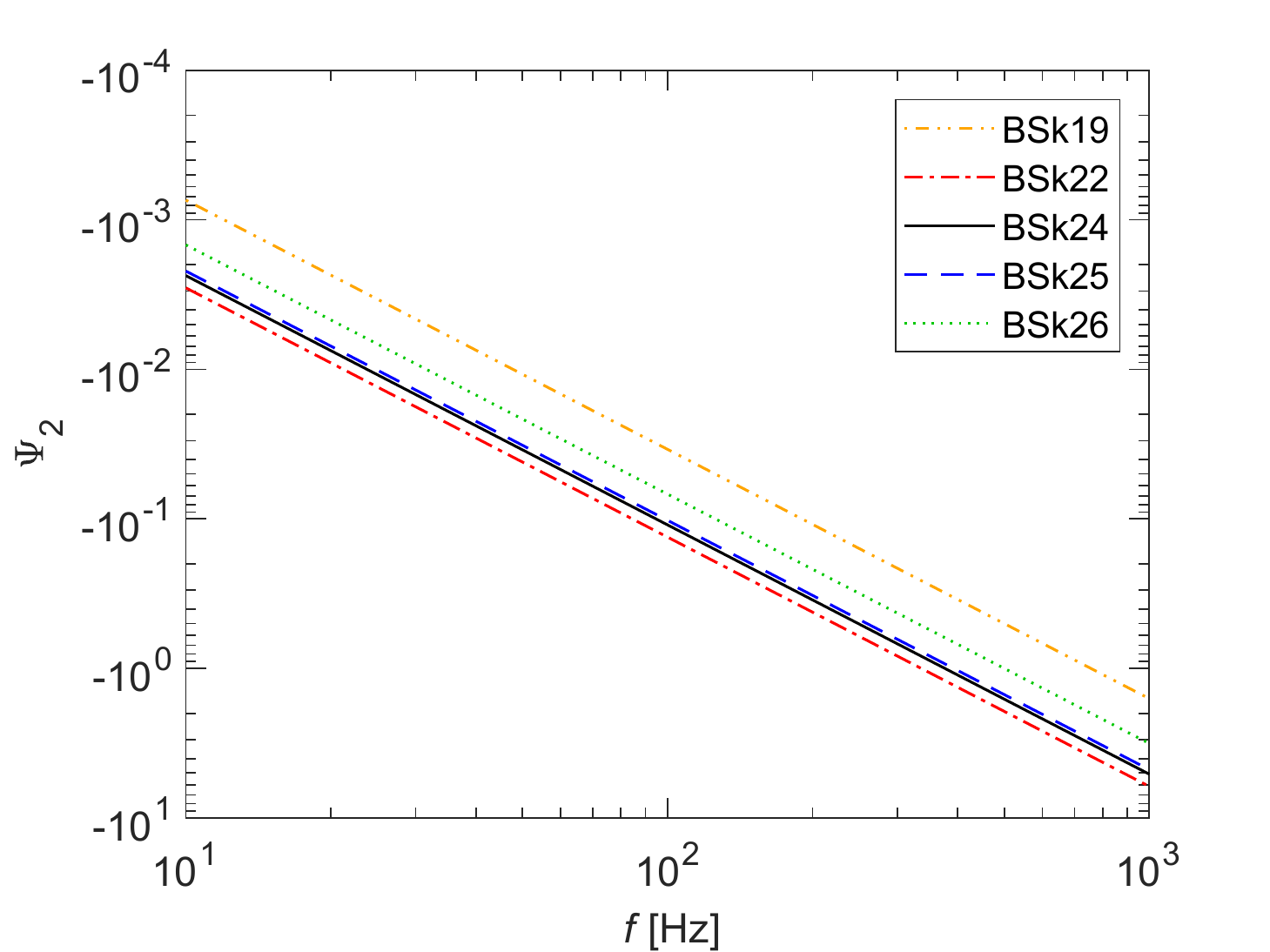}
\caption{(Color online) Contribution from the gravitoelectric Love number $k_2$ to the phase~\eqref{eq:Psil} of the gravitational-wave signal from binary NS inspiral as a function of the frequency $f$ for different EoSs. Calculations were performed for NSs with equal masses of 1.4 $M_\odot$.}
\label{fig:Psi2-BSk}
\end{center}
\end{figure}

\begin{figure}[ht!]
\begin{center}
\includegraphics[width=\textwidth]{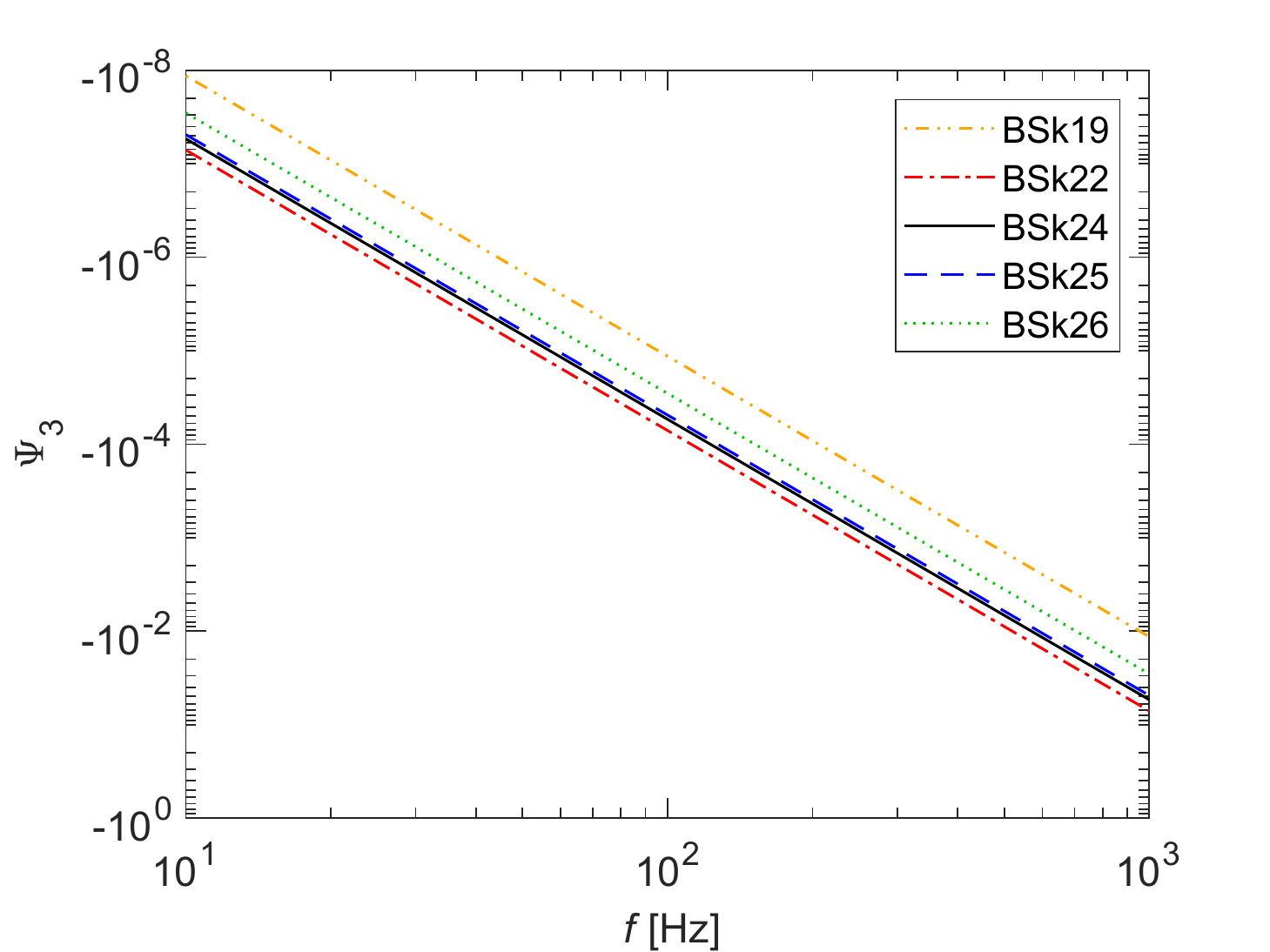}
\caption{(Color online) Same as Fig.~\ref{fig:Psi2-BSk} for the gravitoelectric Love number $k_3$.}
\label{fig:Psi3-BSk}
\end{center}
\end{figure}

\begin{figure}[ht!]
\begin{center}
\includegraphics[width=\textwidth]{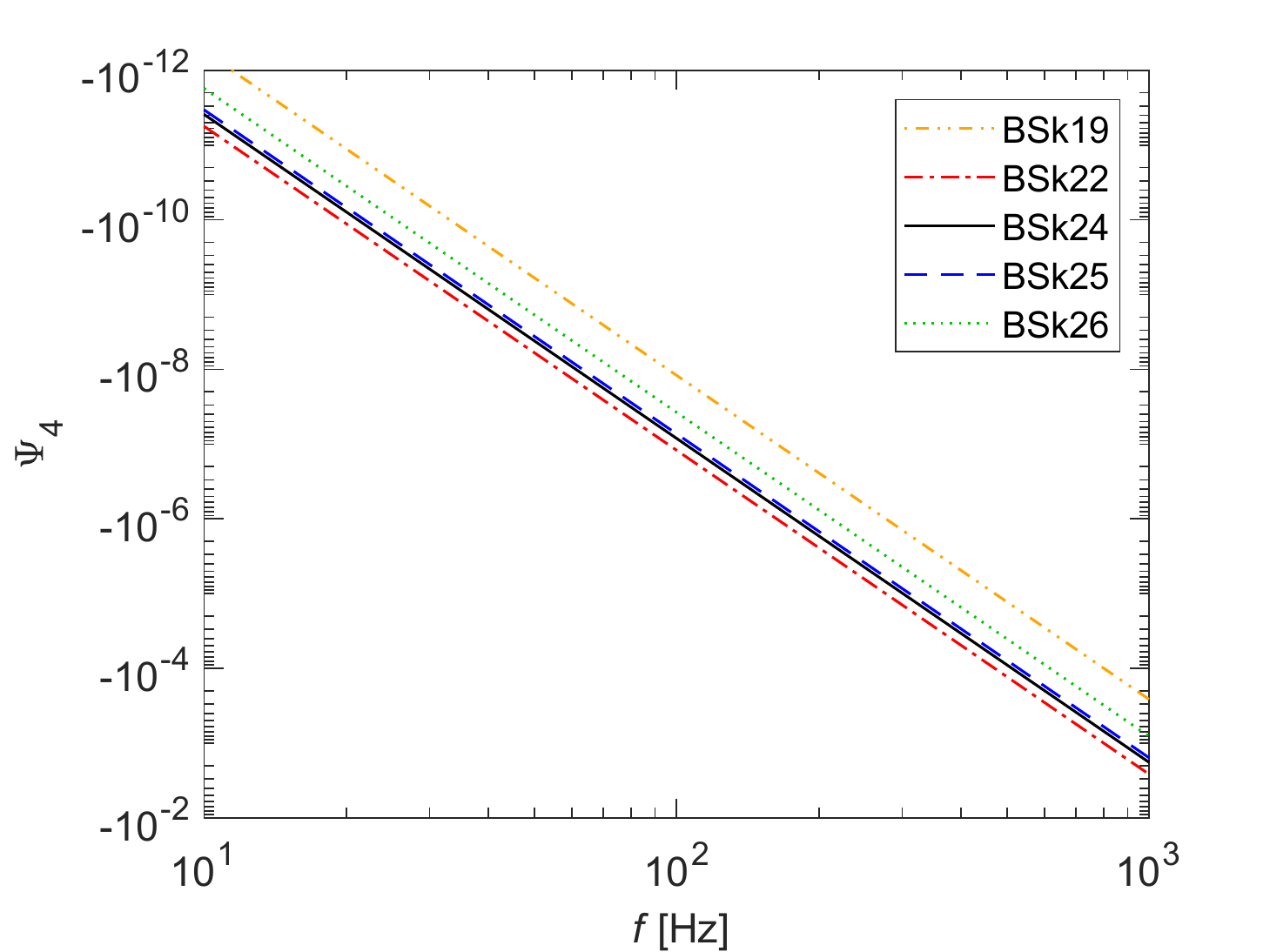}
\caption{(Color online) Same as Fig.~\ref{fig:Psi2-BSk} for the gravitoelectric Love number $k_4$.}
\label{fig:Psi4-BSk}
\end{center}
\end{figure}

\begin{figure}[ht!]
\begin{center}
\includegraphics[width=\textwidth]{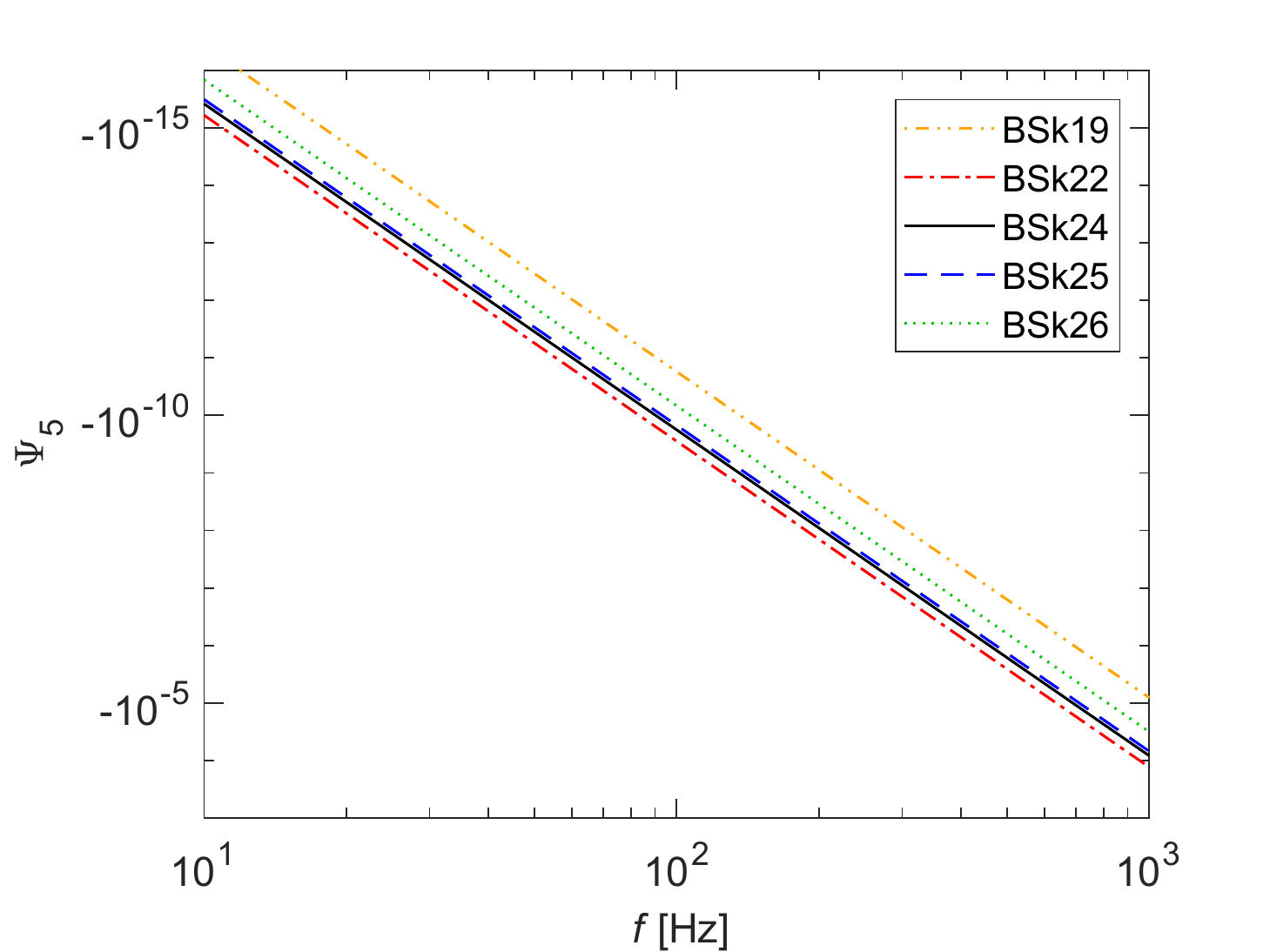}
\caption{(Color online) Same as Fig.~\ref{fig:Psi2-BSk} for the gravitoelectric Love number $k_5$.}
\label{fig:Psi5-BSk}
\end{center}
\end{figure}

\begin{figure}[ht!]
\begin{center}
\includegraphics[width=\textwidth]{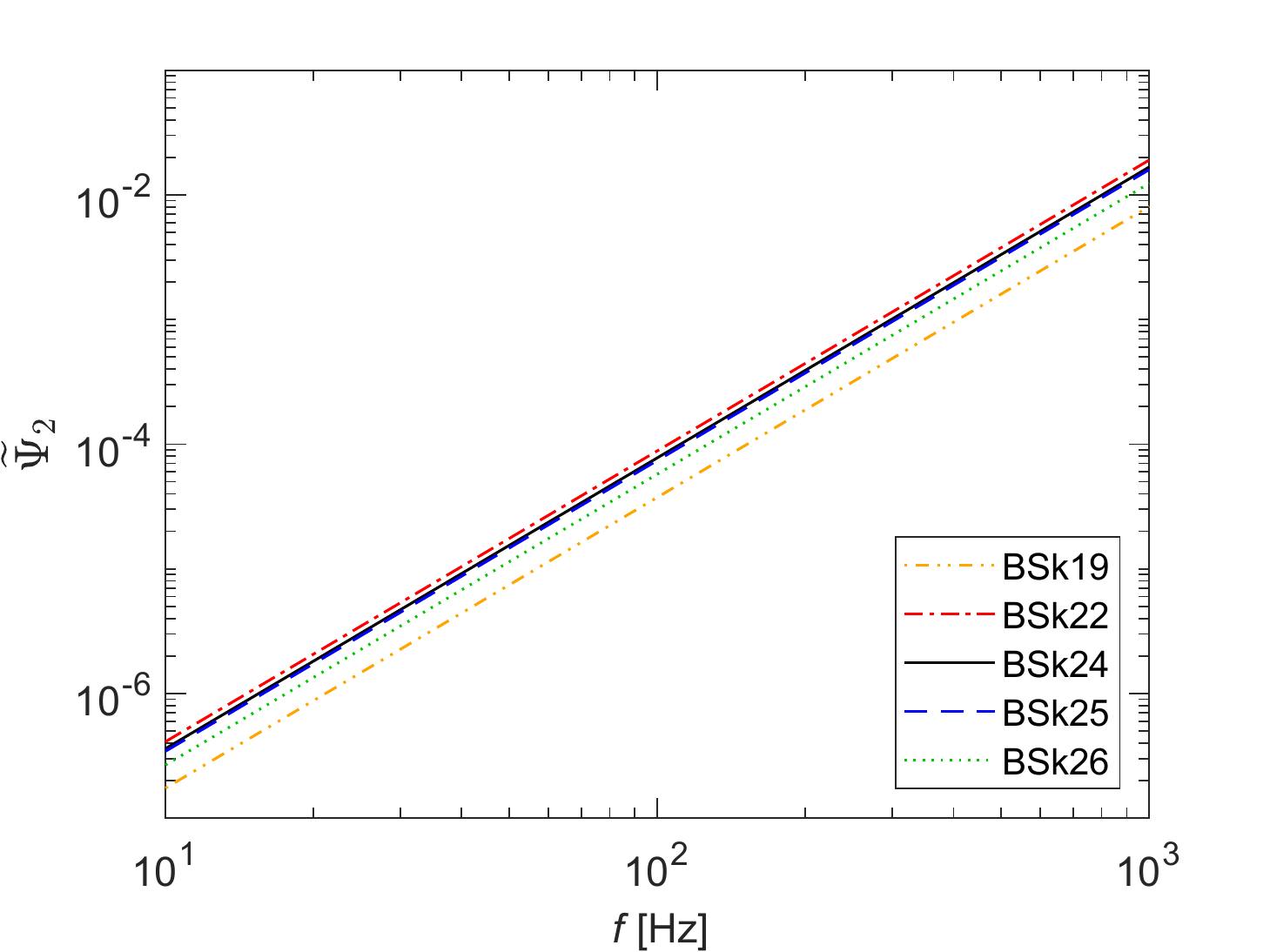}
\caption{(Color online) Contribution from the gravitomagnetic Love number $j_2$ to the phase~\eqref{eq:Psi2tilde} of the gravitational-wave signal from binary NS inspiral as a function of the frequency $f$ for different EoSs. Calculations were performed for an irrotational fluid and NSs with equal masses of 1.4 $M_\odot$.}
\label{fig:Psi2tilde-BSk}
\end{center}
\end{figure}

\begin{figure}[ht!]
\begin{center}
\includegraphics[width=\textwidth]{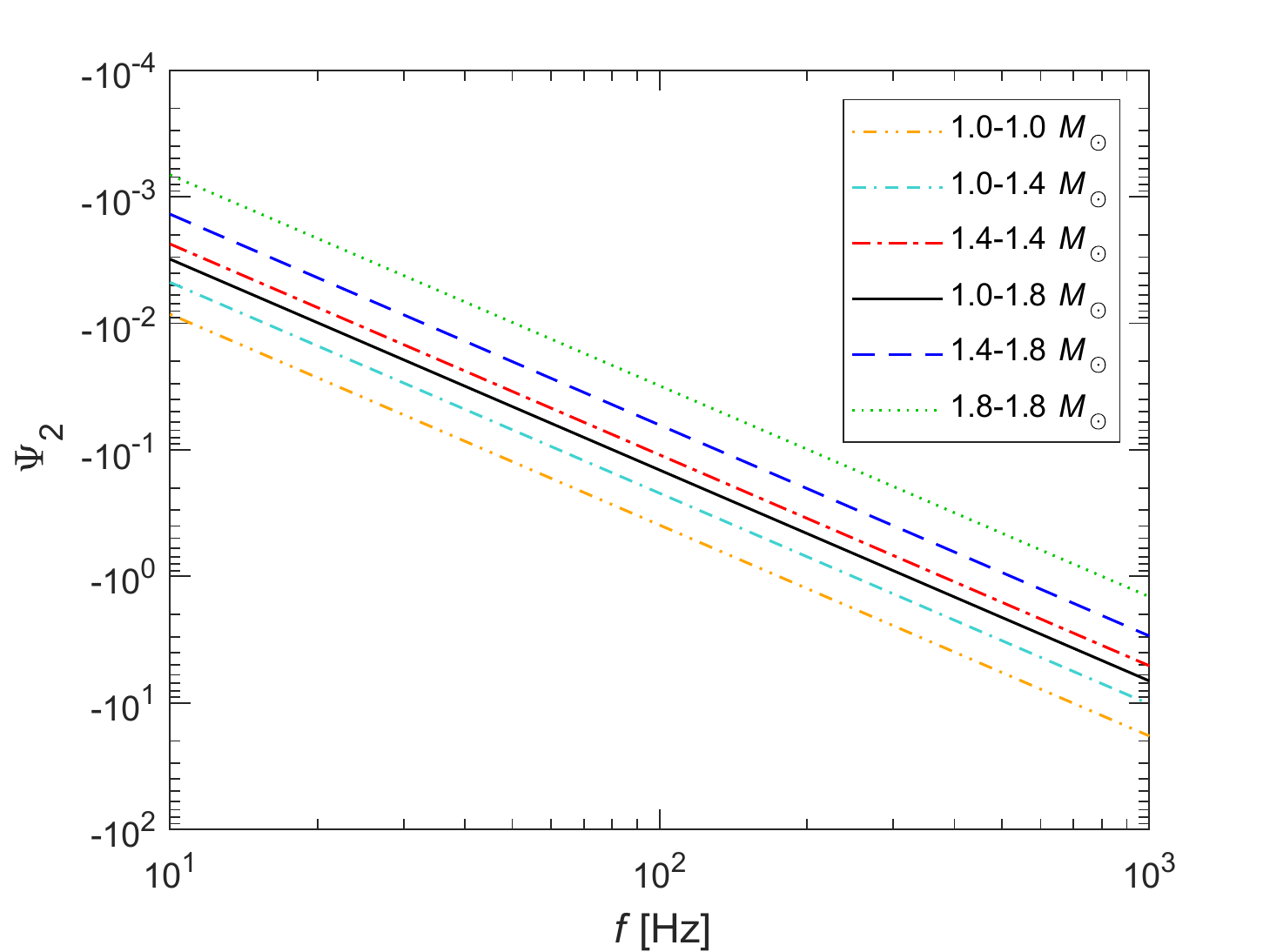}
\caption{(Color online) Contribution from the gravitoelectric Love number $k_2$ to the phase~\eqref{eq:Psil} of the gravitational-wave signal from binary NS inspiral as a function of the frequency $f$ for different NS masses. Calculations were performed using the unified EoS BSk24.}
\label{fig:Psi2}
\end{center}
\end{figure}

\begin{figure}[ht!]
\begin{center}
\includegraphics[width=\textwidth]{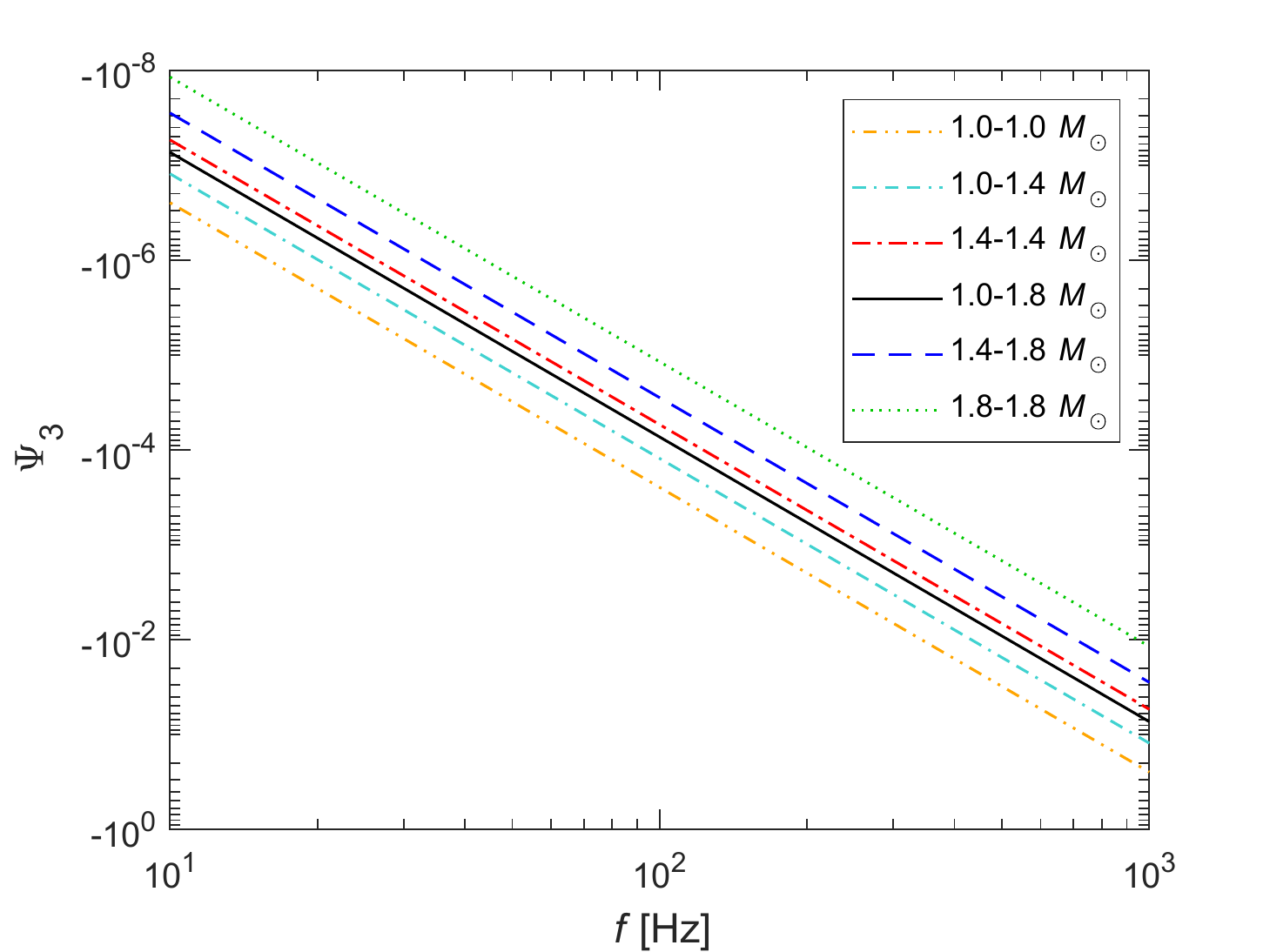}
\caption{(Color online) Same as Fig.~\ref{fig:Psi2} for  the gravitoelectric Love number $k_3$.} 
\label{fig:Psi3}
\end{center}
\end{figure}

\begin{figure}[ht!]
\begin{center}
\includegraphics[width=\textwidth]{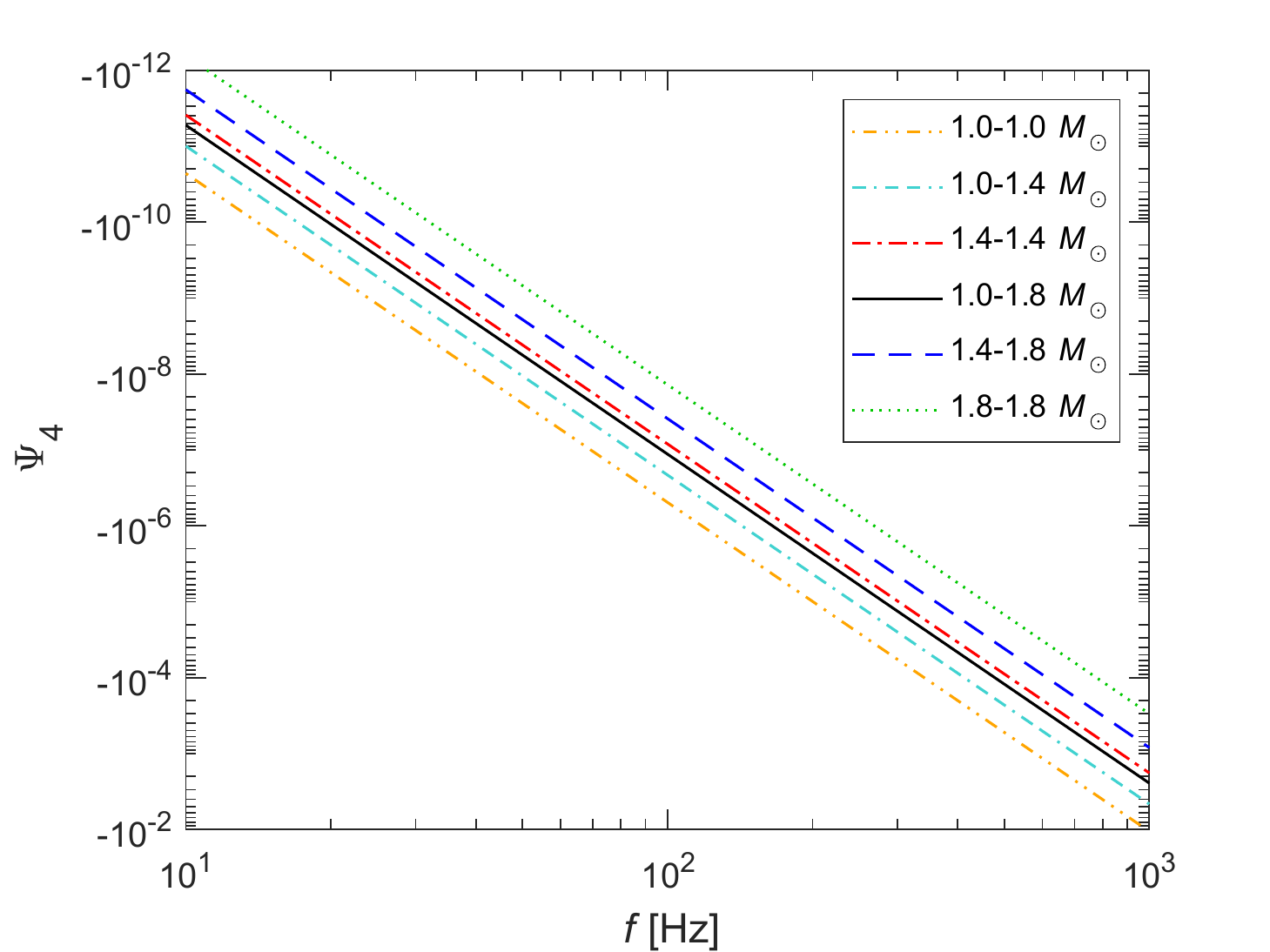}
\caption{(Color online) Same as Fig.~\ref{fig:Psi2} for  the gravitoelectric Love number $k_4$. }
\label{fig:Psi4}
\end{center}
\end{figure}

\begin{figure}[ht!]
\begin{center}
\includegraphics[width=\textwidth]{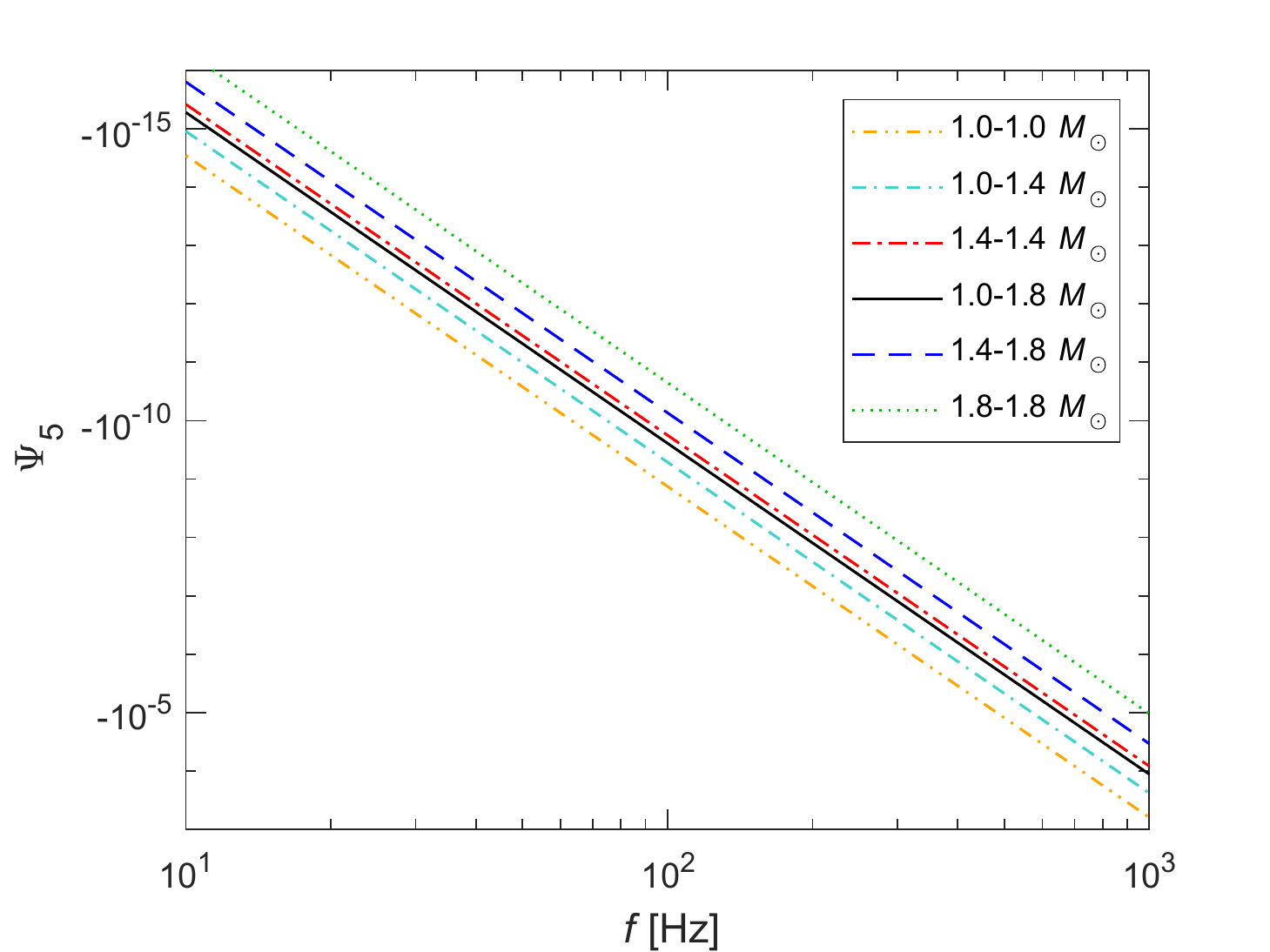}
\caption{(Color online) Same as Fig.~\ref{fig:Psi2} for  the gravitoelectric Love number $k_5$.}
\label{fig:Psi5}
\end{center}
\end{figure}

\begin{figure}[ht!]
\begin{center}
\includegraphics[width=\textwidth]{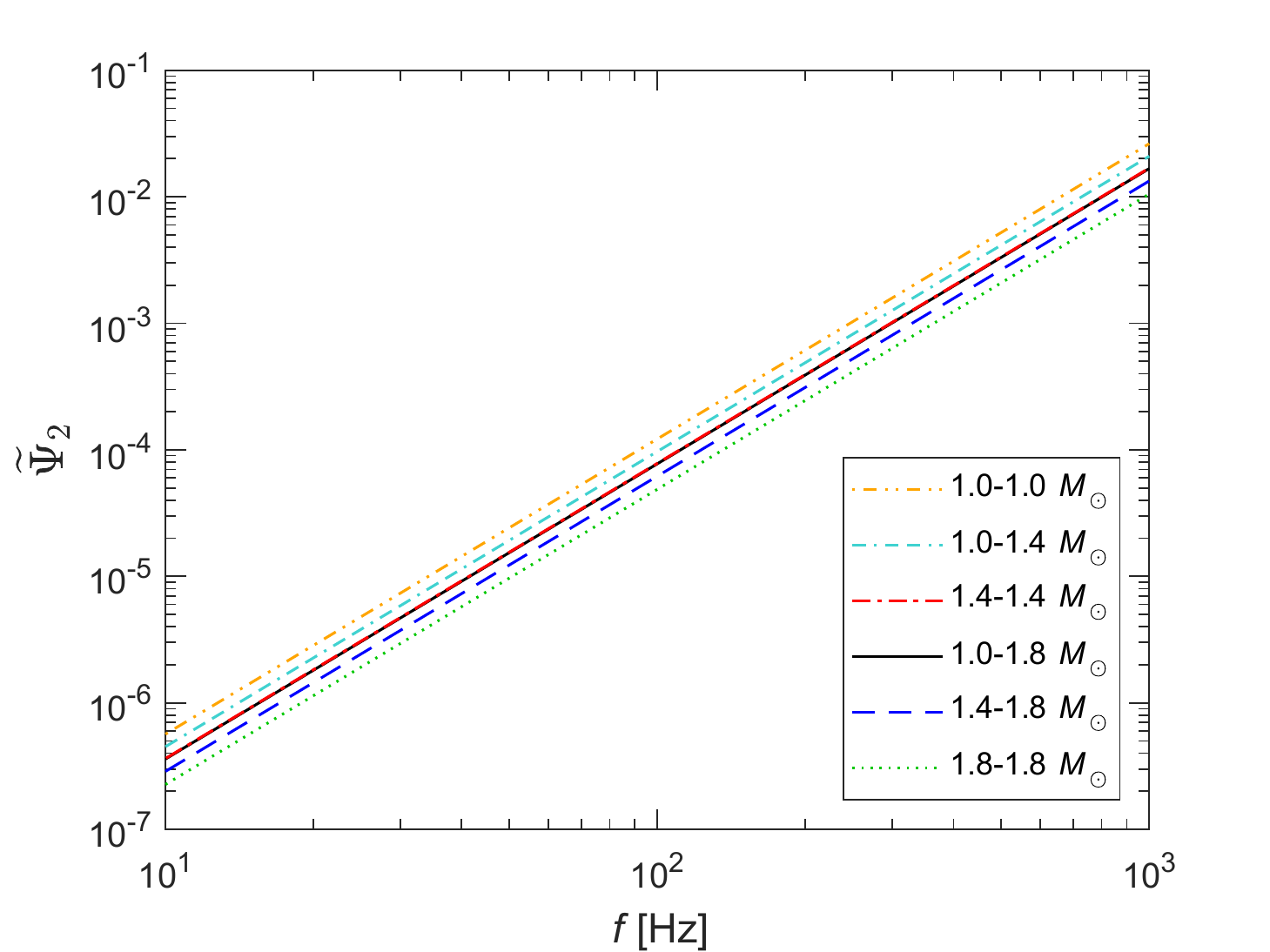}
\caption{(Color online) Contribution from the gravitomagnetic Love number $j_2$ to the phase~\eqref{eq:Psi2tilde} of the gravitational-wave signal  from binary NS inspiral as a function of the frequency $f$ for different NS masses. Calculations were performed using the unified EoS BSk24 for an irrotational fluid.}
\label{fig:Psi2tilde}
\end{center}
\end{figure}

The tidal corrections to the phase of the gravitational waveforms are plotted in Figs~\ref{fig:Psi2-BSk}, \ref{fig:Psi3-BSk}, \ref{fig:Psi4-BSk}, \ref{fig:Psi5-BSk} and \ref{fig:Psi2tilde-BSk} for different frequencies and considering binary systems with both NSs having a mass $M=1.4 M_\odot$. Because the PN approximation breaks down near the merger, the phases are only plotted up to a frequency $f=1000$~Hz. Extracting information about the symmetry energy from the gravitational-wave signal during the inspiral phase will be very difficult, as can be seen by comparing results obtained for EoSs BSk22, BSk24, and BSk25. On the other hand, the stiffness of the neutron-matter EoS leaves a clear imprint on the waveform. The comparison of the  results obtained for EoSs BSk19, BSk24, and BSk26 shows that the softer the EoS is, the more pronounced are the tidal effects. 

As shown in Figs.~~\ref{fig:Psi2}, \ref{fig:Psi3}, \ref{fig:Psi4}, \ref{fig:Psi5} and \ref{fig:Psi2tilde}, the relative importance of the different $\ell$-terms is found to follow the same hierarchy as the Love numbers. In particular, the tidal correction $\Psi_3$ associated with the gravitoelectric Love  number $k_3$ is about two orders of magnitude smaller than the leading tidal term $\Psi_2$. The correction $\widetilde{\Psi}_2$ induced by the gravitomagnetic Love number $j_2$ lies between those associated with $k_3$ and $k_4$. 

\subsection{Role of the crust}

Following our previous study~\cite{perot2020}, we have pursued our analysis of the role of the crust in the tidal deformability of a NS. To this end, we have compared results obtained using the unified EoS BSk24 to those obtained considering a putative NS made entirely of homogeneous matter (see Ref.~\cite{perot2020}). Although the existence of a mantle of nuclear pastas beneath the crust was not considered in Ref.~\cite{pearson2018}, it was later shown to have a negligible impact on the EoS
~\cite{pearson2019}. 

\begin{figure}[ht!]
\begin{center}
\includegraphics[width=\textwidth]{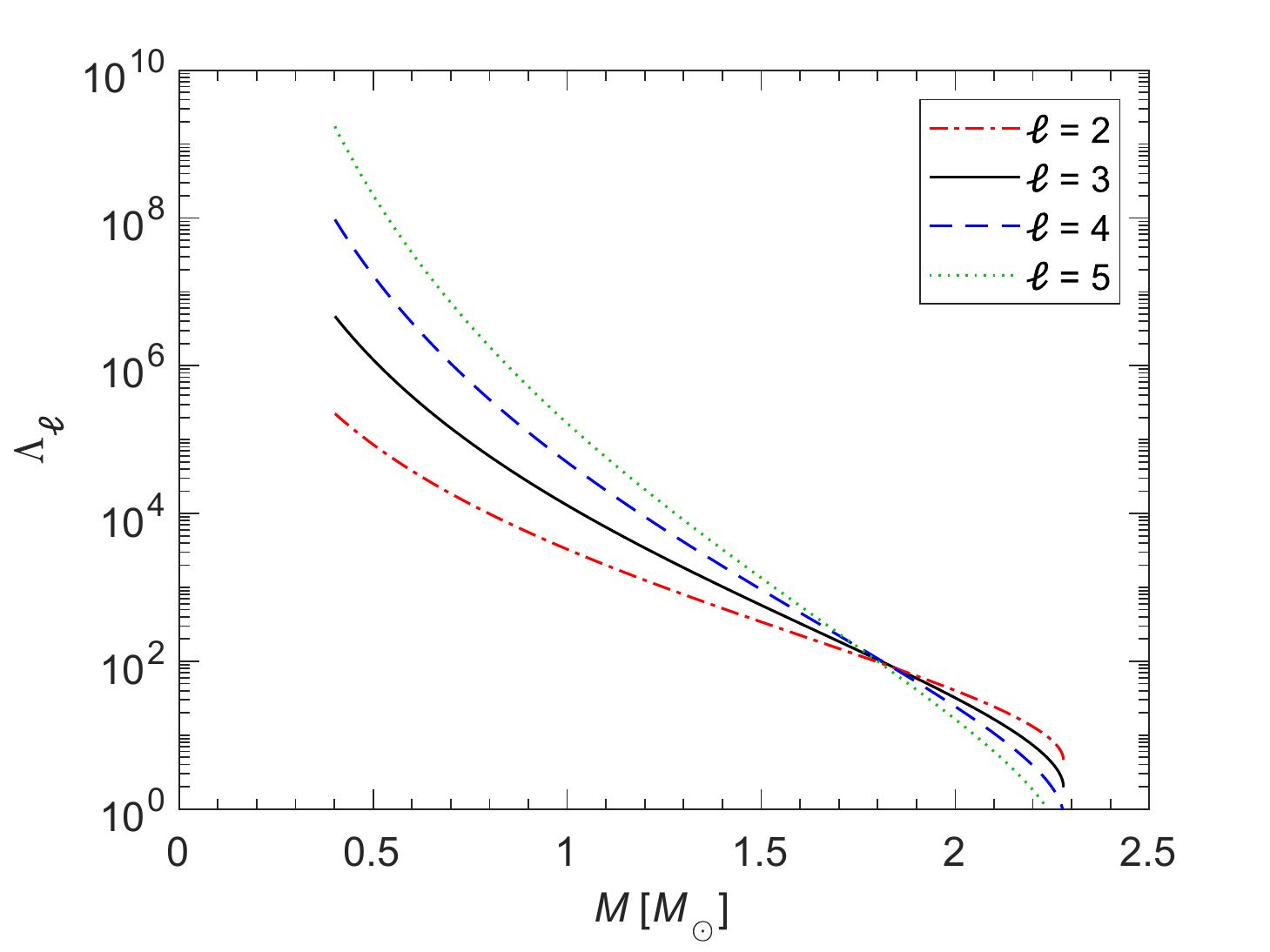}
\caption{(Color online) Dimensionless gravitoelectric tidal deformability coefficients as a function of the gravitational mass $M$ of a NS. Results obtained without crust are indistinguishable from those with crust. Calculations were performed using the Brussels-Montreal nuclear energy-density functional BSk24.}
\label{fig:elec-tidal-crust}
\end{center}
\end{figure}

\begin{figure}[ht!]
\begin{center}
\includegraphics[width=\textwidth]{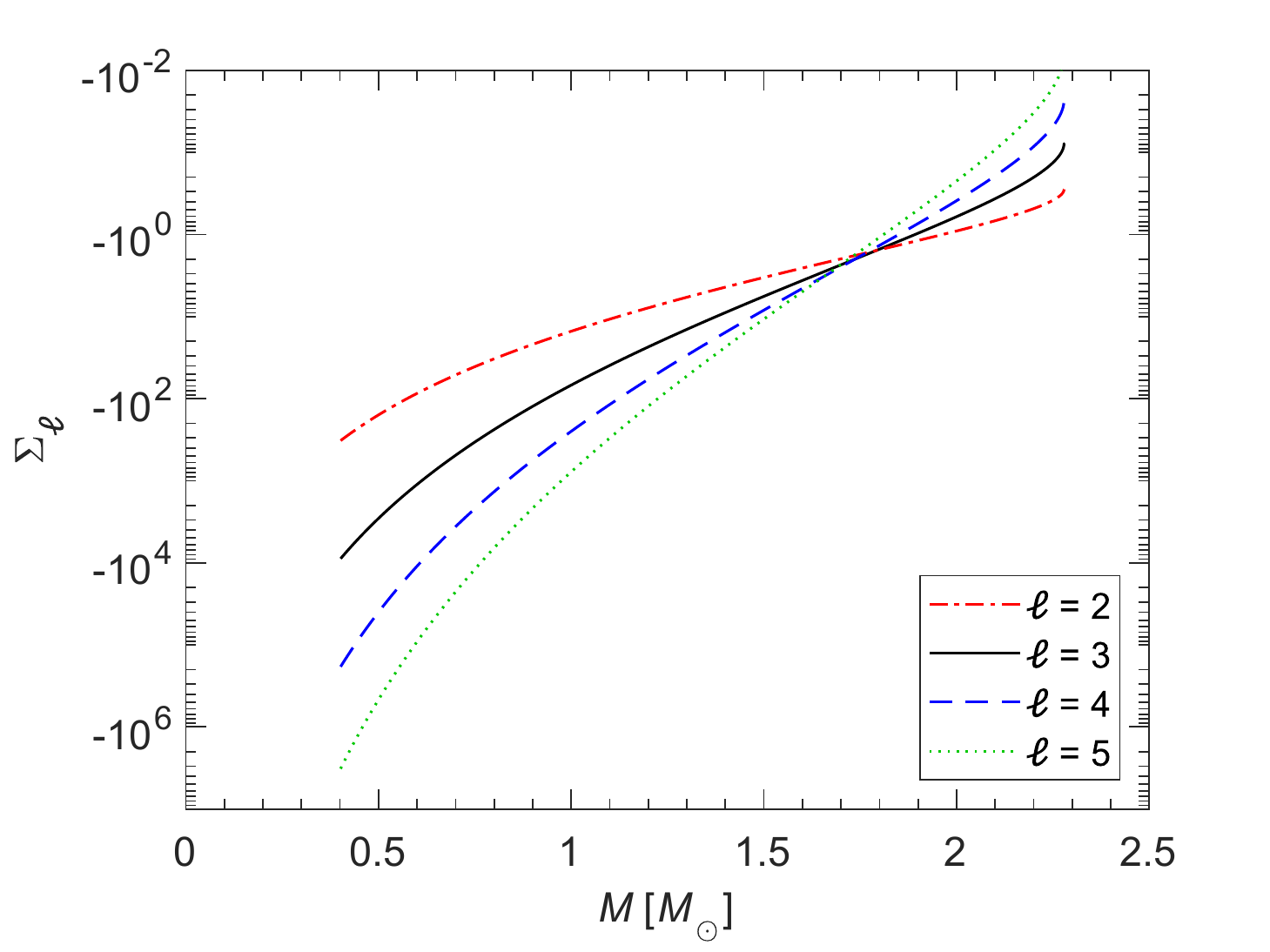}
\caption{(Color online) Dimensionless gravitomagnetic tidal deformability coefficients as a function of the gravitational mass $M$ of a NS (irrotational fluid). Results obtained without crust are indistinguishable from those with crust. Calculations were performed using the Brussels-Montreal nuclear energy-density functional BSk24.}
\label{fig:mag-tidal-crust}
\end{center}
\end{figure}

As shown in Figs.~\ref{fig:k2345} and \ref{fig:j2345}, the presence of the crust reduces systematically the magnitude of the Love numbers. However, this reduction is found to be almost exactly compensated by the increase in the stellar radius so that the dimensionless tidal coefficients $\Lambda_\ell$ and $\Sigma_\ell$, plotted in Figs.~\ref{fig:elec-tidal-crust} and \ref{fig:mag-tidal-crust} respectively,  remain essentially unchanged. This illustrates the importance of consistently determining both the structure of a NS and the tidal deformability coefficients using the same EoS.

\section{Conclusions}

Pursuing our investigations of the role of nuclear-matter properties in the tidal deformability of a cold nonaccreted NS~\cite{perot2019,perot2020}, we have considered higher-order multipoles up to $\ell=5$ for both gravitoelectric and gravitomagnetic tidal perturbations. 

Using the family of Brussels-Montreal unified EoSs~\cite{potekhin2013,pearson2018,pearson2019}, we have found that the associated Love numbers are mainly sensitive to the stiffness of the neutron-matter EoS, and to a lesser extent on the symmetry energy. The EoS of the NS crust also plays an important role, however it hardly has any observable signature: the influence of the crust on the Love numbers is almost exactly compensated by that on the NS radius provided the NS structure and tidal deformations are calculated consistently. 

The gravitomagnetic Love numbers $j_\ell$ are found to be about an order of magnitude smaller than their gravitoelectric counterpart $k_\ell$. Their magnitudes decrease rapidly with increasing $\ell$. Tidal corrections to the phase of the gravitational-wave signals from binary NS inspiral, which are mainly sensitive to the stiffness of the neutron-matter EoS, exhibit a similar hierarchy. In particular, the leading correction induced by gravitomagnetic effects is found to be smaller than the octupole gravitoelectric correction but larger than the one from the hexadecupole moment. In addition to probing dense-matter properties, measurement of gravitomagnetic effects could shed light on the internal dynamics of NSs. Merely determining the sign of $j_2$ would thus discriminate between static and irrotational fluids. Moreover,  observations of gravitomagnetic tidal deformations could also potentially provide additional tests of general relativity since such effects are absent in Newtonian theory. 

Although higher-order tidal effects are small, they will be potentially observable with the upcoming third generation gravitational-wave detectors such as the Einstein telescope~\cite{maggiore2020}, thus providing additional information on dense matter.

\begin{acknowledgments}
The authors thank M. Bejger for valuable discussions. This work was financially supported by Fonds de la Recherche Scientifique (Belgium) under Grant No. PDR T.004320 and the European Cooperation in Science and Technology Action (EU) CA16214. L. P. is a FRIA grantee of the Fonds de la Recherche Scientifique (Belgium).   
\end{acknowledgments}

\bibliography{biblio}

\begin{thebibliography}{76}%
\makeatletter
\providecommand \@ifxundefined [1]{%
 \@ifx{#1\undefined}
}%
\providecommand \@ifnum [1]{%
 \ifnum #1\expandafter \@firstoftwo
 \else \expandafter \@secondoftwo
 \fi
}%
\providecommand \@ifx [1]{%
 \ifx #1\expandafter \@firstoftwo
 \else \expandafter \@secondoftwo
 \fi
}%
\providecommand \natexlab [1]{#1}%
\providecommand \enquote  [1]{``#1''}%
\providecommand \bibnamefont  [1]{#1}%
\providecommand \bibfnamefont [1]{#1}%
\providecommand \citenamefont [1]{#1}%
\providecommand \href@noop [0]{\@secondoftwo}%
\providecommand \href [0]{\begingroup \@sanitize@url \@href}%
\providecommand \@href[1]{\@@startlink{#1}\@@href}%
\providecommand \@@href[1]{\endgroup#1\@@endlink}%
\providecommand \@sanitize@url [0]{\catcode `\\12\catcode `\$12\catcode
  `\&12\catcode `\#12\catcode `\^12\catcode `\_12\catcode `\%12\relax}%
\providecommand \@@startlink[1]{}%
\providecommand \@@endlink[0]{}%
\providecommand \url  [0]{\begingroup\@sanitize@url \@url }%
\providecommand \@url [1]{\endgroup\@href {#1}{\urlprefix }}%
\providecommand \urlprefix  [0]{URL }%
\providecommand \Eprint [0]{\href }%
\providecommand \doibase [0]{http://dx.doi.org/}%
\providecommand \selectlanguage [0]{\@gobble}%
\providecommand \bibinfo  [0]{\@secondoftwo}%
\providecommand \bibfield  [0]{\@secondoftwo}%
\providecommand \translation [1]{[#1]}%
\providecommand \BibitemOpen [0]{}%
\providecommand \bibitemStop [0]{}%
\providecommand \bibitemNoStop [0]{.\EOS\space}%
\providecommand \EOS [0]{\spacefactor3000\relax}%
\providecommand \BibitemShut  [1]{\csname bibitem#1\endcsname}%
\let\auto@bib@innerbib\@empty
\bibitem [{\citenamefont {{Guerra Chaves}}\ and\ \citenamefont
  {{Hinderer}}(2019)}]{chaves2019}%
  \BibitemOpen
  \bibfield  {author} {\bibinfo {author} {\bibfnamefont {A.}~\bibnamefont
  {{Guerra Chaves}}}\ and\ \bibinfo {author} {\bibfnamefont {T.}~\bibnamefont
  {{Hinderer}}},\ }\href {\doibase 10.1088/1361-6471/ab45be} {\bibfield
  {journal} {\bibinfo  {journal} {Journal of Physics G Nuclear Physics}\
  }\textbf {\bibinfo {volume} {46}},\ \bibinfo {eid} {123002} (\bibinfo {year}
  {2019})}\BibitemShut {NoStop}%
\bibitem [{\citenamefont {{Abbott}}\ \emph
  {et~al.}(2020{\natexlab{a}})\citenamefont {{Abbott}}, \citenamefont
  {{Abbott}}, \citenamefont {{Abraham}}, \citenamefont {{Acernese}},
  \citenamefont {{Ackley}}, \citenamefont {{Adams}}, \citenamefont
  {{Adhikari}}, \citenamefont {{Adya}}, \citenamefont {{Affeldt}},
  \citenamefont {{Agathos}},\ and\ \citenamefont {et~al.}}]{ligo2020b}%
  \BibitemOpen
  \bibfield  {author} {\bibinfo {author} {\bibfnamefont {R.}~\bibnamefont
  {{Abbott}}}, \bibinfo {author} {\bibfnamefont {T.~D.}\ \bibnamefont
  {{Abbott}}}, \bibinfo {author} {\bibfnamefont {S.}~\bibnamefont {{Abraham}}},
  \bibinfo {author} {\bibfnamefont {F.}~\bibnamefont {{Acernese}}}, \bibinfo
  {author} {\bibfnamefont {K.}~\bibnamefont {{Ackley}}}, \bibinfo {author}
  {\bibfnamefont {C.}~\bibnamefont {{Adams}}}, \bibinfo {author} {\bibfnamefont
  {R.~X.}\ \bibnamefont {{Adhikari}}}, \bibinfo {author} {\bibfnamefont
  {V.~B.}\ \bibnamefont {{Adya}}}, \bibinfo {author} {\bibfnamefont
  {C.}~\bibnamefont {{Affeldt}}}, \bibinfo {author} {\bibfnamefont
  {M.}~\bibnamefont {{Agathos}}}, \ and\ \bibinfo {author} {\bibnamefont
  {et~al.}},\ }\href {\doibase 10.3847/2041-8213/ab960f} {\bibfield  {journal}
  {\bibinfo  {journal} {\apjl}\ }\textbf {\bibinfo {volume} {896}},\ \bibinfo
  {eid} {L44} (\bibinfo {year} {2020}{\natexlab{a}})}\BibitemShut {NoStop}%
\bibitem [{\citenamefont {{Abbott}}\ \emph
  {et~al.}(2018{\natexlab{a}})\citenamefont {{Abbott}}, \citenamefont
  {{Abbott}}, \citenamefont {{Abbott}}, \citenamefont {{Abernathy}},
  \citenamefont {{Acernese}}, \citenamefont {{Ackley}}, \citenamefont
  {{Adams}}, \citenamefont {{Adams}}, \citenamefont {{Addesso}}, \citenamefont
  {{Adhikari}},\ and\ \citenamefont {et~al.}}]{abbott2018}%
  \BibitemOpen
  \bibfield  {author} {\bibinfo {author} {\bibfnamefont {B.~P.}\ \bibnamefont
  {{Abbott}}}, \bibinfo {author} {\bibfnamefont {R.}~\bibnamefont {{Abbott}}},
  \bibinfo {author} {\bibfnamefont {T.~D.}\ \bibnamefont {{Abbott}}}, \bibinfo
  {author} {\bibfnamefont {M.~R.}\ \bibnamefont {{Abernathy}}}, \bibinfo
  {author} {\bibfnamefont {F.}~\bibnamefont {{Acernese}}}, \bibinfo {author}
  {\bibfnamefont {K.}~\bibnamefont {{Ackley}}}, \bibinfo {author}
  {\bibfnamefont {C.}~\bibnamefont {{Adams}}}, \bibinfo {author} {\bibfnamefont
  {T.}~\bibnamefont {{Adams}}}, \bibinfo {author} {\bibfnamefont
  {P.}~\bibnamefont {{Addesso}}}, \bibinfo {author} {\bibfnamefont
  {R.}~\bibnamefont {{Adhikari}}}, \ and\ \bibinfo {author} {\bibnamefont
  {et~al.}},\ }\href {\doibase 10.1007/s41114-018-0012-9} {\bibfield  {journal}
  {\bibinfo  {journal} {Living Reviews in Relativity}\ }\textbf {\bibinfo
  {volume} {21}},\ \bibinfo {eid} {3} (\bibinfo {year}
  {2018}{\natexlab{a}})}\BibitemShut {NoStop}%
\bibitem [{\citenamefont {{Abbott}}\ \emph {et~al.}(2017)\citenamefont
  {{Abbott}}, \citenamefont {{Abbott}}, \citenamefont {{Abbott}}, \citenamefont
  {{Acernese}}, \citenamefont {{Ackley}}, \citenamefont {{Adams}},
  \citenamefont {{Adams}}, \citenamefont {{Addesso}}, \citenamefont
  {{Adhikari}}, \citenamefont {{Adya}},\ and\ \citenamefont
  {et~al.}}]{ligo2017inspiral}%
  \BibitemOpen
  \bibfield  {author} {\bibinfo {author} {\bibfnamefont {B.~P.}\ \bibnamefont
  {{Abbott}}}, \bibinfo {author} {\bibfnamefont {R.}~\bibnamefont {{Abbott}}},
  \bibinfo {author} {\bibfnamefont {T.~D.}\ \bibnamefont {{Abbott}}}, \bibinfo
  {author} {\bibfnamefont {F.}~\bibnamefont {{Acernese}}}, \bibinfo {author}
  {\bibfnamefont {K.}~\bibnamefont {{Ackley}}}, \bibinfo {author}
  {\bibfnamefont {C.}~\bibnamefont {{Adams}}}, \bibinfo {author} {\bibfnamefont
  {T.}~\bibnamefont {{Adams}}}, \bibinfo {author} {\bibfnamefont
  {P.}~\bibnamefont {{Addesso}}}, \bibinfo {author} {\bibfnamefont {R.~X.}\
  \bibnamefont {{Adhikari}}}, \bibinfo {author} {\bibfnamefont {V.~B.}\
  \bibnamefont {{Adya}}}, \ and\ \bibinfo {author} {\bibnamefont {et~al.}},\
  }\href {\doibase 10.1103/PhysRevLett.119.161101} {\bibfield  {journal}
  {\bibinfo  {journal} {\prl}\ }\textbf {\bibinfo {volume} {119}},\ \bibinfo
  {eid} {161101} (\bibinfo {year} {2017})}\BibitemShut {NoStop}%
\bibitem [{\citenamefont {{De}}\ \emph {et~al.}(2018)\citenamefont {{De}},
  \citenamefont {{Finstad}}, \citenamefont {{Lattimer}}, \citenamefont
  {{Brown}}, \citenamefont {{Berger}},\ and\ \citenamefont {{Biwer}}}]{de2018}%
  \BibitemOpen
  \bibfield  {author} {\bibinfo {author} {\bibfnamefont {S.}~\bibnamefont
  {{De}}}, \bibinfo {author} {\bibfnamefont {D.}~\bibnamefont {{Finstad}}},
  \bibinfo {author} {\bibfnamefont {J.~M.}\ \bibnamefont {{Lattimer}}},
  \bibinfo {author} {\bibfnamefont {D.~A.}\ \bibnamefont {{Brown}}}, \bibinfo
  {author} {\bibfnamefont {E.}~\bibnamefont {{Berger}}}, \ and\ \bibinfo
  {author} {\bibfnamefont {C.~M.}\ \bibnamefont {{Biwer}}},\ }\href {\doibase
  10.1103/PhysRevLett.121.091102} {\bibfield  {journal} {\bibinfo  {journal}
  {\prl}\ }\textbf {\bibinfo {volume} {121}},\ \bibinfo {eid} {091102}
  (\bibinfo {year} {2018})}\BibitemShut {NoStop}%
\bibitem [{\citenamefont {Sathyaprakash}\ and\ \citenamefont
  {Dhurandhar}(1991)}]{sathyaprakash1991}%
  \BibitemOpen
  \bibfield  {author} {\bibinfo {author} {\bibfnamefont {B.~S.}\ \bibnamefont
  {Sathyaprakash}}\ and\ \bibinfo {author} {\bibfnamefont {S.~V.}\ \bibnamefont
  {Dhurandhar}},\ }\href {\doibase 10.1103/PhysRevD.44.3819} {\bibfield
  {journal} {\bibinfo  {journal} {Phys. Rev. D}\ }\textbf {\bibinfo {volume}
  {44}},\ \bibinfo {pages} {3819} (\bibinfo {year} {1991})}\BibitemShut
  {NoStop}%
\bibitem [{\citenamefont {Mik\'oczi}\ \emph {et~al.}(2005)\citenamefont
  {Mik\'oczi}, \citenamefont {Vas\'uth},\ and\ \citenamefont
  {Gergely}}]{mikoczi2005}%
  \BibitemOpen
  \bibfield  {author} {\bibinfo {author} {\bibfnamefont {B.}~\bibnamefont
  {Mik\'oczi}}, \bibinfo {author} {\bibfnamefont {M.}~\bibnamefont {Vas\'uth}},
  \ and\ \bibinfo {author} {\bibfnamefont {L.~A.}\ \bibnamefont {Gergely}},\
  }\href {\doibase 10.1103/PhysRevD.71.124043} {\bibfield  {journal} {\bibinfo
  {journal} {Phys. Rev. D}\ }\textbf {\bibinfo {volume} {71}},\ \bibinfo
  {pages} {124043} (\bibinfo {year} {2005})}\BibitemShut {NoStop}%
\bibitem [{\citenamefont {Buonanno}\ \emph {et~al.}(2009)\citenamefont
  {Buonanno}, \citenamefont {Iyer}, \citenamefont {Ochsner}, \citenamefont
  {Pan},\ and\ \citenamefont {Sathyaprakash}}]{buonanno2009}%
  \BibitemOpen
  \bibfield  {author} {\bibinfo {author} {\bibfnamefont {A.}~\bibnamefont
  {Buonanno}}, \bibinfo {author} {\bibfnamefont {B.~R.}\ \bibnamefont {Iyer}},
  \bibinfo {author} {\bibfnamefont {E.}~\bibnamefont {Ochsner}}, \bibinfo
  {author} {\bibfnamefont {Y.}~\bibnamefont {Pan}}, \ and\ \bibinfo {author}
  {\bibfnamefont {B.~S.}\ \bibnamefont {Sathyaprakash}},\ }\href {\doibase
  10.1103/PhysRevD.80.084043} {\bibfield  {journal} {\bibinfo  {journal} {Phys.
  Rev. D}\ }\textbf {\bibinfo {volume} {80}},\ \bibinfo {pages} {084043}
  (\bibinfo {year} {2009})}\BibitemShut {NoStop}%
\bibitem [{\citenamefont {Arun}\ \emph {et~al.}(2009)\citenamefont {Arun},
  \citenamefont {Buonanno}, \citenamefont {Faye},\ and\ \citenamefont
  {Ochsner}}]{arun2009}%
  \BibitemOpen
  \bibfield  {author} {\bibinfo {author} {\bibfnamefont {K.~G.}\ \bibnamefont
  {Arun}}, \bibinfo {author} {\bibfnamefont {A.}~\bibnamefont {Buonanno}},
  \bibinfo {author} {\bibfnamefont {G.}~\bibnamefont {Faye}}, \ and\ \bibinfo
  {author} {\bibfnamefont {E.}~\bibnamefont {Ochsner}},\ }\href {\doibase
  10.1103/PhysRevD.79.104023} {\bibfield  {journal} {\bibinfo  {journal} {Phys.
  Rev. D}\ }\textbf {\bibinfo {volume} {79}},\ \bibinfo {pages} {104023}
  (\bibinfo {year} {2009})}\BibitemShut {NoStop}%
\bibitem [{\citenamefont {Vines}\ \emph {et~al.}(2011)\citenamefont {Vines},
  \citenamefont {Flanagan},\ and\ \citenamefont {Hinderer}}]{vines2011}%
  \BibitemOpen
  \bibfield  {author} {\bibinfo {author} {\bibfnamefont {J.}~\bibnamefont
  {Vines}}, \bibinfo {author} {\bibfnamefont {E.~E.}\ \bibnamefont {Flanagan}},
  \ and\ \bibinfo {author} {\bibfnamefont {T.}~\bibnamefont {Hinderer}},\
  }\href {\doibase 10.1103/PhysRevD.83.084051} {\bibfield  {journal} {\bibinfo
  {journal} {Phys. Rev. D}\ }\textbf {\bibinfo {volume} {83}},\ \bibinfo
  {pages} {084051} (\bibinfo {year} {2011})}\BibitemShut {NoStop}%
\bibitem [{\citenamefont {{Boh{\'e}}}\ \emph {et~al.}(2013)\citenamefont
  {{Boh{\'e}}}, \citenamefont {{Marsat}},\ and\ \citenamefont
  {{Blanchet}}}]{bohe2013}%
  \BibitemOpen
  \bibfield  {author} {\bibinfo {author} {\bibfnamefont {A.}~\bibnamefont
  {{Boh{\'e}}}}, \bibinfo {author} {\bibfnamefont {S.}~\bibnamefont
  {{Marsat}}}, \ and\ \bibinfo {author} {\bibfnamefont {L.}~\bibnamefont
  {{Blanchet}}},\ }\href {\doibase 10.1088/0264-9381/30/13/135009} {\bibfield
  {journal} {\bibinfo  {journal} {Classical and Quantum Gravity}\ }\textbf
  {\bibinfo {volume} {30}},\ \bibinfo {eid} {135009} (\bibinfo {year}
  {2013})}\BibitemShut {NoStop}%
\bibitem [{\citenamefont {{Blanchet}}(2014)}]{blanchet2014}%
  \BibitemOpen
  \bibfield  {author} {\bibinfo {author} {\bibfnamefont {L.}~\bibnamefont
  {{Blanchet}}},\ }\href {\doibase 10.12942/lrr-2014-2} {\bibfield  {journal}
  {\bibinfo  {journal} {Living Reviews in Relativity}\ }\textbf {\bibinfo
  {volume} {17}},\ \bibinfo {eid} {2} (\bibinfo {year} {2014})}\BibitemShut
  {NoStop}%
\bibitem [{\citenamefont {{Blanchet}}(2019)}]{blanchet2019}%
  \BibitemOpen
  \bibfield  {author} {\bibinfo {author} {\bibfnamefont {L.}~\bibnamefont
  {{Blanchet}}},\ }\href {\doibase 10.1016/j.crhy.2019.02.004} {\bibfield
  {journal} {\bibinfo  {journal} {Comptes Rendus Physique}\ }\textbf {\bibinfo
  {volume} {20}},\ \bibinfo {pages} {507} (\bibinfo {year} {2019})}\BibitemShut
  {NoStop}%
\bibitem [{\citenamefont {{Abbott}}\ \emph
  {et~al.}(2018{\natexlab{b}})\citenamefont {{Abbott}}, \citenamefont
  {{Abbott}}, \citenamefont {{Abbott}}, \citenamefont {{Acernese}},
  \citenamefont {{Ackley}}, \citenamefont {{Adams}}, \citenamefont {{Adams}},
  \citenamefont {{Addesso}}, \citenamefont {{Adhikari}}, \citenamefont
  {{Adya}},\ and\ \citenamefont {et~al.}}]{ligo2018}%
  \BibitemOpen
  \bibfield  {author} {\bibinfo {author} {\bibfnamefont {B.~P.}\ \bibnamefont
  {{Abbott}}}, \bibinfo {author} {\bibfnamefont {R.}~\bibnamefont {{Abbott}}},
  \bibinfo {author} {\bibfnamefont {T.~D.}\ \bibnamefont {{Abbott}}}, \bibinfo
  {author} {\bibfnamefont {F.}~\bibnamefont {{Acernese}}}, \bibinfo {author}
  {\bibfnamefont {K.}~\bibnamefont {{Ackley}}}, \bibinfo {author}
  {\bibfnamefont {C.}~\bibnamefont {{Adams}}}, \bibinfo {author} {\bibfnamefont
  {T.}~\bibnamefont {{Adams}}}, \bibinfo {author} {\bibfnamefont
  {P.}~\bibnamefont {{Addesso}}}, \bibinfo {author} {\bibfnamefont {R.~X.}\
  \bibnamefont {{Adhikari}}}, \bibinfo {author} {\bibfnamefont {V.~B.}\
  \bibnamefont {{Adya}}}, \ and\ \bibinfo {author} {\bibnamefont {et~al.}},\
  }\href {\doibase 10.1103/PhysRevLett.121.161101} {\bibfield  {journal}
  {\bibinfo  {journal} {\prl}\ }\textbf {\bibinfo {volume} {121}},\ \bibinfo
  {eid} {161101} (\bibinfo {year} {2018}{\natexlab{b}})}\BibitemShut {NoStop}%
\bibitem [{\citenamefont {{Radice}}\ and\ \citenamefont
  {{Dai}}(2019)}]{radice2019}%
  \BibitemOpen
  \bibfield  {author} {\bibinfo {author} {\bibfnamefont {D.}~\bibnamefont
  {{Radice}}}\ and\ \bibinfo {author} {\bibfnamefont {L.}~\bibnamefont
  {{Dai}}},\ }\href {\doibase 10.1140/epja/i2019-12716-4} {\bibfield  {journal}
  {\bibinfo  {journal} {European Physical Journal A}\ }\textbf {\bibinfo
  {volume} {55}},\ \bibinfo {eid} {50} (\bibinfo {year} {2019})}\BibitemShut
  {NoStop}%
\bibitem [{\citenamefont {Schmidt}\ \emph {et~al.}(2012)\citenamefont
  {Schmidt}, \citenamefont {Hannam},\ and\ \citenamefont {Husa}}]{schmidt2012}%
  \BibitemOpen
  \bibfield  {author} {\bibinfo {author} {\bibfnamefont {P.}~\bibnamefont
  {Schmidt}}, \bibinfo {author} {\bibfnamefont {M.}~\bibnamefont {Hannam}}, \
  and\ \bibinfo {author} {\bibfnamefont {S.}~\bibnamefont {Husa}},\ }\href
  {\doibase 10.1103/PhysRevD.86.104063} {\bibfield  {journal} {\bibinfo
  {journal} {Phys. Rev. D}\ }\textbf {\bibinfo {volume} {86}},\ \bibinfo
  {pages} {104063} (\bibinfo {year} {2012})}\BibitemShut {NoStop}%
\bibitem [{\citenamefont {Hannam}\ \emph {et~al.}(2014)\citenamefont {Hannam},
  \citenamefont {Schmidt}, \citenamefont {Boh\'e}, \citenamefont {Haegel},
  \citenamefont {Husa}, \citenamefont {Ohme}, \citenamefont {Pratten},\ and\
  \citenamefont {P\"urrer}}]{hannam2014}%
  \BibitemOpen
  \bibfield  {author} {\bibinfo {author} {\bibfnamefont {M.}~\bibnamefont
  {Hannam}}, \bibinfo {author} {\bibfnamefont {P.}~\bibnamefont {Schmidt}},
  \bibinfo {author} {\bibfnamefont {A.}~\bibnamefont {Boh\'e}}, \bibinfo
  {author} {\bibfnamefont {L.}~\bibnamefont {Haegel}}, \bibinfo {author}
  {\bibfnamefont {S.}~\bibnamefont {Husa}}, \bibinfo {author} {\bibfnamefont
  {F.}~\bibnamefont {Ohme}}, \bibinfo {author} {\bibfnamefont {G.}~\bibnamefont
  {Pratten}}, \ and\ \bibinfo {author} {\bibfnamefont {M.}~\bibnamefont
  {P\"urrer}},\ }\href {\doibase 10.1103/PhysRevLett.113.151101} {\bibfield
  {journal} {\bibinfo  {journal} {Phys. Rev. Lett.}\ }\textbf {\bibinfo
  {volume} {113}},\ \bibinfo {pages} {151101} (\bibinfo {year}
  {2014})}\BibitemShut {NoStop}%
\bibitem [{\citenamefont {Schmidt}\ \emph {et~al.}(2015)\citenamefont
  {Schmidt}, \citenamefont {Ohme},\ and\ \citenamefont {Hannam}}]{schmidt2015}%
  \BibitemOpen
  \bibfield  {author} {\bibinfo {author} {\bibfnamefont {P.}~\bibnamefont
  {Schmidt}}, \bibinfo {author} {\bibfnamefont {F.}~\bibnamefont {Ohme}}, \
  and\ \bibinfo {author} {\bibfnamefont {M.}~\bibnamefont {Hannam}},\ }\href
  {\doibase 10.1103/PhysRevD.91.024043} {\bibfield  {journal} {\bibinfo
  {journal} {Phys. Rev. D}\ }\textbf {\bibinfo {volume} {91}},\ \bibinfo
  {pages} {024043} (\bibinfo {year} {2015})}\BibitemShut {NoStop}%
\bibitem [{\citenamefont {Husa}\ \emph {et~al.}(2016)\citenamefont {Husa},
  \citenamefont {Khan}, \citenamefont {Hannam}, \citenamefont {P\"urrer},
  \citenamefont {Ohme}, \citenamefont {Forteza},\ and\ \citenamefont
  {Boh\'e}}]{husa2016}%
  \BibitemOpen
  \bibfield  {author} {\bibinfo {author} {\bibfnamefont {S.}~\bibnamefont
  {Husa}}, \bibinfo {author} {\bibfnamefont {S.}~\bibnamefont {Khan}}, \bibinfo
  {author} {\bibfnamefont {M.}~\bibnamefont {Hannam}}, \bibinfo {author}
  {\bibfnamefont {M.}~\bibnamefont {P\"urrer}}, \bibinfo {author}
  {\bibfnamefont {F.}~\bibnamefont {Ohme}}, \bibinfo {author} {\bibfnamefont
  {X.~J.}\ \bibnamefont {Forteza}}, \ and\ \bibinfo {author} {\bibfnamefont
  {A.}~\bibnamefont {Boh\'e}},\ }\href {\doibase 10.1103/PhysRevD.93.044006}
  {\bibfield  {journal} {\bibinfo  {journal} {Phys. Rev. D}\ }\textbf {\bibinfo
  {volume} {93}},\ \bibinfo {pages} {044006} (\bibinfo {year}
  {2016})}\BibitemShut {NoStop}%
\bibitem [{\citenamefont {Khan}\ \emph {et~al.}(2016)\citenamefont {Khan},
  \citenamefont {Husa}, \citenamefont {Hannam}, \citenamefont {Ohme},
  \citenamefont {P\"urrer}, \citenamefont {Forteza},\ and\ \citenamefont
  {Boh\'e}}]{khan2016}%
  \BibitemOpen
  \bibfield  {author} {\bibinfo {author} {\bibfnamefont {S.}~\bibnamefont
  {Khan}}, \bibinfo {author} {\bibfnamefont {S.}~\bibnamefont {Husa}}, \bibinfo
  {author} {\bibfnamefont {M.}~\bibnamefont {Hannam}}, \bibinfo {author}
  {\bibfnamefont {F.}~\bibnamefont {Ohme}}, \bibinfo {author} {\bibfnamefont
  {M.}~\bibnamefont {P\"urrer}}, \bibinfo {author} {\bibfnamefont {X.~J.}\
  \bibnamefont {Forteza}}, \ and\ \bibinfo {author} {\bibfnamefont
  {A.}~\bibnamefont {Boh\'e}},\ }\href {\doibase 10.1103/PhysRevD.93.044007}
  {\bibfield  {journal} {\bibinfo  {journal} {Phys. Rev. D}\ }\textbf {\bibinfo
  {volume} {93}},\ \bibinfo {pages} {044007} (\bibinfo {year}
  {2016})}\BibitemShut {NoStop}%
\bibitem [{\citenamefont {Dietrich}\ \emph {et~al.}(2017)\citenamefont
  {Dietrich}, \citenamefont {Bernuzzi},\ and\ \citenamefont
  {Tichy}}]{dietrich2017}%
  \BibitemOpen
  \bibfield  {author} {\bibinfo {author} {\bibfnamefont {T.}~\bibnamefont
  {Dietrich}}, \bibinfo {author} {\bibfnamefont {S.}~\bibnamefont {Bernuzzi}},
  \ and\ \bibinfo {author} {\bibfnamefont {W.}~\bibnamefont {Tichy}},\ }\href
  {\doibase 10.1103/PhysRevD.96.121501} {\bibfield  {journal} {\bibinfo
  {journal} {Phys. Rev. D}\ }\textbf {\bibinfo {volume} {96}},\ \bibinfo
  {pages} {121501} (\bibinfo {year} {2017})}\BibitemShut {NoStop}%
\bibitem [{\citenamefont {Dietrich}\ \emph {et~al.}(2019)\citenamefont
  {Dietrich}, \citenamefont {Khan}, \citenamefont {Dudi}, \citenamefont
  {Kapadia}, \citenamefont {Kumar}, \citenamefont {Nagar}, \citenamefont
  {Ohme}, \citenamefont {Pannarale}, \citenamefont {Samajdar}, \citenamefont
  {Bernuzzi}, \citenamefont {Carullo}, \citenamefont {Del~Pozzo}, \citenamefont
  {Haney}, \citenamefont {Markakis}, \citenamefont {P\"urrer}, \citenamefont
  {Riemenschneider}, \citenamefont {Setyawati}, \citenamefont {Tsang},\ and\
  \citenamefont {Van Den~Broeck}}]{dietrich2019}%
  \BibitemOpen
  \bibfield  {author} {\bibinfo {author} {\bibfnamefont {T.}~\bibnamefont
  {Dietrich}}, \bibinfo {author} {\bibfnamefont {S.}~\bibnamefont {Khan}},
  \bibinfo {author} {\bibfnamefont {R.}~\bibnamefont {Dudi}}, \bibinfo {author}
  {\bibfnamefont {S.~J.}\ \bibnamefont {Kapadia}}, \bibinfo {author}
  {\bibfnamefont {P.}~\bibnamefont {Kumar}}, \bibinfo {author} {\bibfnamefont
  {A.}~\bibnamefont {Nagar}}, \bibinfo {author} {\bibfnamefont
  {F.}~\bibnamefont {Ohme}}, \bibinfo {author} {\bibfnamefont {F.}~\bibnamefont
  {Pannarale}}, \bibinfo {author} {\bibfnamefont {A.}~\bibnamefont {Samajdar}},
  \bibinfo {author} {\bibfnamefont {S.}~\bibnamefont {Bernuzzi}}, \bibinfo
  {author} {\bibfnamefont {G.}~\bibnamefont {Carullo}}, \bibinfo {author}
  {\bibfnamefont {W.}~\bibnamefont {Del~Pozzo}}, \bibinfo {author}
  {\bibfnamefont {M.}~\bibnamefont {Haney}}, \bibinfo {author} {\bibfnamefont
  {C.}~\bibnamefont {Markakis}}, \bibinfo {author} {\bibfnamefont
  {M.}~\bibnamefont {P\"urrer}}, \bibinfo {author} {\bibfnamefont
  {G.}~\bibnamefont {Riemenschneider}}, \bibinfo {author} {\bibfnamefont
  {Y.~E.}\ \bibnamefont {Setyawati}}, \bibinfo {author} {\bibfnamefont {K.~W.}\
  \bibnamefont {Tsang}}, \ and\ \bibinfo {author} {\bibfnamefont
  {C.}~\bibnamefont {Van Den~Broeck}},\ }\href {\doibase
  10.1103/PhysRevD.99.024029} {\bibfield  {journal} {\bibinfo  {journal} {Phys.
  Rev. D}\ }\textbf {\bibinfo {volume} {99}},\ \bibinfo {pages} {024029}
  (\bibinfo {year} {2019})}\BibitemShut {NoStop}%
\bibitem [{\citenamefont {Buonanno}\ and\ \citenamefont
  {Damour}(1999)}]{buonanno1999}%
  \BibitemOpen
  \bibfield  {author} {\bibinfo {author} {\bibfnamefont {A.}~\bibnamefont
  {Buonanno}}\ and\ \bibinfo {author} {\bibfnamefont {T.}~\bibnamefont
  {Damour}},\ }\href {\doibase 10.1103/PhysRevD.59.084006} {\bibfield
  {journal} {\bibinfo  {journal} {Phys. Rev. D}\ }\textbf {\bibinfo {volume}
  {59}},\ \bibinfo {pages} {084006} (\bibinfo {year} {1999})}\BibitemShut
  {NoStop}%
\bibitem [{\citenamefont {Buonanno}\ and\ \citenamefont
  {Damour}(2000)}]{buonanno2000}%
  \BibitemOpen
  \bibfield  {author} {\bibinfo {author} {\bibfnamefont {A.}~\bibnamefont
  {Buonanno}}\ and\ \bibinfo {author} {\bibfnamefont {T.}~\bibnamefont
  {Damour}},\ }\href {\doibase 10.1103/PhysRevD.62.064015} {\bibfield
  {journal} {\bibinfo  {journal} {Phys. Rev. D}\ }\textbf {\bibinfo {volume}
  {62}},\ \bibinfo {pages} {064015} (\bibinfo {year} {2000})}\BibitemShut
  {NoStop}%
\bibitem [{\citenamefont {{Damour}}(2014)}]{damour2014}%
  \BibitemOpen
  \bibfield  {author} {\bibinfo {author} {\bibfnamefont {T.}~\bibnamefont
  {{Damour}}},\ }\enquote {\bibinfo {title} {{The General Relativistic Two Body
  Problem and the Effective One Body Formalism}},}\ in\ \href {\doibase
  10.1007/978-3-319-06349-2_5} {\emph {\bibinfo {booktitle} {General
  Relativity, Cosmology and Astrophysics}}},\ Vol.\ \bibinfo {volume} {177},\
  \bibinfo {editor} {edited by\ \bibinfo {editor} {\bibfnamefont
  {J.}~\bibnamefont {{Bi{\v{c}}{\'a}k}}}\ and\ \bibinfo {editor} {\bibfnamefont
  {T.}~\bibnamefont {{Ledvinka}}}}\ (\bibinfo {year} {2014})\ p.\ \bibinfo
  {pages} {111}\BibitemShut {NoStop}%
\bibitem [{\citenamefont {{Abbott}}\ \emph {et~al.}(2019)\citenamefont
  {{Abbott}}, \citenamefont {{Abbott}}, \citenamefont {{Abbott}}, \citenamefont
  {{Acernese}}, \citenamefont {{Ackley}}, \citenamefont {{Adams}},
  \citenamefont {{Adams}}, \citenamefont {{Addesso}}, \citenamefont
  {{Adhikari}}, \citenamefont {{Adya}},\ and\ \citenamefont
  {et~al.}}]{ligo2019prx}%
  \BibitemOpen
  \bibfield  {author} {\bibinfo {author} {\bibfnamefont {B.~P.}\ \bibnamefont
  {{Abbott}}}, \bibinfo {author} {\bibfnamefont {R.}~\bibnamefont {{Abbott}}},
  \bibinfo {author} {\bibfnamefont {T.~D.}\ \bibnamefont {{Abbott}}}, \bibinfo
  {author} {\bibfnamefont {F.}~\bibnamefont {{Acernese}}}, \bibinfo {author}
  {\bibfnamefont {K.}~\bibnamefont {{Ackley}}}, \bibinfo {author}
  {\bibfnamefont {C.}~\bibnamefont {{Adams}}}, \bibinfo {author} {\bibfnamefont
  {T.}~\bibnamefont {{Adams}}}, \bibinfo {author} {\bibfnamefont
  {P.}~\bibnamefont {{Addesso}}}, \bibinfo {author} {\bibfnamefont {R.~X.}\
  \bibnamefont {{Adhikari}}}, \bibinfo {author} {\bibfnamefont {V.~B.}\
  \bibnamefont {{Adya}}}, \ and\ \bibinfo {author} {\bibnamefont {et~al.}},\
  }\href {\doibase 10.1103/PhysRevX.9.011001} {\bibfield  {journal} {\bibinfo
  {journal} {\prx}\ }\textbf {\bibinfo {volume} {9}},\ \bibinfo {eid} {011001}
  (\bibinfo {year} {2019})}\BibitemShut {NoStop}%
\bibitem [{\citenamefont {{Abbott}}\ \emph
  {et~al.}(2020{\natexlab{b}})\citenamefont {{Abbott}}, \citenamefont
  {{Abbott}}, \citenamefont {{Abbott}}, \citenamefont {{Abraham}},
  \citenamefont {{Acernese}}, \citenamefont {{Ackley}}, \citenamefont
  {{Adams}}, \citenamefont {{Adya}}, \citenamefont {{Affeldt}}, \citenamefont
  {{Agathos}},\ and\ \citenamefont {et~al.}}]{ligo2020}%
  \BibitemOpen
  \bibfield  {author} {\bibinfo {author} {\bibfnamefont {B.~P.}\ \bibnamefont
  {{Abbott}}}, \bibinfo {author} {\bibfnamefont {R.}~\bibnamefont {{Abbott}}},
  \bibinfo {author} {\bibfnamefont {T.~D.}\ \bibnamefont {{Abbott}}}, \bibinfo
  {author} {\bibfnamefont {S.}~\bibnamefont {{Abraham}}}, \bibinfo {author}
  {\bibfnamefont {F.}~\bibnamefont {{Acernese}}}, \bibinfo {author}
  {\bibfnamefont {K.}~\bibnamefont {{Ackley}}}, \bibinfo {author}
  {\bibfnamefont {C.}~\bibnamefont {{Adams}}}, \bibinfo {author} {\bibfnamefont
  {V.~B.}\ \bibnamefont {{Adya}}}, \bibinfo {author} {\bibfnamefont
  {C.}~\bibnamefont {{Affeldt}}}, \bibinfo {author} {\bibfnamefont
  {M.}~\bibnamefont {{Agathos}}}, \ and\ \bibinfo {author} {\bibnamefont
  {et~al.}},\ }\href {\doibase 10.1088/1361-6382/ab5f7c} {\bibfield  {journal}
  {\bibinfo  {journal} {Classical and Quantum Gravity}\ }\textbf {\bibinfo
  {volume} {37}},\ \bibinfo {eid} {045006} (\bibinfo {year}
  {2020}{\natexlab{b}})}\BibitemShut {NoStop}%
\bibitem [{\citenamefont {Boh\'e}\ \emph {et~al.}(2017)\citenamefont {Boh\'e},
  \citenamefont {Shao}, \citenamefont {Taracchini}, \citenamefont {Buonanno},
  \citenamefont {Babak}, \citenamefont {Harry}, \citenamefont {Hinder},
  \citenamefont {Ossokine}, \citenamefont {P\"urrer}, \citenamefont {Raymond},
  \citenamefont {Chu}, \citenamefont {Fong}, \citenamefont {Kumar},
  \citenamefont {Pfeiffer}, \citenamefont {Boyle}, \citenamefont {Hemberger},
  \citenamefont {Kidder}, \citenamefont {Lovelace}, \citenamefont {Scheel},\
  and\ \citenamefont {Szil\'agyi}}]{bohe2017}%
  \BibitemOpen
  \bibfield  {author} {\bibinfo {author} {\bibfnamefont {A.}~\bibnamefont
  {Boh\'e}}, \bibinfo {author} {\bibfnamefont {L.}~\bibnamefont {Shao}},
  \bibinfo {author} {\bibfnamefont {A.}~\bibnamefont {Taracchini}}, \bibinfo
  {author} {\bibfnamefont {A.}~\bibnamefont {Buonanno}}, \bibinfo {author}
  {\bibfnamefont {S.}~\bibnamefont {Babak}}, \bibinfo {author} {\bibfnamefont
  {I.~W.}\ \bibnamefont {Harry}}, \bibinfo {author} {\bibfnamefont
  {I.}~\bibnamefont {Hinder}}, \bibinfo {author} {\bibfnamefont
  {S.}~\bibnamefont {Ossokine}}, \bibinfo {author} {\bibfnamefont
  {M.}~\bibnamefont {P\"urrer}}, \bibinfo {author} {\bibfnamefont
  {V.}~\bibnamefont {Raymond}}, \bibinfo {author} {\bibfnamefont
  {T.}~\bibnamefont {Chu}}, \bibinfo {author} {\bibfnamefont {H.}~\bibnamefont
  {Fong}}, \bibinfo {author} {\bibfnamefont {P.}~\bibnamefont {Kumar}},
  \bibinfo {author} {\bibfnamefont {H.~P.}\ \bibnamefont {Pfeiffer}}, \bibinfo
  {author} {\bibfnamefont {M.}~\bibnamefont {Boyle}}, \bibinfo {author}
  {\bibfnamefont {D.~A.}\ \bibnamefont {Hemberger}}, \bibinfo {author}
  {\bibfnamefont {L.~E.}\ \bibnamefont {Kidder}}, \bibinfo {author}
  {\bibfnamefont {G.}~\bibnamefont {Lovelace}}, \bibinfo {author}
  {\bibfnamefont {M.~A.}\ \bibnamefont {Scheel}}, \ and\ \bibinfo {author}
  {\bibfnamefont {B.}~\bibnamefont {Szil\'agyi}},\ }\href {\doibase
  10.1103/PhysRevD.95.044028} {\bibfield  {journal} {\bibinfo  {journal} {Phys.
  Rev. D}\ }\textbf {\bibinfo {volume} {95}},\ \bibinfo {pages} {044028}
  (\bibinfo {year} {2017})}\BibitemShut {NoStop}%
\bibitem [{\citenamefont {{P{\"u}rrer}}(2014)}]{purrer2014}%
  \BibitemOpen
  \bibfield  {author} {\bibinfo {author} {\bibfnamefont {M.}~\bibnamefont
  {{P{\"u}rrer}}},\ }\href {\doibase 10.1088/0264-9381/31/19/195010} {\bibfield
   {journal} {\bibinfo  {journal} {Classical and Quantum Gravity}\ }\textbf
  {\bibinfo {volume} {31}},\ \bibinfo {eid} {195010} (\bibinfo {year}
  {2014})}\BibitemShut {NoStop}%
\bibitem [{\citenamefont {Nagar}\ \emph {et~al.}(2018)\citenamefont {Nagar},
  \citenamefont {Bernuzzi}, \citenamefont {Del~Pozzo}, \citenamefont
  {Riemenschneider}, \citenamefont {Akcay}, \citenamefont {Carullo},
  \citenamefont {Fleig}, \citenamefont {Babak}, \citenamefont {Tsang},
  \citenamefont {Colleoni}, \citenamefont {Messina}, \citenamefont {Pratten},
  \citenamefont {Radice}, \citenamefont {Rettegno}, \citenamefont {Agathos},
  \citenamefont {Fauchon-Jones}, \citenamefont {Hannam}, \citenamefont {Husa},
  \citenamefont {Dietrich}, \citenamefont {Cerd\'a-Duran}, \citenamefont
  {Font}, \citenamefont {Pannarale}, \citenamefont {Schmidt},\ and\
  \citenamefont {Damour}}]{TEOBRESUMS}%
  \BibitemOpen
  \bibfield  {author} {\bibinfo {author} {\bibfnamefont {A.}~\bibnamefont
  {Nagar}}, \bibinfo {author} {\bibfnamefont {S.}~\bibnamefont {Bernuzzi}},
  \bibinfo {author} {\bibfnamefont {W.}~\bibnamefont {Del~Pozzo}}, \bibinfo
  {author} {\bibfnamefont {G.}~\bibnamefont {Riemenschneider}}, \bibinfo
  {author} {\bibfnamefont {S.}~\bibnamefont {Akcay}}, \bibinfo {author}
  {\bibfnamefont {G.}~\bibnamefont {Carullo}}, \bibinfo {author} {\bibfnamefont
  {P.}~\bibnamefont {Fleig}}, \bibinfo {author} {\bibfnamefont
  {S.}~\bibnamefont {Babak}}, \bibinfo {author} {\bibfnamefont {K.~W.}\
  \bibnamefont {Tsang}}, \bibinfo {author} {\bibfnamefont {M.}~\bibnamefont
  {Colleoni}}, \bibinfo {author} {\bibfnamefont {F.}~\bibnamefont {Messina}},
  \bibinfo {author} {\bibfnamefont {G.}~\bibnamefont {Pratten}}, \bibinfo
  {author} {\bibfnamefont {D.}~\bibnamefont {Radice}}, \bibinfo {author}
  {\bibfnamefont {P.}~\bibnamefont {Rettegno}}, \bibinfo {author}
  {\bibfnamefont {M.}~\bibnamefont {Agathos}}, \bibinfo {author} {\bibfnamefont
  {E.}~\bibnamefont {Fauchon-Jones}}, \bibinfo {author} {\bibfnamefont
  {M.}~\bibnamefont {Hannam}}, \bibinfo {author} {\bibfnamefont
  {S.}~\bibnamefont {Husa}}, \bibinfo {author} {\bibfnamefont {T.}~\bibnamefont
  {Dietrich}}, \bibinfo {author} {\bibfnamefont {P.}~\bibnamefont
  {Cerd\'a-Duran}}, \bibinfo {author} {\bibfnamefont {J.~A.}\ \bibnamefont
  {Font}}, \bibinfo {author} {\bibfnamefont {F.}~\bibnamefont {Pannarale}},
  \bibinfo {author} {\bibfnamefont {P.}~\bibnamefont {Schmidt}}, \ and\
  \bibinfo {author} {\bibfnamefont {T.}~\bibnamefont {Damour}},\ }\href
  {\doibase 10.1103/PhysRevD.98.104052} {\bibfield  {journal} {\bibinfo
  {journal} {Phys. Rev. D}\ }\textbf {\bibinfo {volume} {98}},\ \bibinfo
  {pages} {104052} (\bibinfo {year} {2018})}\BibitemShut {NoStop}%
\bibitem [{\citenamefont {Hinderer}\ \emph {et~al.}(2016)\citenamefont
  {Hinderer}, \citenamefont {Taracchini}, \citenamefont {Foucart},
  \citenamefont {Buonanno}, \citenamefont {Steinhoff}, \citenamefont {Duez},
  \citenamefont {Kidder}, \citenamefont {Pfeiffer}, \citenamefont {Scheel},
  \citenamefont {Szilagyi}, \citenamefont {Hotokezaka}, \citenamefont
  {Kyutoku}, \citenamefont {Shibata},\ and\ \citenamefont
  {Carpenter}}]{SEOBNRV4}%
  \BibitemOpen
  \bibfield  {author} {\bibinfo {author} {\bibfnamefont {T.}~\bibnamefont
  {Hinderer}}, \bibinfo {author} {\bibfnamefont {A.}~\bibnamefont
  {Taracchini}}, \bibinfo {author} {\bibfnamefont {F.}~\bibnamefont {Foucart}},
  \bibinfo {author} {\bibfnamefont {A.}~\bibnamefont {Buonanno}}, \bibinfo
  {author} {\bibfnamefont {J.}~\bibnamefont {Steinhoff}}, \bibinfo {author}
  {\bibfnamefont {M.}~\bibnamefont {Duez}}, \bibinfo {author} {\bibfnamefont
  {L.~E.}\ \bibnamefont {Kidder}}, \bibinfo {author} {\bibfnamefont {H.~P.}\
  \bibnamefont {Pfeiffer}}, \bibinfo {author} {\bibfnamefont {M.~A.}\
  \bibnamefont {Scheel}}, \bibinfo {author} {\bibfnamefont {B.}~\bibnamefont
  {Szilagyi}}, \bibinfo {author} {\bibfnamefont {K.}~\bibnamefont
  {Hotokezaka}}, \bibinfo {author} {\bibfnamefont {K.}~\bibnamefont {Kyutoku}},
  \bibinfo {author} {\bibfnamefont {M.}~\bibnamefont {Shibata}}, \ and\
  \bibinfo {author} {\bibfnamefont {C.~W.}\ \bibnamefont {Carpenter}},\ }\href
  {\doibase 10.1103/PhysRevLett.116.181101} {\bibfield  {journal} {\bibinfo
  {journal} {Phys. Rev. Lett.}\ }\textbf {\bibinfo {volume} {116}},\ \bibinfo
  {pages} {181101} (\bibinfo {year} {2016})}\BibitemShut {NoStop}%
\bibitem [{\citenamefont {Damour}\ \emph {et~al.}(2012)\citenamefont {Damour},
  \citenamefont {Nagar},\ and\ \citenamefont {Villain}}]{damour2012}%
  \BibitemOpen
  \bibfield  {author} {\bibinfo {author} {\bibfnamefont {T.}~\bibnamefont
  {Damour}}, \bibinfo {author} {\bibfnamefont {A.}~\bibnamefont {Nagar}}, \
  and\ \bibinfo {author} {\bibfnamefont {L.}~\bibnamefont {Villain}},\ }\href
  {\doibase 10.1103/PhysRevD.85.123007} {\bibfield  {journal} {\bibinfo
  {journal} {Phys. Rev. D}\ }\textbf {\bibinfo {volume} {85}},\ \bibinfo
  {pages} {123007} (\bibinfo {year} {2012})}\BibitemShut {NoStop}%
\bibitem [{\citenamefont {Bini}\ \emph {et~al.}(2012)\citenamefont {Bini},
  \citenamefont {Damour},\ and\ \citenamefont {Faye}}]{bini2012}%
  \BibitemOpen
  \bibfield  {author} {\bibinfo {author} {\bibfnamefont {D.}~\bibnamefont
  {Bini}}, \bibinfo {author} {\bibfnamefont {T.}~\bibnamefont {Damour}}, \ and\
  \bibinfo {author} {\bibfnamefont {G.}~\bibnamefont {Faye}},\ }\href {\doibase
  10.1103/PhysRevD.85.124034} {\bibfield  {journal} {\bibinfo  {journal} {Phys.
  Rev. D}\ }\textbf {\bibinfo {volume} {85}},\ \bibinfo {pages} {124034}
  (\bibinfo {year} {2012})}\BibitemShut {NoStop}%
\bibitem [{\citenamefont {Samajdar}\ and\ \citenamefont
  {Dietrich}(2018)}]{samajdar2018}%
  \BibitemOpen
  \bibfield  {author} {\bibinfo {author} {\bibfnamefont {A.}~\bibnamefont
  {Samajdar}}\ and\ \bibinfo {author} {\bibfnamefont {T.}~\bibnamefont
  {Dietrich}},\ }\href {\doibase 10.1103/PhysRevD.98.124030} {\bibfield
  {journal} {\bibinfo  {journal} {Phys. Rev. D}\ }\textbf {\bibinfo {volume}
  {98}},\ \bibinfo {pages} {124030} (\bibinfo {year} {2018})}\BibitemShut
  {NoStop}%
\bibitem [{\citenamefont {{Narikawa}}\ \emph {et~al.}(2020)\citenamefont
  {{Narikawa}}, \citenamefont {{Uchikata}}, \citenamefont {{Kawaguchi}},
  \citenamefont {{Kiuchi}}, \citenamefont {{Kyutoku}}, \citenamefont
  {{Shibata}},\ and\ \citenamefont {{Tagoshi}}}]{narikawa2020}%
  \BibitemOpen
  \bibfield  {author} {\bibinfo {author} {\bibfnamefont {T.}~\bibnamefont
  {{Narikawa}}}, \bibinfo {author} {\bibfnamefont {N.}~\bibnamefont
  {{Uchikata}}}, \bibinfo {author} {\bibfnamefont {K.}~\bibnamefont
  {{Kawaguchi}}}, \bibinfo {author} {\bibfnamefont {K.}~\bibnamefont
  {{Kiuchi}}}, \bibinfo {author} {\bibfnamefont {K.}~\bibnamefont {{Kyutoku}}},
  \bibinfo {author} {\bibfnamefont {M.}~\bibnamefont {{Shibata}}}, \ and\
  \bibinfo {author} {\bibfnamefont {H.}~\bibnamefont {{Tagoshi}}},\ }\href
  {\doibase 10.1103/PhysRevResearch.2.043039} {\bibfield  {journal} {\bibinfo
  {journal} {Physical Review Research}\ }\textbf {\bibinfo {volume} {2}},\
  \bibinfo {eid} {043039} (\bibinfo {year} {2020})}\BibitemShut {NoStop}%
\bibitem [{\citenamefont {{Gamba}}\ \emph {et~al.}(2020)\citenamefont
  {{Gamba}}, \citenamefont {{Breschi}}, \citenamefont {{Bernuzzi}},
  \citenamefont {{Agathos}},\ and\ \citenamefont {{Nagar}}}]{gamba2020}%
  \BibitemOpen
  \bibfield  {author} {\bibinfo {author} {\bibfnamefont {R.}~\bibnamefont
  {{Gamba}}}, \bibinfo {author} {\bibfnamefont {M.}~\bibnamefont {{Breschi}}},
  \bibinfo {author} {\bibfnamefont {S.}~\bibnamefont {{Bernuzzi}}}, \bibinfo
  {author} {\bibfnamefont {M.}~\bibnamefont {{Agathos}}}, \ and\ \bibinfo
  {author} {\bibfnamefont {A.}~\bibnamefont {{Nagar}}},\ }\href@noop {}
  {\bibfield  {journal} {\bibinfo  {journal} {arXiv e-prints}\ ,\ \bibinfo
  {eid} {arXiv:2009.08467}} (\bibinfo {year} {2020})},\ \Eprint
  {http://arxiv.org/abs/2009.08467} {arXiv:2009.08467 [gr-qc]} \BibitemShut
  {NoStop}%
\bibitem [{\citenamefont {{Jim{\'e}nez Forteza}}\ \emph
  {et~al.}(2018)\citenamefont {{Jim{\'e}nez Forteza}}, \citenamefont
  {{Abdelsalhin}}, \citenamefont {{Pani}},\ and\ \citenamefont
  {{Gualtieri}}}]{jimenez2018}%
  \BibitemOpen
  \bibfield  {author} {\bibinfo {author} {\bibfnamefont {X.}~\bibnamefont
  {{Jim{\'e}nez Forteza}}}, \bibinfo {author} {\bibfnamefont {T.}~\bibnamefont
  {{Abdelsalhin}}}, \bibinfo {author} {\bibfnamefont {P.}~\bibnamefont
  {{Pani}}}, \ and\ \bibinfo {author} {\bibfnamefont {L.}~\bibnamefont
  {{Gualtieri}}},\ }\href {\doibase 10.1103/PhysRevD.98.124014} {\bibfield
  {journal} {\bibinfo  {journal} {\prd}\ }\textbf {\bibinfo {volume} {98}},\
  \bibinfo {eid} {124014} (\bibinfo {year} {2018})}\BibitemShut {NoStop}%
\bibitem [{\citenamefont {{Piekarewicz}}\ and\ \citenamefont
  {{Fattoyev}}(2019)}]{piekarewicz2019}%
  \BibitemOpen
  \bibfield  {author} {\bibinfo {author} {\bibfnamefont {J.}~\bibnamefont
  {{Piekarewicz}}}\ and\ \bibinfo {author} {\bibfnamefont {F.~J.}\ \bibnamefont
  {{Fattoyev}}},\ }\href {\doibase 10.1103/PhysRevC.99.045802} {\bibfield
  {journal} {\bibinfo  {journal} {\prc}\ }\textbf {\bibinfo {volume} {99}},\
  \bibinfo {eid} {045802} (\bibinfo {year} {2019})},\ \Eprint
  {http://arxiv.org/abs/1812.09974} {arXiv:1812.09974 [nucl-th]} \BibitemShut
  {NoStop}%
\bibitem [{\citenamefont {{Perot}}\ \emph {et~al.}(2020)\citenamefont
  {{Perot}}, \citenamefont {{Chamel}},\ and\ \citenamefont
  {{Sourie}}}]{perot2020}%
  \BibitemOpen
  \bibfield  {author} {\bibinfo {author} {\bibfnamefont {L.}~\bibnamefont
  {{Perot}}}, \bibinfo {author} {\bibfnamefont {N.}~\bibnamefont {{Chamel}}}, \
  and\ \bibinfo {author} {\bibfnamefont {A.}~\bibnamefont {{Sourie}}},\ }\href
  {\doibase 10.1103/PhysRevC.101.015806} {\bibfield  {journal} {\bibinfo
  {journal} {\prc}\ }\textbf {\bibinfo {volume} {101}},\ \bibinfo {eid}
  {015806} (\bibinfo {year} {2020})}\BibitemShut {NoStop}%
\bibitem [{\citenamefont {{Gittins}}\ \emph {et~al.}(2020)\citenamefont
  {{Gittins}}, \citenamefont {{Andersson}},\ and\ \citenamefont
  {{Pereira}}}]{gittins2020}%
  \BibitemOpen
  \bibfield  {author} {\bibinfo {author} {\bibfnamefont {F.}~\bibnamefont
  {{Gittins}}}, \bibinfo {author} {\bibfnamefont {N.}~\bibnamefont
  {{Andersson}}}, \ and\ \bibinfo {author} {\bibfnamefont {J.~P.}\ \bibnamefont
  {{Pereira}}},\ }\href {\doibase 10.1103/PhysRevD.101.103025} {\bibfield
  {journal} {\bibinfo  {journal} {\prd}\ }\textbf {\bibinfo {volume} {101}},\
  \bibinfo {eid} {103025} (\bibinfo {year} {2020})}\BibitemShut {NoStop}%
\bibitem [{\citenamefont {{Maggiore}}\ \emph {et~al.}(2020)\citenamefont
  {{Maggiore}}, \citenamefont {{Van Den Broeck}}, \citenamefont {{Bartolo}},
  \citenamefont {{Belgacem}}, \citenamefont {{Bertacca}}, \citenamefont
  {{Bizouard}}, \citenamefont {{Branchesi}}, \citenamefont {{Clesse}},
  \citenamefont {{Foffa}}, \citenamefont {{Garc{\'\i}a-Bellido}}, \citenamefont
  {{Grimm}}, \citenamefont {{Harms}}, \citenamefont {{Hinderer}}, \citenamefont
  {{Matarrese}}, \citenamefont {{Palomba}}, \citenamefont {{Peloso}},
  \citenamefont {{Ricciardone}},\ and\ \citenamefont
  {{Sakellariadou}}}]{maggiore2020}%
  \BibitemOpen
  \bibfield  {author} {\bibinfo {author} {\bibfnamefont {M.}~\bibnamefont
  {{Maggiore}}}, \bibinfo {author} {\bibfnamefont {C.}~\bibnamefont {{Van Den
  Broeck}}}, \bibinfo {author} {\bibfnamefont {N.}~\bibnamefont {{Bartolo}}},
  \bibinfo {author} {\bibfnamefont {E.}~\bibnamefont {{Belgacem}}}, \bibinfo
  {author} {\bibfnamefont {D.}~\bibnamefont {{Bertacca}}}, \bibinfo {author}
  {\bibfnamefont {M.~A.}\ \bibnamefont {{Bizouard}}}, \bibinfo {author}
  {\bibfnamefont {M.}~\bibnamefont {{Branchesi}}}, \bibinfo {author}
  {\bibfnamefont {S.}~\bibnamefont {{Clesse}}}, \bibinfo {author}
  {\bibfnamefont {S.}~\bibnamefont {{Foffa}}}, \bibinfo {author} {\bibfnamefont
  {J.}~\bibnamefont {{Garc{\'\i}a-Bellido}}}, \bibinfo {author} {\bibfnamefont
  {S.}~\bibnamefont {{Grimm}}}, \bibinfo {author} {\bibfnamefont
  {J.}~\bibnamefont {{Harms}}}, \bibinfo {author} {\bibfnamefont
  {T.}~\bibnamefont {{Hinderer}}}, \bibinfo {author} {\bibfnamefont
  {S.}~\bibnamefont {{Matarrese}}}, \bibinfo {author} {\bibfnamefont
  {C.}~\bibnamefont {{Palomba}}}, \bibinfo {author} {\bibfnamefont
  {M.}~\bibnamefont {{Peloso}}}, \bibinfo {author} {\bibfnamefont
  {A.}~\bibnamefont {{Ricciardone}}}, \ and\ \bibinfo {author} {\bibfnamefont
  {M.}~\bibnamefont {{Sakellariadou}}},\ }\href {\doibase
  10.1088/1475-7516/2020/03/050} {\bibfield  {journal} {\bibinfo  {journal}
  {\jcap}\ }\textbf {\bibinfo {volume} {2020}},\ \bibinfo {eid} {050} (\bibinfo
  {year} {2020})}\BibitemShut {NoStop}%
\bibitem [{\citenamefont {Kumar}\ \emph {et~al.}(2017)\citenamefont {Kumar},
  \citenamefont {Biswal},\ and\ \citenamefont {Patra}}]{Kumar2017}%
  \BibitemOpen
  \bibfield  {author} {\bibinfo {author} {\bibfnamefont {B.}~\bibnamefont
  {Kumar}}, \bibinfo {author} {\bibfnamefont {S.~K.}\ \bibnamefont {Biswal}}, \
  and\ \bibinfo {author} {\bibfnamefont {S.~K.}\ \bibnamefont {Patra}},\ }\href
  {\doibase 10.1103/PhysRevC.95.015801} {\bibfield  {journal} {\bibinfo
  {journal} {Phys. Rev. C}\ }\textbf {\bibinfo {volume} {95}},\ \bibinfo
  {pages} {015801} (\bibinfo {year} {2017})}\BibitemShut {NoStop}%
\bibitem [{\citenamefont {{Perot}}\ \emph {et~al.}(2019)\citenamefont
  {{Perot}}, \citenamefont {{Chamel}},\ and\ \citenamefont
  {{Sourie}}}]{perot2019}%
  \BibitemOpen
  \bibfield  {author} {\bibinfo {author} {\bibfnamefont {L.}~\bibnamefont
  {{Perot}}}, \bibinfo {author} {\bibfnamefont {N.}~\bibnamefont {{Chamel}}}, \
  and\ \bibinfo {author} {\bibfnamefont {A.}~\bibnamefont {{Sourie}}},\ }\href
  {\doibase 10.1103/PhysRevC.100.035801} {\bibfield  {journal} {\bibinfo
  {journal} {\prc}\ }\textbf {\bibinfo {volume} {100}},\ \bibinfo {eid}
  {035801} (\bibinfo {year} {2019})}\BibitemShut {NoStop}%
\bibitem [{\citenamefont {{Potekhin}}\ \emph {et~al.}(2013)\citenamefont
  {{Potekhin}}, \citenamefont {{Fantina}}, \citenamefont {{Chamel}},
  \citenamefont {{Pearson}},\ and\ \citenamefont {{Goriely}}}]{potekhin2013}%
  \BibitemOpen
  \bibfield  {author} {\bibinfo {author} {\bibfnamefont {A.~Y.}\ \bibnamefont
  {{Potekhin}}}, \bibinfo {author} {\bibfnamefont {A.~F.}\ \bibnamefont
  {{Fantina}}}, \bibinfo {author} {\bibfnamefont {N.}~\bibnamefont {{Chamel}}},
  \bibinfo {author} {\bibfnamefont {J.~M.}\ \bibnamefont {{Pearson}}}, \ and\
  \bibinfo {author} {\bibfnamefont {S.}~\bibnamefont {{Goriely}}},\ }\href
  {\doibase 10.1051/0004-6361/201321697} {\bibfield  {journal} {\bibinfo
  {journal} {\aap}\ }\textbf {\bibinfo {volume} {560}},\ \bibinfo {eid} {A48}
  (\bibinfo {year} {2013})}\BibitemShut {NoStop}%
\bibitem [{\citenamefont {{Pearson}}\ \emph {et~al.}(2018)\citenamefont
  {{Pearson}}, \citenamefont {{Chamel}}, \citenamefont {{Potekhin}},
  \citenamefont {{Fantina}}, \citenamefont {{Ducoin}}, \citenamefont
  {{Dutta}},\ and\ \citenamefont {{Goriely}}}]{pearson2018}%
  \BibitemOpen
  \bibfield  {author} {\bibinfo {author} {\bibfnamefont {J.~M.}\ \bibnamefont
  {{Pearson}}}, \bibinfo {author} {\bibfnamefont {N.}~\bibnamefont {{Chamel}}},
  \bibinfo {author} {\bibfnamefont {A.~Y.}\ \bibnamefont {{Potekhin}}},
  \bibinfo {author} {\bibfnamefont {A.~F.}\ \bibnamefont {{Fantina}}}, \bibinfo
  {author} {\bibfnamefont {C.}~\bibnamefont {{Ducoin}}}, \bibinfo {author}
  {\bibfnamefont {A.~K.}\ \bibnamefont {{Dutta}}}, \ and\ \bibinfo {author}
  {\bibfnamefont {S.}~\bibnamefont {{Goriely}}},\ }\href {\doibase
  10.1093/mnras/sty2413} {\bibfield  {journal} {\bibinfo  {journal} {\mnras}\
  }\textbf {\bibinfo {volume} {481}},\ \bibinfo {pages} {2994} (\bibinfo {year}
  {2018})}\BibitemShut {NoStop}%
\bibitem [{\citenamefont {Pearson}\ \emph {et~al.}(2020)\citenamefont
  {Pearson}, \citenamefont {Chamel},\ and\ \citenamefont
  {Potekhin}}]{pearson2019}%
  \BibitemOpen
  \bibfield  {author} {\bibinfo {author} {\bibfnamefont {J.~M.}\ \bibnamefont
  {Pearson}}, \bibinfo {author} {\bibfnamefont {N.}~\bibnamefont {Chamel}}, \
  and\ \bibinfo {author} {\bibfnamefont {A.~Y.}\ \bibnamefont {Potekhin}},\
  }\href {\doibase 10.1103/PhysRevC.101.015802} {\bibfield  {journal} {\bibinfo
   {journal} {Phys. Rev. C}\ }\textbf {\bibinfo {volume} {101}},\ \bibinfo
  {pages} {015802} (\bibinfo {year} {2020})}\BibitemShut {NoStop}%
\bibitem [{\citenamefont {{Flanagan}}\ and\ \citenamefont
  {{Hinderer}}(2008)}]{flanagan08}%
  \BibitemOpen
  \bibfield  {author} {\bibinfo {author} {\bibfnamefont {{\'E}.~{\'E}.}\
  \bibnamefont {{Flanagan}}}\ and\ \bibinfo {author} {\bibfnamefont
  {T.}~\bibnamefont {{Hinderer}}},\ }\href {\doibase
  10.1103/PhysRevD.77.021502} {\bibfield  {journal} {\bibinfo  {journal}
  {\prd}\ }\textbf {\bibinfo {volume} {77}},\ \bibinfo {eid} {021502(R)}
  (\bibinfo {year} {2008})}\BibitemShut {NoStop}%
\bibitem [{\citenamefont {{Hinderer}}(2008)}]{hinderer08}%
  \BibitemOpen
  \bibfield  {author} {\bibinfo {author} {\bibfnamefont {T.}~\bibnamefont
  {{Hinderer}}},\ }\href {\doibase 10.1086/533487} {\bibfield  {journal}
  {\bibinfo  {journal} {\apj}\ }\textbf {\bibinfo {volume} {677}},\ \bibinfo
  {pages} {1216} (\bibinfo {year} {2008})}\BibitemShut {NoStop}%
\bibitem [{\citenamefont {Damour}\ and\ \citenamefont
  {Nagar}(2009)}]{damour2009}%
  \BibitemOpen
  \bibfield  {author} {\bibinfo {author} {\bibfnamefont {T.}~\bibnamefont
  {Damour}}\ and\ \bibinfo {author} {\bibfnamefont {A.}~\bibnamefont {Nagar}},\
  }\href {\doibase 10.1103/PhysRevD.80.084035} {\bibfield  {journal} {\bibinfo
  {journal} {Phys. Rev. D}\ }\textbf {\bibinfo {volume} {80}},\ \bibinfo
  {pages} {084035} (\bibinfo {year} {2009})}\BibitemShut {NoStop}%
\bibitem [{\citenamefont {{Binnington}}\ and\ \citenamefont
  {{Poisson}}(2009)}]{poisson2009}%
  \BibitemOpen
  \bibfield  {author} {\bibinfo {author} {\bibfnamefont {T.}~\bibnamefont
  {{Binnington}}}\ and\ \bibinfo {author} {\bibfnamefont {E.}~\bibnamefont
  {{Poisson}}},\ }\href {\doibase 10.1103/PhysRevD.80.084018} {\bibfield
  {journal} {\bibinfo  {journal} {\prd}\ }\textbf {\bibinfo {volume} {80}},\
  \bibinfo {eid} {084018} (\bibinfo {year} {2009})}\BibitemShut {NoStop}%
\bibitem [{\citenamefont {{Landry}}\ and\ \citenamefont
  {{Poisson}}(2015)}]{poisson2015}%
  \BibitemOpen
  \bibfield  {author} {\bibinfo {author} {\bibfnamefont {P.}~\bibnamefont
  {{Landry}}}\ and\ \bibinfo {author} {\bibfnamefont {E.}~\bibnamefont
  {{Poisson}}},\ }\href {\doibase 10.1103/PhysRevD.91.104026} {\bibfield
  {journal} {\bibinfo  {journal} {\prd}\ }\textbf {\bibinfo {volume} {91}},\
  \bibinfo {eid} {104026} (\bibinfo {year} {2015})}\BibitemShut {NoStop}%
\bibitem [{\citenamefont {{Pani}}\ \emph {et~al.}(2018)\citenamefont {{Pani}},
  \citenamefont {{Gualtieri}}, \citenamefont {{Abdelsalhin}},\ and\
  \citenamefont {{Jim{\'e}nez-Forteza}}}]{pani2018}%
  \BibitemOpen
  \bibfield  {author} {\bibinfo {author} {\bibfnamefont {P.}~\bibnamefont
  {{Pani}}}, \bibinfo {author} {\bibfnamefont {L.}~\bibnamefont {{Gualtieri}}},
  \bibinfo {author} {\bibfnamefont {T.}~\bibnamefont {{Abdelsalhin}}}, \ and\
  \bibinfo {author} {\bibfnamefont {X.}~\bibnamefont {{Jim{\'e}nez-Forteza}}},\
  }\href {\doibase 10.1103/PhysRevD.98.124023} {\bibfield  {journal} {\bibinfo
  {journal} {\prd}\ }\textbf {\bibinfo {volume} {98}},\ \bibinfo {eid} {124023}
  (\bibinfo {year} {2018})}\BibitemShut {NoStop}%
\bibitem [{\citenamefont {{Tolman}}(1939)}]{tolman1939}%
  \BibitemOpen
  \bibfield  {author} {\bibinfo {author} {\bibfnamefont {R.~C.}\ \bibnamefont
  {{Tolman}}},\ }\href {\doibase 10.1103/PhysRev.55.364} {\bibfield  {journal}
  {\bibinfo  {journal} {\pr}\ }\textbf {\bibinfo {volume} {55}},\ \bibinfo
  {pages} {364} (\bibinfo {year} {1939})}\BibitemShut {NoStop}%
\bibitem [{\citenamefont {{Oppenheimer}}\ and\ \citenamefont
  {{Volkoff}}(1939)}]{oppenheimer1939}%
  \BibitemOpen
  \bibfield  {author} {\bibinfo {author} {\bibfnamefont {J.~R.}\ \bibnamefont
  {{Oppenheimer}}}\ and\ \bibinfo {author} {\bibfnamefont {G.~M.}\ \bibnamefont
  {{Volkoff}}},\ }\href {\doibase 10.1103/PhysRev.55.374} {\bibfield  {journal}
  {\bibinfo  {journal} {\pr}\ }\textbf {\bibinfo {volume} {55}},\ \bibinfo
  {pages} {374} (\bibinfo {year} {1939})}\BibitemShut {NoStop}%
\bibitem [{\citenamefont {Yagi}(2014)}]{yagi2014}%
  \BibitemOpen
  \bibfield  {author} {\bibinfo {author} {\bibfnamefont {K.}~\bibnamefont
  {Yagi}},\ }\href {\doibase 10.1103/PhysRevD.89.043011} {\bibfield  {journal}
  {\bibinfo  {journal} {Phys. Rev. D}\ }\textbf {\bibinfo {volume} {89}},\
  \bibinfo {pages} {043011} (\bibinfo {year} {2014})}\BibitemShut {NoStop}%
\bibitem [{\citenamefont {{Yagi}}(2017)}]{yagi2017}%
  \BibitemOpen
  \bibfield  {author} {\bibinfo {author} {\bibfnamefont {K.}~\bibnamefont
  {{Yagi}}},\ }\href {\doibase 10.1103/PhysRevD.96.129904} {\bibfield
  {journal} {\bibinfo  {journal} {\prd}\ }\textbf {\bibinfo {volume} {96}},\
  \bibinfo {eid} {129904} (\bibinfo {year} {2017})}\BibitemShut {NoStop}%
\bibitem [{\citenamefont {{Ferreira}}\ and\ \citenamefont
  {{Provid{\^e}ncia}}(2020)}]{ferreira2020}%
  \BibitemOpen
  \bibfield  {author} {\bibinfo {author} {\bibfnamefont {M.}~\bibnamefont
  {{Ferreira}}}\ and\ \bibinfo {author} {\bibfnamefont {C.}~\bibnamefont
  {{Provid{\^e}ncia}}},\ }\href {\doibase 10.3390/universe6110220} {\bibfield
  {journal} {\bibinfo  {journal} {Universe}\ }\textbf {\bibinfo {volume} {6}},\
  \bibinfo {pages} {220} (\bibinfo {year} {2020})}\BibitemShut {NoStop}%
\bibitem [{\citenamefont {{Lalazissis}}\ \emph {et~al.}(1997)\citenamefont
  {{Lalazissis}}, \citenamefont {{K{\"o}nig}},\ and\ \citenamefont
  {{Ring}}}]{lalazissis1997}%
  \BibitemOpen
  \bibfield  {author} {\bibinfo {author} {\bibfnamefont {G.~A.}\ \bibnamefont
  {{Lalazissis}}}, \bibinfo {author} {\bibfnamefont {J.}~\bibnamefont
  {{K{\"o}nig}}}, \ and\ \bibinfo {author} {\bibfnamefont {P.}~\bibnamefont
  {{Ring}}},\ }\href {\doibase 10.1103/PhysRevC.55.540} {\bibfield  {journal}
  {\bibinfo  {journal} {\prc}\ }\textbf {\bibinfo {volume} {55}},\ \bibinfo
  {pages} {540} (\bibinfo {year} {1997})}\BibitemShut {NoStop}%
\bibitem [{\citenamefont {{Danielewicz}}\ \emph {et~al.}(2002)\citenamefont
  {{Danielewicz}}, \citenamefont {{Lacey}},\ and\ \citenamefont
  {{Lynch}}}]{danielewicz2002}%
  \BibitemOpen
  \bibfield  {author} {\bibinfo {author} {\bibfnamefont {P.}~\bibnamefont
  {{Danielewicz}}}, \bibinfo {author} {\bibfnamefont {R.}~\bibnamefont
  {{Lacey}}}, \ and\ \bibinfo {author} {\bibfnamefont {W.~G.}\ \bibnamefont
  {{Lynch}}},\ }\href {\doibase 10.1126/science.1078070} {\bibfield  {journal}
  {\bibinfo  {journal} {Science}\ }\textbf {\bibinfo {volume} {298}},\ \bibinfo
  {pages} {1592} (\bibinfo {year} {2002})}\BibitemShut {NoStop}%
\bibitem [{\citenamefont {{Dutra}}\ \emph {et~al.}(2014)\citenamefont
  {{Dutra}}, \citenamefont {{Louren{\c{c}}o}}, \citenamefont {{Avancini}},
  \citenamefont {{Carlson}}, \citenamefont {{Delfino}}, \citenamefont
  {{Menezes}}, \citenamefont {{Provid{\^e}ncia}}, \citenamefont {{Typel}},\
  and\ \citenamefont {{Stone}}}]{dutra2014}%
  \BibitemOpen
  \bibfield  {author} {\bibinfo {author} {\bibfnamefont {M.}~\bibnamefont
  {{Dutra}}}, \bibinfo {author} {\bibfnamefont {O.}~\bibnamefont
  {{Louren{\c{c}}o}}}, \bibinfo {author} {\bibfnamefont {S.~S.}\ \bibnamefont
  {{Avancini}}}, \bibinfo {author} {\bibfnamefont {B.~V.}\ \bibnamefont
  {{Carlson}}}, \bibinfo {author} {\bibfnamefont {A.}~\bibnamefont
  {{Delfino}}}, \bibinfo {author} {\bibfnamefont {D.~P.}\ \bibnamefont
  {{Menezes}}}, \bibinfo {author} {\bibfnamefont {C.}~\bibnamefont
  {{Provid{\^e}ncia}}}, \bibinfo {author} {\bibfnamefont {S.}~\bibnamefont
  {{Typel}}}, \ and\ \bibinfo {author} {\bibfnamefont {J.~R.}\ \bibnamefont
  {{Stone}}},\ }\href {\doibase 10.1103/PhysRevC.90.055203} {\bibfield
  {journal} {\bibinfo  {journal} {\prc}\ }\textbf {\bibinfo {volume} {90}},\
  \bibinfo {eid} {055203} (\bibinfo {year} {2014})}\BibitemShut {NoStop}%
\bibitem [{\citenamefont {{Audi}}\ \emph {et~al.}(2012)\citenamefont {{Audi}},
  \citenamefont {{Wang}}, \citenamefont {{Wapstra}}, \citenamefont {{Kondev}},
  \citenamefont {{MacCormick}}, \citenamefont {{Xu}},\ and\ \citenamefont
  {{Pfeiffer}}}]{AME2012}%
  \BibitemOpen
  \bibfield  {author} {\bibinfo {author} {\bibfnamefont {G.}~\bibnamefont
  {{Audi}}}, \bibinfo {author} {\bibfnamefont {M.}~\bibnamefont {{Wang}}},
  \bibinfo {author} {\bibfnamefont {A.~H.}\ \bibnamefont {{Wapstra}}}, \bibinfo
  {author} {\bibfnamefont {F.~G.}\ \bibnamefont {{Kondev}}}, \bibinfo {author}
  {\bibfnamefont {M.}~\bibnamefont {{MacCormick}}}, \bibinfo {author}
  {\bibfnamefont {X.}~\bibnamefont {{Xu}}}, \ and\ \bibinfo {author}
  {\bibfnamefont {B.}~\bibnamefont {{Pfeiffer}}},\ }\href {\doibase
  10.1088/1674-1137/36/12/002} {\bibfield  {journal} {\bibinfo  {journal}
  {\cpc}\ }\textbf {\bibinfo {volume} {36}},\ \bibinfo {pages} {002} (\bibinfo
  {year} {2012})}\BibitemShut {NoStop}%
\bibitem [{\citenamefont {{Goriely}}\ \emph {et~al.}(2013)\citenamefont
  {{Goriely}}, \citenamefont {{Chamel}},\ and\ \citenamefont
  {{Pearson}}}]{gcp2013}%
  \BibitemOpen
  \bibfield  {author} {\bibinfo {author} {\bibfnamefont {S.}~\bibnamefont
  {{Goriely}}}, \bibinfo {author} {\bibfnamefont {N.}~\bibnamefont {{Chamel}}},
  \ and\ \bibinfo {author} {\bibfnamefont {J.~M.}\ \bibnamefont {{Pearson}}},\
  }\href {\doibase 10.1103/PhysRevC.88.024308} {\bibfield  {journal} {\bibinfo
  {journal} {\prc}\ }\textbf {\bibinfo {volume} {88}},\ \bibinfo {eid} {024308}
  (\bibinfo {year} {2013})}\BibitemShut {NoStop}%
\bibitem [{\citenamefont {{Col{\`o}}}\ \emph {et~al.}(2004)\citenamefont
  {{Col{\`o}}}, \citenamefont {{van Giai}}, \citenamefont {{Meyer}},
  \citenamefont {{Bennaceur}},\ and\ \citenamefont {{Bonche}}}]{colo2004}%
  \BibitemOpen
  \bibfield  {author} {\bibinfo {author} {\bibfnamefont {G.}~\bibnamefont
  {{Col{\`o}}}}, \bibinfo {author} {\bibfnamefont {N.}~\bibnamefont {{van
  Giai}}}, \bibinfo {author} {\bibfnamefont {J.}~\bibnamefont {{Meyer}}},
  \bibinfo {author} {\bibfnamefont {K.}~\bibnamefont {{Bennaceur}}}, \ and\
  \bibinfo {author} {\bibfnamefont {P.}~\bibnamefont {{Bonche}}},\ }\href
  {\doibase 10.1103/PhysRevC.70.024307} {\bibfield  {journal} {\bibinfo
  {journal} {\prc}\ }\textbf {\bibinfo {volume} {70}},\ \bibinfo {eid} {024307}
  (\bibinfo {year} {2004})}\BibitemShut {NoStop}%
\bibitem [{\citenamefont {{Li}}\ and\ \citenamefont
  {{Schulze}}(2008)}]{ls2008}%
  \BibitemOpen
  \bibfield  {author} {\bibinfo {author} {\bibfnamefont {Z.~H.}\ \bibnamefont
  {{Li}}}\ and\ \bibinfo {author} {\bibfnamefont {H.-J.}\ \bibnamefont
  {{Schulze}}},\ }\href {\doibase 10.1103/PhysRevC.78.028801} {\bibfield
  {journal} {\bibinfo  {journal} {\prc}\ }\textbf {\bibinfo {volume} {78}},\
  \bibinfo {eid} {028801} (\bibinfo {year} {2008})}\BibitemShut {NoStop}%
\bibitem [{\citenamefont {{Akmal}}\ \emph {et~al.}(1998)\citenamefont
  {{Akmal}}, \citenamefont {{Pandharipande}},\ and\ \citenamefont
  {{Ravenhall}}}]{apr1998}%
  \BibitemOpen
  \bibfield  {author} {\bibinfo {author} {\bibfnamefont {A.}~\bibnamefont
  {{Akmal}}}, \bibinfo {author} {\bibfnamefont {V.~R.}\ \bibnamefont
  {{Pandharipande}}}, \ and\ \bibinfo {author} {\bibfnamefont {D.~G.}\
  \bibnamefont {{Ravenhall}}},\ }\href {\doibase 10.1103/PhysRevC.58.1804}
  {\bibfield  {journal} {\bibinfo  {journal} {\prc}\ }\textbf {\bibinfo
  {volume} {58}},\ \bibinfo {pages} {1804} (\bibinfo {year}
  {1998})}\BibitemShut {NoStop}%
\bibitem [{\citenamefont {{Pearson}}\ \emph {et~al.}(2011)\citenamefont
  {{Pearson}}, \citenamefont {{Goriely}},\ and\ \citenamefont
  {{Chamel}}}]{pearson2011}%
  \BibitemOpen
  \bibfield  {author} {\bibinfo {author} {\bibfnamefont {J.~M.}\ \bibnamefont
  {{Pearson}}}, \bibinfo {author} {\bibfnamefont {S.}~\bibnamefont
  {{Goriely}}}, \ and\ \bibinfo {author} {\bibfnamefont {N.}~\bibnamefont
  {{Chamel}}},\ }\href {\doibase 10.1103/PhysRevC.83.065810} {\bibfield
  {journal} {\bibinfo  {journal} {\prc}\ }\textbf {\bibinfo {volume} {83}},\
  \bibinfo {eid} {065810} (\bibinfo {year} {2011})}\BibitemShut {NoStop}%
\bibitem [{\citenamefont {{Pearson}}\ \emph {et~al.}(2012)\citenamefont
  {{Pearson}}, \citenamefont {{Chamel}}, \citenamefont {{Goriely}},\ and\
  \citenamefont {{Ducoin}}}]{pcgd2012}%
  \BibitemOpen
  \bibfield  {author} {\bibinfo {author} {\bibfnamefont {J.~M.}\ \bibnamefont
  {{Pearson}}}, \bibinfo {author} {\bibfnamefont {N.}~\bibnamefont {{Chamel}}},
  \bibinfo {author} {\bibfnamefont {S.}~\bibnamefont {{Goriely}}}, \ and\
  \bibinfo {author} {\bibfnamefont {C.}~\bibnamefont {{Ducoin}}},\ }\href
  {\doibase 10.1103/PhysRevC.85.065803} {\bibfield  {journal} {\bibinfo
  {journal} {\prc}\ }\textbf {\bibinfo {volume} {85}},\ \bibinfo {eid} {065803}
  (\bibinfo {year} {2012})}\BibitemShut {NoStop}%
\bibitem [{\citenamefont {{Goriely}}\ \emph {et~al.}(2010)\citenamefont
  {{Goriely}}, \citenamefont {{Chamel}},\ and\ \citenamefont
  {{Pearson}}}]{gcp2010}%
  \BibitemOpen
  \bibfield  {author} {\bibinfo {author} {\bibfnamefont {S.}~\bibnamefont
  {{Goriely}}}, \bibinfo {author} {\bibfnamefont {N.}~\bibnamefont {{Chamel}}},
  \ and\ \bibinfo {author} {\bibfnamefont {J.~M.}\ \bibnamefont {{Pearson}}},\
  }\href {\doibase 10.1103/PhysRevC.82.035804} {\bibfield  {journal} {\bibinfo
  {journal} {\prc}\ }\textbf {\bibinfo {volume} {82}},\ \bibinfo {eid} {035804}
  (\bibinfo {year} {2010})}\BibitemShut {NoStop}%
\bibitem [{\citenamefont {{Friedman}}\ and\ \citenamefont
  {{Pandharipande}}(1981)}]{Friedman1981}%
  \BibitemOpen
  \bibfield  {author} {\bibinfo {author} {\bibfnamefont {B.}~\bibnamefont
  {{Friedman}}}\ and\ \bibinfo {author} {\bibfnamefont {V.~R.}\ \bibnamefont
  {{Pandharipande}}},\ }\href {\doibase 10.1016/0375-9474(81)90649-7}
  {\bibfield  {journal} {\bibinfo  {journal} {\nphysa}\ }\textbf {\bibinfo
  {volume} {361}},\ \bibinfo {pages} {502} (\bibinfo {year}
  {1981})}\BibitemShut {NoStop}%
\bibitem [{\citenamefont {{Chamel}}\ \emph {et~al.}(2011)\citenamefont
  {{Chamel}}, \citenamefont {{Fantina}}, \citenamefont {{Pearson}},\ and\
  \citenamefont {{Goriely}}}]{chamel2011}%
  \BibitemOpen
  \bibfield  {author} {\bibinfo {author} {\bibfnamefont {N.}~\bibnamefont
  {{Chamel}}}, \bibinfo {author} {\bibfnamefont {A.~F.}\ \bibnamefont
  {{Fantina}}}, \bibinfo {author} {\bibfnamefont {J.~M.}\ \bibnamefont
  {{Pearson}}}, \ and\ \bibinfo {author} {\bibfnamefont {S.}~\bibnamefont
  {{Goriely}}},\ }\href {\doibase 10.1103/PhysRevC.84.062802} {\bibfield
  {journal} {\bibinfo  {journal} {\prc}\ }\textbf {\bibinfo {volume} {84}},\
  \bibinfo {eid} {062802} (\bibinfo {year} {2011})}\BibitemShut {NoStop}%
\bibitem [{\citenamefont {{Fuchs}}\ \emph {et~al.}(2001)\citenamefont
  {{Fuchs}}, \citenamefont {{Faessler}}, \citenamefont {{Zabrodin}},\ and\
  \citenamefont {{Zheng}}}]{fuchs2001}%
  \BibitemOpen
  \bibfield  {author} {\bibinfo {author} {\bibfnamefont {C.}~\bibnamefont
  {{Fuchs}}}, \bibinfo {author} {\bibfnamefont {A.}~\bibnamefont {{Faessler}}},
  \bibinfo {author} {\bibfnamefont {E.}~\bibnamefont {{Zabrodin}}}, \ and\
  \bibinfo {author} {\bibfnamefont {Y.-M.}\ \bibnamefont {{Zheng}}},\ }\href
  {\doibase 10.1103/PhysRevLett.86.1974} {\bibfield  {journal} {\bibinfo
  {journal} {\prl}\ }\textbf {\bibinfo {volume} {86}},\ \bibinfo {pages} {1974}
  (\bibinfo {year} {2001})}\BibitemShut {NoStop}%
\bibitem [{\citenamefont {{Sturm}}\ \emph {et~al.}(2001)\citenamefont
  {{Sturm}}, \citenamefont {{B{\"o}ttcher}}, \citenamefont {{D{\c{e}}bowski}},
  \citenamefont {{F{\"o}rster}}, \citenamefont {{Grosse}}, \citenamefont
  {{Koczo{\'n}}}, \citenamefont {{Kohlmeyer}}, \citenamefont {{Laue}},
  \citenamefont {{Mang}}, \citenamefont {{Naumann}}, \citenamefont
  {{Oeschler}}, \citenamefont {{P{\"u}hlhofer}}, \citenamefont {{Schwab}},
  \citenamefont {{Senger}}, \citenamefont {{Shin}}, \citenamefont {{Speer}},
  \citenamefont {{Str{\"o}bele}}, \citenamefont {{Sur{\'o}wka}}, \citenamefont
  {{Uhlig}}, \citenamefont {{Wagner}},\ and\ \citenamefont
  {{Walu{\'s}}}}]{sturm2001}%
  \BibitemOpen
  \bibfield  {author} {\bibinfo {author} {\bibfnamefont {C.}~\bibnamefont
  {{Sturm}}}, \bibinfo {author} {\bibfnamefont {I.}~\bibnamefont
  {{B{\"o}ttcher}}}, \bibinfo {author} {\bibfnamefont {M.}~\bibnamefont
  {{D{\c{e}}bowski}}}, \bibinfo {author} {\bibfnamefont {A.}~\bibnamefont
  {{F{\"o}rster}}}, \bibinfo {author} {\bibfnamefont {E.}~\bibnamefont
  {{Grosse}}}, \bibinfo {author} {\bibfnamefont {P.}~\bibnamefont
  {{Koczo{\'n}}}}, \bibinfo {author} {\bibfnamefont {B.}~\bibnamefont
  {{Kohlmeyer}}}, \bibinfo {author} {\bibfnamefont {F.}~\bibnamefont {{Laue}}},
  \bibinfo {author} {\bibfnamefont {M.}~\bibnamefont {{Mang}}}, \bibinfo
  {author} {\bibfnamefont {L.}~\bibnamefont {{Naumann}}}, \bibinfo {author}
  {\bibfnamefont {H.}~\bibnamefont {{Oeschler}}}, \bibinfo {author}
  {\bibfnamefont {F.}~\bibnamefont {{P{\"u}hlhofer}}}, \bibinfo {author}
  {\bibfnamefont {E.}~\bibnamefont {{Schwab}}}, \bibinfo {author}
  {\bibfnamefont {P.}~\bibnamefont {{Senger}}}, \bibinfo {author}
  {\bibfnamefont {Y.}~\bibnamefont {{Shin}}}, \bibinfo {author} {\bibfnamefont
  {J.}~\bibnamefont {{Speer}}}, \bibinfo {author} {\bibfnamefont
  {H.}~\bibnamefont {{Str{\"o}bele}}}, \bibinfo {author} {\bibfnamefont
  {G.}~\bibnamefont {{Sur{\'o}wka}}}, \bibinfo {author} {\bibfnamefont
  {F.}~\bibnamefont {{Uhlig}}}, \bibinfo {author} {\bibfnamefont
  {A.}~\bibnamefont {{Wagner}}}, \ and\ \bibinfo {author} {\bibfnamefont
  {W.}~\bibnamefont {{Walu{\'s}}}},\ }\href {\doibase
  10.1103/PhysRevLett.86.39} {\bibfield  {journal} {\bibinfo  {journal} {\prl}\
  }\textbf {\bibinfo {volume} {86}},\ \bibinfo {pages} {39} (\bibinfo {year}
  {2001})}\BibitemShut {NoStop}%
\bibitem [{\citenamefont {{Hartnack}}\ \emph {et~al.}(2006)\citenamefont
  {{Hartnack}}, \citenamefont {{Oeschler}},\ and\ \citenamefont
  {{Aichelin}}}]{hartnack2006}%
  \BibitemOpen
  \bibfield  {author} {\bibinfo {author} {\bibfnamefont {C.}~\bibnamefont
  {{Hartnack}}}, \bibinfo {author} {\bibfnamefont {H.}~\bibnamefont
  {{Oeschler}}}, \ and\ \bibinfo {author} {\bibfnamefont {J.}~\bibnamefont
  {{Aichelin}}},\ }\href {\doibase 10.1103/PhysRevLett.96.012302} {\bibfield
  {journal} {\bibinfo  {journal} {\prl}\ }\textbf {\bibinfo {volume} {96}},\
  \bibinfo {eid} {012302} (\bibinfo {year} {2006})}\BibitemShut {NoStop}%
\bibitem [{\citenamefont {{Xiao}}\ \emph {et~al.}(2009)\citenamefont {{Xiao}},
  \citenamefont {{Li}}, \citenamefont {{Chen}}, \citenamefont {{Yong}},\ and\
  \citenamefont {{Zhang}}}]{xiao2009}%
  \BibitemOpen
  \bibfield  {author} {\bibinfo {author} {\bibfnamefont {Z.}~\bibnamefont
  {{Xiao}}}, \bibinfo {author} {\bibfnamefont {B.-A.}\ \bibnamefont {{Li}}},
  \bibinfo {author} {\bibfnamefont {L.-W.}\ \bibnamefont {{Chen}}}, \bibinfo
  {author} {\bibfnamefont {G.-C.}\ \bibnamefont {{Yong}}}, \ and\ \bibinfo
  {author} {\bibfnamefont {M.}~\bibnamefont {{Zhang}}},\ }\href {\doibase
  10.1103/PhysRevLett.102.062502} {\bibfield  {journal} {\bibinfo  {journal}
  {\prl}\ }\textbf {\bibinfo {volume} {102}},\ \bibinfo {eid} {062502}
  (\bibinfo {year} {2009})}\BibitemShut {NoStop}%
\bibitem [{\citenamefont {{Lynch}}\ \emph {et~al.}(2009)\citenamefont
  {{Lynch}}, \citenamefont {{Tsang}}, \citenamefont {{Zhang}}, \citenamefont
  {{Danielewicz}}, \citenamefont {{Famiano}}, \citenamefont {{Li}},\ and\
  \citenamefont {{Steiner}}}]{lynch2009}%
  \BibitemOpen
  \bibfield  {author} {\bibinfo {author} {\bibfnamefont {W.~G.}\ \bibnamefont
  {{Lynch}}}, \bibinfo {author} {\bibfnamefont {M.~B.}\ \bibnamefont
  {{Tsang}}}, \bibinfo {author} {\bibfnamefont {Y.}~\bibnamefont {{Zhang}}},
  \bibinfo {author} {\bibfnamefont {P.}~\bibnamefont {{Danielewicz}}}, \bibinfo
  {author} {\bibfnamefont {M.}~\bibnamefont {{Famiano}}}, \bibinfo {author}
  {\bibfnamefont {Z.}~\bibnamefont {{Li}}}, \ and\ \bibinfo {author}
  {\bibfnamefont {A.~W.}\ \bibnamefont {{Steiner}}},\ }\href {\doibase
  10.1016/j.ppnp.2009.01.001} {\bibfield  {journal} {\bibinfo  {journal}
  {Progress in Particle and Nuclear Physics}\ }\textbf {\bibinfo {volume}
  {62}},\ \bibinfo {pages} {427} (\bibinfo {year} {2009})}\BibitemShut
  {NoStop}%
\bibitem [{\citenamefont {{Douchin}}\ and\ \citenamefont
  {{Haensel}}(2001)}]{douchin2001}%
  \BibitemOpen
  \bibfield  {author} {\bibinfo {author} {\bibfnamefont {F.}~\bibnamefont
  {{Douchin}}}\ and\ \bibinfo {author} {\bibfnamefont {P.}~\bibnamefont
  {{Haensel}}},\ }\href {\doibase 10.1051/0004-6361:20011402} {\bibfield
  {journal} {\bibinfo  {journal} {\aap}\ }\textbf {\bibinfo {volume} {380}},\
  \bibinfo {pages} {151} (\bibinfo {year} {2001})}\BibitemShut {NoStop}%
\end{thebibliography}%

\end{document}